\documentclass[showpacs,showkeys,amsmath,amssymb,floatfix,aps,twocolumn]{revtex4}
\usepackage{graphicx}% Include figure files
\usepackage{dcolumn}% Align table columns on decimal point
\usepackage{bm}% bold math
\usepackage{upgreek}
\usepackage{times}
\usepackage{color}

\newcommand{\bOmega}{\bm \Omega}
\newcommand{\bomega}{\bm \omega}
\newcommand{\bPhi}{\bm \Phi}
\newcommand{\btau}{\bm \tau}
\newcommand{\bbeta}{\bm \beta}
\newcommand{\balpha}{\bm \alpha}
\newcommand{\bnabla}{{\bm \nabla}}
\newcommand{\bDelta}{{\bm \Delta}}
\newcommand{\Deltaf}{\overline{\Delta}}
\newcommand{\Deltah}{\widehat{\Delta}}
\newcommand{\pd}[2]{\frac{\partial #1}{\partial #2}}
\newcommand{\bu}{\mathbf u}
\newcommand{\uf}{\overline{\mathbf u}}
\newcommand{\vf}{\overline{u}}
\newcommand{\uff}{\overline{\overline{\mathbf u}}}
\newcommand{\uprime}{{\mathbf u}'}
\newcommand{\fuf}{\underline{\overline{\mathbf u}}}
\newcommand{\uuf}{\overline{\mathbf u \mathbf u}}
\newcommand{\pf}{\overline{p}}
\newcommand{\fpf}{\underline{\overline{p}}}
\newcommand{\Pf}{\overline{P}}
\newcommand{\nl}{nonlinear}

\newcommand{\dd}{\text{d}}
\newcommand{\bM}{{\mathbf M}}
\newcommand{\bC}{{\mathbf C}}

\newcommand{\bD}{{\mathbf D}}
\newcommand{\bK}{{\mathbf K}}
\newcommand{\bS}{{\mathbf S}}
\newcommand{\bSf}{\overline{\mathbf S}}
\newcommand{\Sf}{\overline{S}}
\newcommand{\bT}{{\mathbf T}}
\newcommand{\bL}{{\mathbf L}}

\newcommand{\cL}{{\mathcal L}}
\newcommand{\cR}{{\mathcal R}}
\newcommand{\cC}{{\mathcal C}}
\newcommand{\cH}{{\mathcal H}}
\newcommand{\cG}{{\mathcal G}}
\newcommand{\cK}{{\mathcal K}}
\newcommand{\cV}{{\mathcal V}}

\newcommand{\cAs}{{\mathcal A}_{\textrm{sgs}}}

\newcommand{\ddp}[2]{\frac{\partial #1}{\partial #2}}
\newcommand{\mU}{\langle U\rangle}
\newcommand{\mV}{\langle V\rangle}
\newcommand{\mW}{\langle W\rangle}
\newcommand{\mP}{\langle P\rangle}
\newcommand{\urms}{\sqrt{\langle u^2\rangle}}
\newcommand{\vrms}{\sqrt{\langle v^2\rangle}}
\newcommand{\wrms}{\sqrt{\langle w^2\rangle}}

\newcommand{\CS}{C_{\text{S}}}
\newcommand{\Cd}{C_{\dd}}
\newcommand{\m}[1]{\langle #1 \rangle}
\newcommand{\eps}{\varepsilon}
\newcommand{\xx}{{\mathbf x}}
\newcommand{\mnusgs}{\langle \nu_{\textrm{sgs}} \rangle}
\newcommand{\nusgs}{\nu_{\textrm{sgs}}}
\newcommand{\etal}{\textit{et al.}}

\begin{document}
\preprint{APS/123-QED}

\title{Large-eddy simulation of the flow in a lid-driven cubical cavity}

\author{Roland Bouffanais}
\email{roland.bouffanais@epfl.ch}
\author{Michel O. Deville}
\email{michel.deville@epfl.ch}
\affiliation{%
\'Ecole Polytechnique F\'ed\'erale de Lausanne\\
STI--ISE--LIN\\
Station 9\\
CH-1015 Lausanne (Switzerland)
}%

\author{Emmanuel Leriche}
\email{emmanuel.leriche@univ-st-etienne.fr}
\affiliation{%
Laboratoire de Math\'ematiques de l'Universit\'e de Saint-\'Etienne (LaMUSE)\\
Facult\'e des Sciences et Techniques -- Universit\'e Jean-Monnet\\
23 rue du Docteur Paul Michelon\\
F-42023, Saint-\'Etienne (France)
}%

\date{\today}% It is always \today, today,

\begin{abstract}
Large-eddy simulations of the turbulent flow in a lid-driven cubical cavity have been carried out at a Reynolds number of $12\,000$ using spectral element methods. Two distinct subgrid-scales models, namely a dynamic Smagorinsky model and a dynamic mixed model, have been both implemented and used to perform long-lasting simulations required by the relevant time scales of the flow. All filtering levels make use of explicit filters applied in the physical space (on an element-by-element approach) and spectral (modal) spaces. The two subgrid-scales models are validated and compared to available experimental and numerical reference results, showing very good agreement. Specific features of lid-driven cavity flow in the turbulent regime, such as inhomogeneity of turbulence, turbulence production near the downstream corner eddy, small-scales localization and helical properties are investigated and discussed in the large-eddy simulation framework. Time histories of quantities such as the total energy, total turbulent kinetic energy or helicity exhibit different evolutions but only after a relatively long transient period. However, the average values remain extremely close.

\end{abstract}

\pacs{47.25, 47.55}% PACS, the Physics and Astronomy  % Classification Scheme.
\keywords{LES, lid-driven cavity, spectral element, explicit filtering, helicity}%Use showkeys class option if keyword
\maketitle

%================ 1 Introduction =============================================================
\section{Introduction}
\label{sec:introduction}
The study of a lid-driven flow of a Newtonian fluid in a rectangular three-dimensional cavity is of particular interest in view not only of the simplicity of the flow geometry but also the richness of the fluid flow physics manifested by multiple counter-rotating recirculating regions at the corners of the cavity depending on the Reynolds number, Taylor-G\"ortler-like (TGL) vortices, flow bifurcations and transition to turbulence. This flow structure is now well documented thanks to a relatively rich literature reporting both computational and experimental studies. A comprehensive review of the fluid mechanics of driven cavities is provided by Shankar and Deshpande in \cite{shankar00:_fluid_mechan_driven_cavit}. 

In the present paper, our focus resides in relatively high-Reynolds-number and three-dimensional lid-driven cubical cavity flows. At Reynolds number higher than a critical value comprised between $2\,000$ and $3\,000$, an instability appears in the vicinity of the downstream corner eddy \cite{iwatsu89:_numer,leriche99:_direc_cheby,albensoeder05:_accur}. As the Reynolds number further increases, turbulence develops near the cavity walls, and at Reynolds number higher than $10\,000$, the flow near the downstream corner eddy becomes fully turbulent. The highest Reynolds number attained was $\textrm{12\,000}$ by direct numerical simulation (DNS) performed by Leriche and Gavrilakis \cite{leriche00:_direc} and $\textrm{10\,000}$ experimentally by Koseff and Street \cite{koseff84} and Prasad and Koseff \cite{prasad89:_reynol}. In the literature, papers using the lid-driven cavity problem as a benchmark test case to evaluate the performance of numerical algorithms are proliferating, but are often limited to two space dimensions or to Reynolds numbers below $10\,000$. More recently, one may however notice the important developments of novel and more physical numerical methods applied to the lid-driven cavity flow such as molecular dynamics by Chen and Lin in \cite{chen05:_tip4p} and also the lattice--Boltzmann model applied by He \etal\ in \cite{he04:_bgk}.

The results reported herein correspond to the numerical simulation of the flow in a lid-driven cubical cavity at the Reynolds number of $12\,000$ placing us in the locally-turbulent regime. The spatial discretization relies on spectral element methods (SEM) which have been mainly applied to the DNS of fluid flow problems at low and moderate Reynolds numbers. With the advent of more powerful computers, especially through cluster technology, higher Re values seem to fall within the realm of feasibility. However, despite their high accuracy, spectral element methods are still far from reaching industrial applications that involve developed turbulence at Re values of the order of $10^6-10^7$. The reason for that dismal performance is that a resolved DNS including all scales from the largest structures to Kolmogorov scales, needs a number of degrees of freedom that grows like $\textrm{Re}^{9/4}$. Therefore with increasing Re, one has to increase the number of elements, $E$, and the degree, $N$, of the polynomial spaces. This places the computational load far out of the reach of present day computers. Large-eddy simulation (LES) represents an alternative to DNS insofar that it involves less degrees of freedom because the behavior of the small scales are modeled. The numerical simulations presented in this paper encompass two different LES based on two distinct subgrid-scales modeling both using an eddy-viscosity assumption, and one using in addition a mixed model relying on the scale-similarity hypothesis, similarly to Zang \etal\ in \cite{zang92:_applic,zang93} for $\textrm{Re}=10\,000$. Compared to the previous works of Zang \etal, the two LES reported here offer simulation length ten times larger therefore increasing the accuracy of the ensemble averaging and more importantly allowing to capture intermittent turbulent production. These events lead to the determination of large eddies suggested to be mainly responsible for the turbulence production near the downstream corner eddy.

Unlike low-order methods such as finite volumes or finite differences, spectral and spectral element methods allow a complete decoupling between the mathematical formulation, the subgrid modeling, the numerical technique and the filtering technique, which are introduced successively in Sec. \ref{sec:model}. Specifically, we are first seeking to validate the two subgrid-scale models introduced in Sec. \ref{sec:model} which rely on explicit filtering techniques specific to spectral element spatial discretization. Sec. \ref{sec:phys-comp-param} presents a short, but comprehensive validation procedure. In Sec. \ref{sec:turbulence} emphasis is put on characterizing the turbulent flow in its locally-turbulent regime. Fundamental features are qualitatively and quantitatively investigated such as the inhomogeneity of the turbulence, the turbulence production in the downstream-corner-eddy region, the small-scales turbulent structures in the cavity flow and finally the peculiar helical properties.

%================ 2 The model and the numerical technique ====================================
\section{The model and numerical technique}
\label{sec:model}

%-------- 1 Mathematical Modeling -------------
\subsection{Mathematical modeling}
\label{sec:math-model}
\begin{figure}[htbp]
  \hspace{-6cm}
  \input{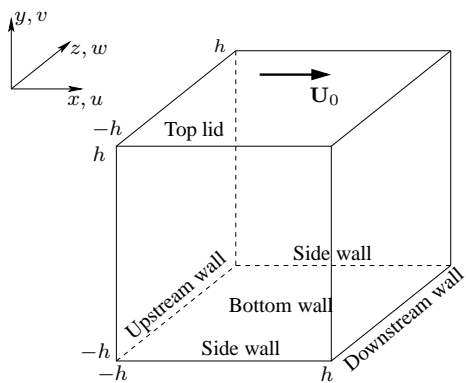}\hspace{10cm}
    \caption{Sketch of the geometry of the lid-driven cubical cavity.}\label{fig:sketch}
\end{figure}
The fluid enclosed in the cavity is assumed to be incompressible, Newtonian with uniform density and temperature. The flow is governed by the Navier--Stokes equations inside the fluid domain denoted by $\cV=[-h,+h]^3$ with no-slip boundary conditions on every cavity walls, except on the top, see Fig. \ref{fig:sketch}. The flow is driven by imposing a prescribed velocity distribution with nonzero mean on the ``top'' wall---named lid in the sequel---with the velocity field maintained everywhere parallel to a given direction. The details regarding the imposition of this Dirichlet boundary condition for the velocity field at the lid is discussed in Sec. \ref{sec:lid}. As the flow presents turbulent zones coexisting with laminar regions, the numerical simulation incorporates the mathematical models involved by the large-eddy simulation method in order to resolve the complex dynamics of the flow. As a consequence, the governing equations of the large-eddy simulations are the filtered Navier--Stokes equations. Large-scale quantities, designated in the sequel by an ``overbar'', are obtained by a filtering procedure on the computational domain $\widehat{\cV}=[-1,+1]^3$ using $h$ for the non-dimensionalization of lengths. The application of a low-pass inhomogeneous and anisotropic spatial filter to the Navier--Stokes equations in the Eulerian velocity--pressure formulation and in convective form for the \nl{} term yields
\begin{align}
  \label{eq:filtered_NS}
  \pd{\uf}{t} + \uf \cdot \bnabla \uf &= -\bnabla \pf +\nu \bDelta \uf - \bnabla \cdot \btau,\\
  \bnabla \cdot \uf &=0,
\end{align}
in which $\uf$ is the filtered velocity field, $t$ denotes the time, $\pf=\Pf/\rho$ is the filtered static pressure and $\nu$ the assuredly constant, uniform kinematic viscosity. The symbols $\bnabla$ and $\bDelta$ represent the nabla and Laplacian operators, respectively. The subgrid-scale (SGS) stress tensor $\btau$ is given by
\begin{equation}
  \label{eq:SGS_definition}
  \btau = \uuf - \uf\, \uf,
\end{equation}
and accounts for the effects of the unresolved- or small-scales on the dynamics of the resolved- or large-scales \cite{sagaut03:_large}.

%-------- 2 Subgrid-scale models -------------
\subsection{Subgrid-scale models}
\label{sec:subgrid-modeling}

\subsubsection{Under-resolved direct numerical simulation}
In the same framework as it prevails among the practioners, one can resort to the DNS computations without any LES model, but with the nodal filtering technique described in Sec. \ref{sec:nodal-filter} to let the numerical method dissipate locally the high-wave-number modes introduced by the insufficient space discretization. Such an approach corresponds to an under-resolved DNS (UDNS).
\subsubsection{Smagorinsky model}
\label{sec:smagorinsky-model}
The SGS Smagorinsky model (SM) \cite{smagorinsky63:_gener} uses the concept of turbulent viscosity and assumes that the small scales are in equilibrium, balancing energy production and dissipation. This yields the following expression for the eddy viscosity
\begin{equation} 
\nu_{\textrm{sgs}} = (\CS\Deltaf)^2  |\bSf|, \label{eq:nut_smag}
\end{equation}
where $|\bSf|=(2\Sf_{ij}\Sf_{ij})^{1/2}$ is the magnitude of the filtered strain-rate-tensor with $\Sf_{ij}=1/2(\partial \vf_i/\partial x_j + \partial \vf_j/\partial x_i)$, $\CS$ is the Smagorinsky constant and $\Deltaf$ the  filter width. The Smagorinsky model has several drawbacks. The most severe one is the constant value of $\CS$ during the computation which produces too much dissipation. Furthermore the SM does not provide the modeler with backscattering where kinetic energy is transferred from small scales to larger scales in an inverse-cascading process.

\subsubsection{Dynamic Smagorinsky model}
\label{sec:dynamic-model}
The dynamic Smagorinsky model (DSM) proposed by Germano \etal\ \cite{germano91} overcomes the difficulty of constant $\CS$, by allowing it to become dependent of space and time. Now we have a dynamic parameter $\Cd=\Cd({\xx}, t)$. Let us introduce a test-filter length scale $\Deltah$ that is larger than the grid length scale $\Deltaf$ (e.g. $\Deltah=2\Deltaf$). Using the information provided by those two filters and assuming that in the inertial range of the turbulence energy spectrum, the statistical self-similarity applies, we can better determine the features of the SGS stress. As detailed in \cite{bouffanais05:_large}, with the test filter, the former LES Eq. \eqref{eq:filtered_NS} yields a dynamic parameter having the expression

\begin{equation}
\Cd=\frac{(\balpha-\widehat{\bbeta}):\bL^\dd}{(\balpha-\widehat{\bbeta}):(\balpha-\widehat{\bbeta})}, \label{eq:Cdyn}
\end{equation}
where
\begin{align}
 \bL     &=\widehat{\uf\,\uf}-\widehat{\uf} \, \widehat{\uf}, \\
 \balpha &= -2 \widehat{\Deltaf}^2|\widehat{\bSf}| \widehat{\bSf}, \\
 \bbeta  &= -2 \Deltaf^2 |\bSf| \bSf.
\end{align}
 The notation ``$:$'' in Eq. \eqref{eq:Cdyn} is used for inner tensor product (double contraction), and the upper index d denotes the deviatoric part of the tensor.

\subsubsection{Dynamic mixed model}
\label{sec:dynamic-mixed-model}
The dynamic mixed model \cite{zang93} introduced to tackle cavity flows is a blend of the mixed model of Bardina \etal\ \cite{bardina83:_improv} and the former dynamic Smagorinsky model. We notice that Bardina's scale similarity model is not an eddy-viscosity based model. Instead it belongs to the class of structural models \cite{sagaut03:_large} and relies on the scale-similarity principle. It produces almost no dissipation and for that reason needs to be used jointly with dissipative models such as the Smagorinsky model---Bardina's mixed model---or with the dynamic Smagorinsky model. The approach of Zang \etal\ \cite{zang93} was extended by Liu \etal\ \cite{liu94} who proposed a new similarity subgrid-scale model for incompressible flows, in which the subgrid stress tensor is assumed to be proportional to the resolved stress tensor. Vreman \etal\ \cite{vreman94} later modified the DMM formulation to remove a mathematical inconsistency by expressing the scale-similarity part of the sub-test-scale stress $\bT$---see \cite{bouffanais05:_large}---using only $\widehat{\uf}$. Salvetti and Banerjee \cite{salvetti97} and Horiuti \cite{horiuti97} extended the DMM to two distinct dynamic two-parameter models. Morinishi and Vasilyev \cite{morinishi01} recommended a modification to the dynamic two-parameter mixed model of Salvetti and Banerjee \cite{salvetti97} for large-eddy simulation of wall bounded turbulent flow. The works of Vreman \etal\ \cite{vreman97:_large} and Winckelmans \etal\ \cite{winckelmans01:_explic_smagor} also closely relate to the DMM approach. As mentioned by Morinishi and Vasilyev \cite{morinishi01} and Ghosal \cite{ghosal96}, the reliability of the results of large-eddy simulation is strongly affected by both the effectiveness of the subgrid scale model and the accuracy of the numerical method, particularly in the approximation of the non-linear convective term. As mentioned in Sec. \ref{sec:introduction}, the SEM is decoupled from the subgrid modeling and offers a high accuracy characteristics of spectral methods. Therefore, the present work focuses on the one-parameter type of dynamic mixed model DMM as introduced by Zang \etal\ \cite{zang93} for the lid-driven cavity flow. The modification suggested by Vreman \etal\ \cite{vreman94} was not implemented; \textit{a priori} tests with their modified DMM using samples from the DNS results by Leriche and Gavrilakis \cite{leriche00:_direc} showed no noticeable improvement over the DMM of Zang \etal\ in the subgrid stress correlations. Therefore increasing the computational expense by adding an additional filtering level operation as required by the modification of Vreman \etal \cite{vreman94}, seemed unjustified.

By decomposing the velocity field as
\begin{equation}
\bu =\uf+\uprime \label{eq:decomp},
\end{equation}
where $\uprime$ represents the subgrid-scale velocity field and by inserting in Eq. \eqref{eq:SGS_definition}, we can redefine the SGS stress as proposed by Germano \cite{germano86:_navier}
\begin{equation}
\btau=\cL+\cC+\cR, \label{eq:redef}
\end{equation}
where
\begin{align}
\cL & = \overline{\uf\,\,\uf}- \uff \,\,\uff, \nonumber \\
\cC & = \overline{\uf\, \uprime + \uprime\, \uf}-(\uff\,\, \overline{\uprime}+\overline{\uprime}\,\,\uff),\\
\cR & = \overline{\uprime\, \uprime}-\overline{\uprime}\,\,\overline{\uprime},\nonumber
\end{align}
are designated as the modified Leonard stress, the SGS cross term, and the modified SGS Reynolds stress, respectively. The modified Leonard term can be calculated by resolved quantities and corresponds essentially to the mixed model. The two other terms are unresolved residual stresses and are treated through the Smagorinsky model, see \cite{zang93} for greater details. Following the same dynamic procedure as in \cite{bouffanais05:_large} and Sec. \ref{sec:dynamic-model}, and with the same notations, one obtains a dynamic coefficient which reads
\begin{equation}
\Cd=\frac{(\balpha-\widehat{\bbeta}):(\cG^\dd-\bL^\dd)}{(\balpha-\widehat{\bbeta}):(\balpha-\widehat{\bbeta})}, \label{eq:Cdynmix}
\end{equation}
where
\begin{equation}
\cG  = \widehat{\uff\,\,\uff}-\widehat{\uff}\,\,\widehat{\uff}.
\end{equation}
The expression of the dynamic coefficient given in Eq. \eqref{eq:Cdynmix} for the dynamic mixed model is similar to the one for the dynamic model---see Eq. \eqref{eq:Cdyn}---the tensor $\bL^\dd$ being replaced by $\cG^\dd-\bL^\dd$.

%-------- 3 Numerical Technique -------------
\subsection{Numerical technique}
\label{sec:numerical-technique}
\subsubsection{Space discretization}
The numerical method treats Eqs. \eqref{eq:filtered_NS}--\eqref{eq:SGS_definition} within the weak Galerkin formulation framework.
The spatial discretization uses
Lagrange--Legendre polynomial interpolants. The reader is referred to the monograph by Deville \etal\ \cite{deville02:_high} for full details. The velocity and pressure are expressed in the $\mathbb{P}_N-\mathbb{P}_{N-2}$ functional spaces where $\mathbb{P}_N$ is the set of polynomials of degree lower than $N$ in each space direction. This spectral element method avoids the presence of spurious pressure modes as it was proved by Maday and Patera \cite{maday92:_nimes_n_stokes,maday89:_spect_navier_stokes}. The quadrature rules are based on a Gauss--Lobatto--Legendre (GLL) grid for the velocity nodes and a Gauss--Legendre grid (GL) for the pressure nodes. 

Borrowing the notation from \cite{deville02:_high}, the semi-discrete filtered Navier--Stokes equations resulting from space discretization are 
\begin{align}
\bM \frac{\dd \fuf}{\dd t}+ \bC \fuf +\nu \bK \fuf -\bD^T \fpf+\bD \underline{\btau}&=0,\label{eq:odes}\\
-\bD\fuf &=0\label{eq:constr}.
\end{align}
The diagonal mass matrix $\bM$ is composed of three blocks, namely the mass matrices $M$. The global vector $\fuf$ contains all the nodal velocity components while $\fpf$ is made of all nodal pressures. The matrices $\bK$, $\bD^T$, $\bD$ are the discrete Laplacian, gradient and divergence operators, respectively. The matrix operator $\bC$ represents the action of the non-linear term written in convective form $\fuf\cdot \bnabla$, on the velocity field and depends on $\fuf$ itself. The semi-discrete equations constitute a set of non-linear ordinary differential equations \eqref{eq:odes} subject to the incompressibility condition \eqref{eq:constr}.
\subsubsection{Time integration}
The state-of-the-art time integrators in spectral methods handle the viscous linear term and the pressure implicitly by a backward differentiation formula of order $2$ to avoid stability restrictions such that
\begin{equation}
 \nu \Delta t \leq C/N^4,
\end{equation}
while all non-linearities are computed explicitly, e.g. by a second order extrapolation method, under the CFL restriction $\overline u_{\text{max}}\Delta t \leq C/N^2$. Nonetheless, as the LES viscosity is not constant, we modify the standard time scheme in such a way that this space varying viscosity be handled explicitly as this was done e.g. in \cite{bouffanais05:_large,karamanos99:_large,blackburn03:_spect}. Let us define the effective viscosity as
\begin{equation}
\nu_{\text{eff}}=\nu +\nu_{\textrm{sgs}}=\nu_{\text{cst}}+(\nu_{\text{eff}}-\nu_{\text{cst}}),
\end{equation}
where $\nu_{\text{cst}}$ is the sum of the physical viscosity $\nu$ and the average of $\nu_{\textrm{sgs}}$ over the computational domain. The filtered semi-discrete Navier--Stokes equations become
\begin{align}
\bM \frac {\dd \fuf}{\dd t} +\nu_{\text{cst}}\bK \fuf -\bD^T \fpf&=-\bC \fuf +2\bD (\nu_{\text{eff}}-\nu_{\text{cst}}){\overline {\underline{\bS}}},\label{eq:odefin}\\
-\bD\fuf&=0,
\end{align}
 and the previous time splitting still applies. The  viscous explicit term on the right-hand side does not harm stability as the magnitude of the term $2\bD(\nu_{\text{eff}}-\nu_{\text{cst}}){\overline {\underline \bS}}$ is less than that of $\bC\fuf$.

The implicit part is solved by a generalized block LU decomposition with a pressure correction algorithm \cite{perot93,deville02:_high}. 

\subsubsection{The lid-filtered velocity distribution}
\label{sec:lid}
As already mentioned by Leriche and Gavrilakis in \cite{leriche00:_direc}, imposing a given velocity distribution on the lid of a cavity is neither an easy task experimentally nor numerically. Indeed imposing a constant lid velocity profile leads to a singularity (discontinuous behavior in the velocity boundary conditions) at the edges and at the corners of the lid, see Fig. \ref{fig:sketch}. Without adequate treatment, this discontinuous behavior will undermine the convergence and the accuracy of any numerical method in the vicinity of the lid. For the two-dimensional case, a well known solution (but with no physical relevance) is to subtract the most singular terms of the analytical expression of the local stream-function expansion near the lid corners. The extension of such procedure to three-dimensional cases is still missing even though several recent attempts are reported, see \cite{gomilko03:_stokes,albensoeder05:_accur}. In order to explicitly filter the discontinuous behavior, the constant lid velocity profile is regularized by the use of a high-order polynomial expansion which vanishes along its first derivatives at the lid edges and corners
\begin{align}
u(x,y=h,z,t)&= U_0\left[1-\left(x/h\right)^{18}\right]^2 \left[1-\left(z/h\right)^{18}\right]^2,\\\label{eq:bclid}
v(x,y=h,z,t)&=w(x,y=h,z,t)=0.\nonumber
\end{align}
This profile flattens very quickly near the lid edges and corners while away from them, it grows rapidly to a constant value over a short distance. The exact form and the polynomial order of the profile is
discussed in \cite{leriche99:_direc_cheby,leriche00:_direc}. The highest polynomial order of this distribution in both $x$- and $z$-direction is 36. Such high-order polynomial expansions lead to steep velocity gradients in the vicinity of the edges of the lid. The grid refinement, in terms of spectral element distribution near the lid will be presented in greater details in Sec. \ref{sec:phys-comp-param}. One of the constraint in the grid design is to ensure the proper resolution of the lid velocity distribution by the spectral element decomposition.

%-------- 4 Filtering Techniques -------------
\subsection{Filtering techniques}
\label{sec:filtering-techniques}
As spectral elements offer high spatial accuracy, we construct explicitly the filters using two spectral techniques. The first one is a nodal filter acting in physical space on the nodal velocity components (and pressure) to render the computations stable in the long range integration. The second method is designed as a modal filter and is carried out in spectral space in an element by element fashion. That filter corresponds specifically to the convolution kernel of the low-pass LES filtering.  
\subsubsection{Nodal filter}\label{sec:nodal-filter}
The nodal filter due to Fischer and Mullen \cite{fischer01:_filter_based_stabil_spect_elemen_method} is adequately suited to parallel spectral element computation. Introducing $h_{N,j}, j=0,\ldots,N$ the set of Lagrange--Legendre interpolant polynomials of degree $N$ based on the GLL grid nodes $\xi_{N,k}, k=0,\ldots,N$, the rectangular matrix operator $I^M_N$ of size $(M+1)\times(N+1)$ is such that 
\begin{equation}
(I^M_N)_{ij}=h_{N,j}(\xi_{M,i}).
\end{equation}
Therefore, the  matrix operator of order $N-1$
\begin{equation}
\Pi_{N-1}=I^N_{N-1}\,I^{N-1}_N,
\end{equation}
interpolates on the GLL grid of degree $N-1$ a function defined on the GLL grid of degree $N$ and transfers the data back to the original grid. This process eliminates the highest modes of the polynomial representation. The one-dimensional (1D) filter is given by the relation
\begin{equation}
\overline{u}=[\alpha \Pi_{N-1}+(1-\alpha)I^N_N]u.
\end{equation}
The LES version of the filter sets $\alpha=1$ and is given by
\begin{equation}
{\overline u}=I^N_M\,I^M_Nu,
\end{equation}
where $M$ is equal to $N-2$ or $N-3$. The three-dimensional (3D) extension results easily from the matrix tensor product properties of the filter. It is worth noting that by construction such nodal filter constitutes a projective filter, i.e $\uff=\uf$.
\subsubsection{Modal filter}\label{sec:modal-filter}
Here, the variable $u$ is approximated by a modal basis first proposed in the $p$-version of the finite elements and used by Boyd \cite{boyd98:_two_cheby_legen} as a filter technique. It is built up on the reference parent element as
\begin{align}\label{eq:modal-basis}
\phi_0&=\frac{1-\xi}{2},\quad \phi_1=\frac{1+\xi}{2},\\
\phi_k&=L_k(\xi)-L_{k-2}(\xi),\quad \nonumber
2 \leq  k \leq N.
\end{align} 
Conversely to the Lagrange--Legendre nodal basis used in our spectral element calculations, this modal basis \eqref{eq:modal-basis} forms a hierarchical set of polynomials allowing to define in an explicit and straightforward manner a low-pass filtering procedure. The one-to-one correspondence between the nodal Lagrangian basis and the $p$-representation yields 
\begin{equation}
u(\xi_i)=\sum^N_{k=0}{\widehat u_k} \phi_k(\xi_i),\label{eq:oto}
\end{equation}
which in matrix notation reads
\begin{equation}
\bu=\bPhi {\widehat {\bu}}.
\end{equation}
The low-pass filtering operation is performed in spectral space through a diagonal matrix $\mathbb T$ with components such that
\begin{equation}
T_0=T_1=1\qquad \text{and}\qquad T_k=\frac{1}{(1+(k/k_c)^2)}\qquad 2\leq k \leq N,
\end{equation}
where the cut-off value $k_c$ corresponds to $T_k=1/2$. The entire filtering process for a one-dimensional problem is given by 
\begin{equation}
\uf =G\star \bu=\bPhi{\mathbb T}\bPhi^{-1}\bu.
\end{equation}
The three-dimensional extension is again trivial by the matrix tensor product properties.

As noted in \cite{blackburn03:_spect}, the effect of such modal filter onto a given field expanded in the modal basis \eqref{eq:modal-basis} presents the interesting feature of maintaining the inter-element $\cC^0$-continuity. More rigorously, such $\cC^0$-continuity is enforced if and only if both $\phi_0$ and $\phi_1$ are not at all affected by the low-pass filtering, in other words if and only if $T_0=T_1=1$. Nevertheless, it has been observed that such $\cC^0$-breakage does not constitute a major issue for our simulations as it only affects the eddy viscosity field and other terms present into the modeling of the effect of the SGS tensor \eqref{eq:SGS_definition}, and which are not used directly for constructing a solution retaining the $\cC^0$-continuity feature. 

Such modal filter is invertible and consequently is not projective, i.e $\uff \neq \uf$.
\subsubsection{The filter length}
The decomposition of the computational domain into spectral elements of given sizes, within which a GLL distribution of grid points based on the polynomial degree is chosen, requires a specific definition of the filter length $\Delta$. In order to account for both the size of each spectral element and its value of the polynomial order, and following \cite{karamanos00:_spect_vanis_viscos_method_large_eddy_simul}, the filter length for a 1D spectral element method is chosen as
\begin{equation}\label{eq:1D_filter_length}
\Delta=\frac{s}{p},
\end{equation}
where $s$ is the element size and $p$ the highest polynomial degree in the spectral decomposition Eq. \eqref{eq:oto} that is the closest to the cut-off frequency $k_c$. In the particular context of the modal filter previously introduced, $p$ is such that
\begin{equation}
p=k,\,  \mbox{ such that } \inf_k (T_k) <T_{k_c}=1/2, \,\, k=0, \ldots, N.
\end{equation}
We notice that the filter length decreases when the element is refined. The straightforward extension of Eq. \eqref{eq:1D_filter_length} to our 3D problem using rectilinear elements leads to
\begin{equation} \label{eq:filter-length}
\Delta(x,y,z)=(\Delta_1(x)\Delta_2(y)\Delta_3(z))^{1/3}=\left(\frac{s_1}{p_1}\frac{s_2}{p_2}\frac{s_3}{p_3}\right)^{1/3}.
\end{equation}

%================ 3 Physical and computational parameters ====================================
\section{Physical and computational parameters}
\label{sec:phys-comp-param}
The different large-eddy simulations presented here refer to the same geometry---see Fig. \ref{fig:sketch}---and physical parameters as the direct numerical simulation (DNS) performed by Leriche and Gavrilakis \cite{leriche00:_direc}. The details relative to these parameters are gathered in Table \ref{tab:parameters}. The Reynolds number based on the maximum velocity on the lid was chosen to be $\textrm{Re}=U_02h/\nu=12\,000$. 

The kinetic energy is provided to the flow by the shear stress at the top lid through viscous diffusion. The amplitude of the Reynolds stress below the lid is negligible indicating that the flow under the lid is mainly laminar but transient. The momentum transfer from the lid induces a region of strong pressure in the upper corner of the downstream wall as the flow, mainly horizontal prior the corner, has to change direction and moves vertically downwards. This sharp turn dissipates energy in that region. Along the downstream wall the plunging flow behaves like a wall jet with a variable thickness. Near the symmetry plane the jet thickness  is reduced while it increases away from this plane. This jet, laminar and unsteady at the very beginning,  separates from the cavity wall at mid-height and grows as two elliptical jets on both sides of the symmetry plane. They hit the bottom wall where they produce turbulence. This turbulence is convected away by the main central vortex towards the upstream wall where the flow slows down and relaminarizes during the fluid rise.

\begin{figure}[htbp]
\includegraphics[width=6cm]{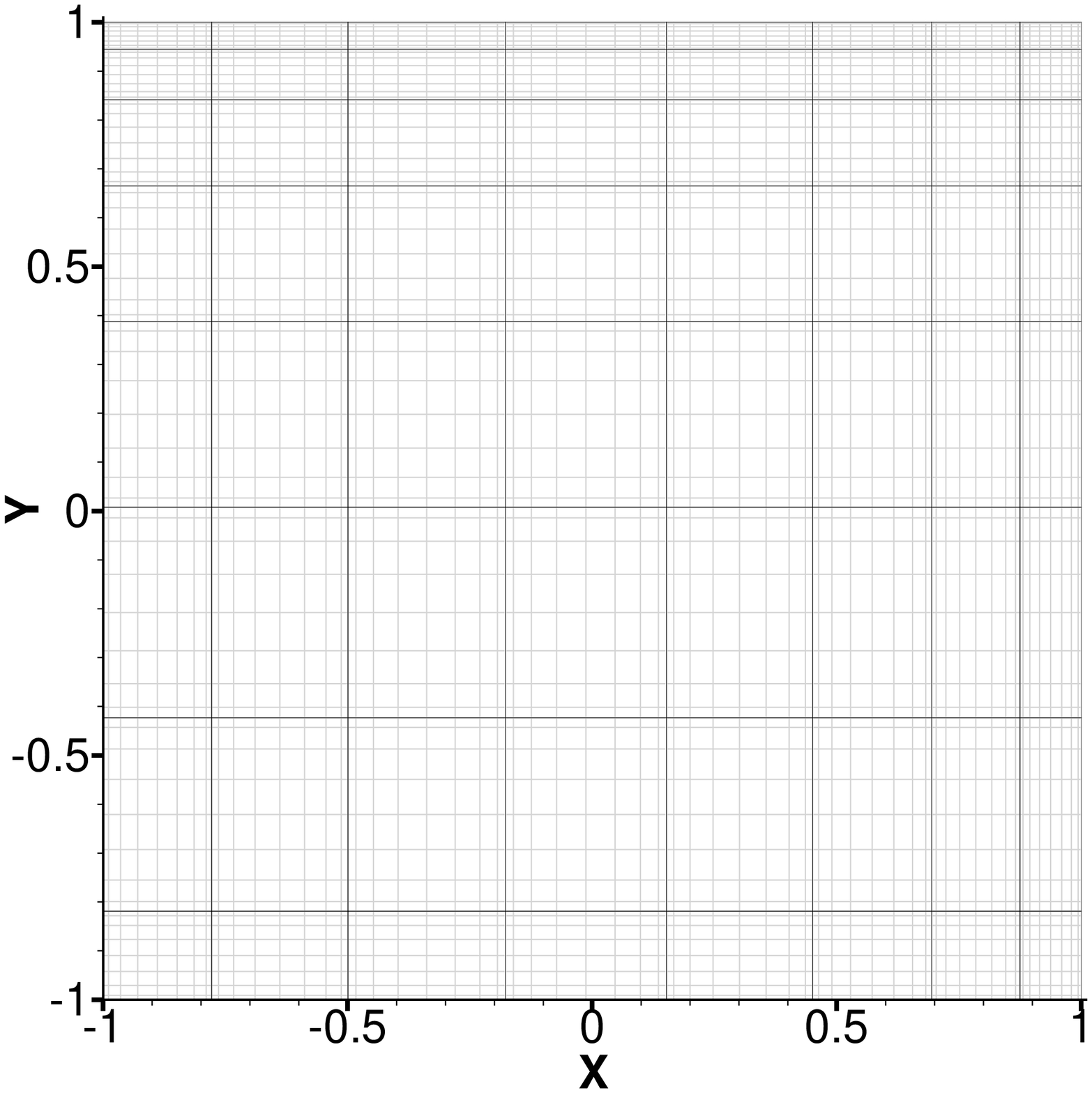}
\caption{Spectral element grid in the mid-plane $z/h=0$.}\label{fig:grid-LES}
\end{figure}

\begin{table}[htbp]
\begin{tabular}{ll}
  \hline\hline
    Domain size $(x,y,z)$ & $(2h,2h,2h)$ \\
    Wall positions & $x,\, y,\, z=\pm h$ \\
    Reynolds number $\textrm{Re}=U_02h/\nu$ & $12\,000$ \\
    No. of spectral elements $(E_x,E_y,E_z)$ & $(8,8,8)$ \\
    Polynomial orders $(N_x,N_y,N_z)$ & $(8,8,8)$ \\
    Time step & $0.002\,h/U_0$ \\
    No. of time iterations & $387\,000$ \\
    Dynamic range & $774h/U_0$ \\
    Nodal filtering -- DSM \& DMM & $M=N-2=6$ \\
    Modal filtering -- DSM \& DMM (1$^{\textrm{st}}$ level)& $k_c=N-2=6$ \\
    Modal filtering -- DSM \& DMM (2$^{\textrm{nd}}$ level)& $k_c'=N-3=5$ \\
    \hline
\end{tabular}
\caption{Numerical and physical parameters of the simulations}\label{tab:parameters}
\end{table}
In order to resolve the boundary layers along the lid and the downstream wall, the spectral elements are unevenly distributed as can be seen in Fig. \ref{fig:grid-LES}. The spatial discretization has $E_x=E_y=E_z=8$ elements in the three space directions with $N_x=N_y=N_z=8$ polynomial degree, equivalent to $65^3$ grid points in total. The spectral element calculation has two times less points per space direction than the DNS of Leriche and Gavrilakis \cite{leriche00:_direc} who employed a $129^3$ Chebyshev discretization.
Both nodal and modal filters were used in our LES computations based on DSM and DMM; the former with $M=N-2$ to stabilize the velocity field at each time step and the latter with $k_c=N-2$ (resp. $k_c'=N-3$) to filter the highest modes in the modal Legendre space at the first level (resp. second level) of explicit filtering. These filtering levels refer to the overbar and the test filtering respectively. It is noteworthy recalling here that the modal filter introduced in Sec. \ref{sec:filtering-techniques} is not projective. The computations are particularly sensitive to the values of $M$ and $k_c$; smaller values will affect spectral convergence whereas higher values will have very little effect on the smallest scales of the problem. The reference results are the DNS data of Leriche \cite{leriche00:_direc,leriche99:_direc_cheby} and the experimental ones from Koseff and Street \cite{koseff84}, corresponding to $1\,000$ and $145.5$ time units respectively. In the cavity flow, the average is obtained by time averaging. 

The LES-DSM and LES-DMM were both started from the same initial condition, namely the velocity field obtained from the DNS by Leriche and Gavrilakis and re-interpolated from the Chebyshev grid onto the spectral-element GLL grid.

Non-dimensionalization is performed using $h$ as length scale, $h/U_0$ as time scale and $U_0$ as velocity scale. All the results and data presented in the sequel will be based on this non-dimensionalization.

%-------- 1 Statistical ensemble averaging -------------
\subsection{Statistical ensemble averaging}\label{sec:statistical-ensemble-averaging}
For any variable, the Reynolds statistical splitting introduces the average value denoted by a capital letter into brackets ``$\langle \cdot \rangle $'' whereas a lower case letter will be used to denote its fluctuating part. It is noteworthy reminding here the filter splitting introduced in Eq. \eqref{eq:decomp} using the overbar and prime notations to denote respectively the resolved and subgrid scales. To simplify notations, and unless otherwise stated, the overbar will be omitted in the sequel as most of the fields considered are resolved fields derived from solutions of the filtered Navier--Stokes equations \eqref{eq:filtered_NS}--\eqref{eq:SGS_definition}. More precisely considering any variable $X$ can be decomposed as follows
\begin{equation}
X =\m{X}+x= (\m{\overline{X}}+\overline{x})+(\m{X'}+x'),
\end{equation}
where $\m{\overline{X}}$ (resp. $\m{X'}$) is the average resolved (resp. subgrid) part of $X$ and $\overline{x}$ (resp. $x'$) is the fluctuating resolved (resp. subgrid) part of $X$. The subgrid scales being unknown, the term $\m{X'}+x'$ cannot be directly computed from the simulation. All the results presented in this article refer to resolved quantities be them average $\m{\overline{X}}$ or fluctuating $\overline{x}$. For the sake of simplicity, these quantities are directly and respectively compared to $\m{X}$ and $x$, obtained from reference results, see Sec. \ref{sec:available-results}.

\begin{figure}[htbp]
\input{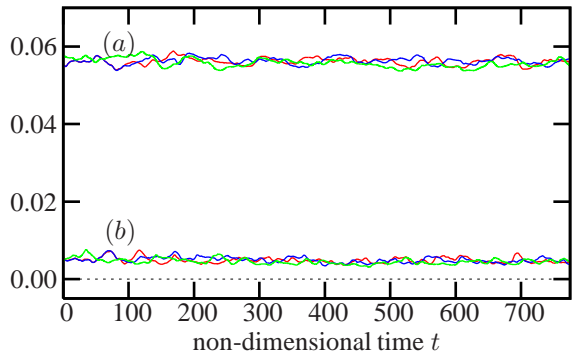}
\caption{Time histories of $\cK(t)$ (graphs $(a)$) and of $\kappa (t)$ (graphs $(b)$) for the DNS (green lines), the LES-DSM (red lines) and the LES-DMM (blue lines)}
\label{fig:time-history-energy}
\end{figure}

We assume that a statistically-steady state is attained and time averaging will be taken as ensemble averaging. The whole dynamic range---cf. Table \ref{tab:parameters}--- corresponding to $1\,290$ equally spaced samples has been considered when averaging. As the starting point of all LES is the same DNS sample taken from a statistically-steady state, it is reasonable to also assume that these simulations will reach a statistically-steady state very quickly. These assumptions can be verified in several number of ways. First, we present in Fig. \ref{fig:time-history-energy} the time histories of the volume integral of the total kinetic energy $\cK(t)$ and the volume integral of the fluctuating energy $\kappa (t)$ for the DNS and both the LES-DSM and LES-DMM. On this figure, one can observe that after approximately $80\,h/U_0$ time units, the two LES models DSM and DMM start being effective and providing different macroscopic results. Both $\cK(t)$ and $\kappa (t)$ have different time evolutions but within the same range of fluctuations and with very close average values, see Table \ref{tab:cKkappa}.

\begin{table}[htbp]
\begin{tabular}{cc }
  Average integral terms & Magnitude in $U_0^2$ units\\
  \hline\hline
  $\m{\cK(t)}_{\textrm{DNS}}$ & $0.055527$ \\
  $\m{\cK(t)}_{\textrm{DSM}}$ & $0.056296$ \\
  $\m{\cK(t)}_{\textrm{DMM}}$ & $0.056194$ \\
  \\
  $\m{\kappa(t)}_{\textrm{DNS}}$ & $0.004529$ \\
  $\m{\kappa(t)}_{\textrm{DSM}}$ & $0.004960$ \\
 $\m{\kappa(t)}_{\textrm{DMM}}$ & $0.004864$\\
    \hline
\end{tabular}
\caption{Average values of $\cK$ and $\kappa$ for the DNS, LES-DSM and LES-DMM}\label{tab:cKkappa}
\end{table}

A second way to assess the accuracy of the ensemble averaging is done by testing the property of symmetry (resp. antisymmetry) with respect to the mid-plane $z/h=0$, of some first- and second-order statistics of the resolved velocity and pressure fields. For each grid point, the relative difference between the nodal value at this point and the corresponding nodal value at the symmetric grid point is calculated. In the antisymmetric case, the opposite nodal value is considered at the symmetric grid point. The $z$-component of the average resolved velocity field $\m{W}$ is the only field presented being antisymmetric with respect to the mid-plane $z/h=0$. The results of the maximum errors on the grid are gathered in Table \ref{tab:symmetry} and are showing to be of the order of the error introduced by the space and time discretizations.

\begin{table}[htbp]
\begin{tabular}{crrc}
  Variable & Rel. diff. DSM & Rel. diff. DMM & Anti-/Symmetry\\
  \hline\hline
    $\mU$          & \tt 4.807e-04 & \tt 6.696e-05 & S\\
    $\mV$          & \tt 4.591e-04 & \tt 3.014e-04 & S\\
    $\mW$          & \tt 6.966e-05 & \tt 4.129e-04 & A\\
    $\mP$          & \tt 1.120e-04 & \tt 7.333e-05 & S\\
    $\urms$        & \tt 8.758e-05 & \tt 8.899e-05 & S\\
    $\vrms$        & \tt 1.696e-03 & \tt 7.764e-04 & S\\
    $\wrms$        & \tt 4.501e-04 & \tt 8.447e-05 & S\\
    $\m{uv}$       & \tt 1.107e-04 & \tt 2.236e-04 & S\\
    \hline
\end{tabular}
\caption{Quantitative assessment of the symmetry and anti-symmetry properties of some resolved average fields in the cavity; ``Rel. diff.'' stands for maximum relative difference}\label{tab:symmetry}
\end{table}

%-------- 2 Basic LES modeling: Under-resolved DNS and Smagorinsky model -------------
\subsection{Under-resolved DNS and Smagorinsky model}\label{sec:basic-les-modeling}

Before providing the reader with a comprehensive review of results obtained for the two models LES-DSM and LES-DMM, partial results for the UDNS and the LES-SM are presented in this section. These results correspond to the same parameters as the one in Table \ref{tab:parameters}, except that the number of iterations is $33\,000$ corresponding to a simulation length of $66\,h/U_0$---approximately one tenth of the total simulation time of the LES-DSM and LES-DMM. Moreover, for the LES-SM the value of the Smagorinsky constant $\CS$ defined in Eq. \eqref{eq:nut_smag} was taken equal to its theoretical value $0.18$, see \cite{sagaut03:_large} for greater details, and no wall-damping procedure was implemented for these preliminary simulations. The reference result is the DNS by Leriche and Gavrilakis \cite{leriche00:_direc} and is represented by the solid line in the profiles in Figs. \ref{fig:mU-mV-33000} and \ref{fig:muu-mvv-33000}, whereas dashed (resp. dotted) lines refer to the UDNS (resp. LES-SM). The results in Figs. \ref{fig:mU-mV-33000} and \ref{fig:muu-mvv-33000} are one-dimensional profiles of the average velocity field and its rms-fluctuations in the mid-plane $z/h=0$. General conclusions can be drawn from all these figures. First, UDNS is totally inoperative in the particular context of this simulation. Even first-order statistics such as $\m{U}$ and $\m{V}$ are far from being well predicted, not to mention rms-fluctuations. Second, the Smagorinsky model LES-SM results show a real improvement in predicting the fields compared to the UDNS but as already known, the simplicity of this model does not allow to correctly predict the stiff physics of this simulation. These results justify the need for a more complex LES modeling such as LES-DSM and LES-DMM presented in the sequel.

\begin{figure}[htbp]
  \begin{center}
    \input{mU_MidY_LineX_33000.pslatex}
    \vspace{-0.6cm}
    $$ $$
    \input{mV_MidX_LineY_33000.pslatex}
  \end{center}
  \caption{In the mid-plane $z/h=0$: $\mU$ on the horizontal centerline $y/h=0$ (top), $\mV$ on the vertical centerline $x/h=0$ (bottom): DNS (solid line), MILES (dashed line) and LES-SM (dotted line)}\label{fig:mU-mV-33000}
\end{figure}

\begin{figure}[htbp]
  \begin{center}
  \input{muu_MidY_LineX_33000.pslatex}
    \vspace{-0.6cm}
    $$ $$
    \input{mvv_MidX_LineY_33000.pslatex}
  \end{center}
%    \caption{In the mid-plane $z/h=0$: $\urms$ on the horizontal centerline $y=0$ (top), on the vertical centerline $x=0$ (bottom)}\label{fig:muu-33000}
  \caption{In the mid-plane $z/h=0$: $\urms$ on the horizontal centerline $y/h=0$ (top), $\vrms$ on the vertical centerline $x/h=0$ (bottom): DNS (solid line), MILES (dashed line) and LES-SM (dotted line)}\label{fig:muu-mvv-33000}
\end{figure}

%-------- 3 Comparisons with pre-existing results -------------
\subsection{Comparisons with available results}\label{sec:available-results}
In this section, results of the LES-DSM and the LES-DMM are compared with the available reference experimental and numerical results. 

\subsubsection{One-dimensional profiles}\label{sec:1D-profiles}
Of the previous work available in the literature on the lid-driven cubical cavity flow, the numerical DNS data from Leriche and Gavrilakis \cite{leriche00:_direc}, Leriche \cite{leriche05:_direc_numer_simul_lid_driven} and the experimental data of Prasad and Koseff \cite{prasad89:_reynol} constitute the two main references. This work being an extension of the one by Leriche, it borrows from \cite{leriche00:_direc} the values of the main physical parameters---see Table \ref{tab:parameters}. The work from Prasad and Koseff \cite{prasad89:_reynol} includes data from a flow at Reynolds number similar to that of the present LES. The measurements that these authors reported were taken in the mid-plane $z/h=0$, which is a statistical symmetry plane of the flow domain. As it will be shown in the sequel, the flow near the downstream secondary eddy---see Fig. \ref{fig:sketch}---is not homogeneous in the $z$-direction. In the ``turbulent'' part of the cavity, the mid-plane is found to cut through surfaces of local minima in the intensity field with rapid changes occurring on both sides of it.

The set of experimental data corresponding to a Reynolds number $\textrm{Re}=10\,000$ is used for the comparisons of the one-dimensional average velocity profiles along the vertical and horizontal symmetry axes. It is important to note that no experimental error-bars were given for any data. The only information related to the local experimental measurement error is provided by the two crosses corresponding to two different measurements in the middle ($x/h=0$ or $y/h=0$) of each centerline---the velocity probing system going back and forth from this point \cite{prasad:}. In addition the experimental data of Prasad and Koseff \cite{prasad89:_reynol,prasad:} are obtained over a non-dimensional averaging time of 145.5 whereas the DNS (resp. LES) results were obtained over an averaging time of $1\,000$ (resp. $774$). In absence of local error-bars in the measurements, this may explain the scattering (and possible non-convergence) of some experimental data, together with practical difficulties of accurately measuring fluctuating fields in region of low or almost constant velocity.  A detailed analysis of the disparity between the numerical results and some experimental data can be found in \cite{leriche00:_direc,leriche99:_direc_cheby}.

For the sets of experimental and DNS data, the total velocity field is considered whereas in the case of LES, only its resolved part is presented. The legend for Figs. \ref{fig:1D-mU}--\ref{fig:1D-muv} is as follows: crosses refer to the experimental data of Prasad and Koseff, the solid lines to the DNS by Leriche and Gavrilakis, the dashed lines to the LES-DSM and the dotted lines to the LES-DMM. All the data related to average and rms-fluctuations of the velocity field are expressed in terms of the velocity scale $U_0$ and $\m{uv}$ in terms of $U_0^2$.

\begin{figure}[htbp]
  \input{mU_MidY_LineX_387000.pslatex}
  \vspace{-0.6cm}
  $$ $$
  \input{mU_MidX_LineY_387000.pslatex}
  \caption{In the mid-plane $z/h=0$: $\mU$ on the horizontal centerline $y/h=0$ (top), $\mU$ on the vertical centerline $x/h=0$ (bottom); Experiment (crosses), DNS (solid line), LES-DSM (dashed line), LES-DMM (dotted line)}\label{fig:1D-mU}
\end{figure}

\begin{figure}[htbp]
  \input{mV_MidY_LineX_387000.pslatex}
  \vspace{-0.6cm}
  $$ $$
  \input{mV_MidX_LineY_387000.pslatex}
  \caption{In the mid-plane $z/h=0$: $\mV$ on the horizontal centerline $y/h=0$ (top), $\mV$ on the vertical centerline $x/h=0$ (bottom); Experiment (crosses), DNS (solid line), LES-DSM (dashed line), LES-DMM (dotted line)}\label{fig:1D-mV}
\end{figure}

\begin{figure}[htbp]
  \input{muu_MidY_LineX_387000.pslatex}
  \vspace{-0.6cm}
  $$ $$
  \input{muu_MidX_LineY_387000.pslatex}
  \caption{In the mid-plane $z/h=0$: $\urms$ on the horizontal centerline $y/h=0$ (top), $\urms$ on the vertical centerline $x/h=0$ (bottom); Experiment (crosses), DNS (solid line), LES-DSM (dashed line), LES-DMM (dotted line)}\label{fig:1D-muu}
\end{figure}

\begin{figure}[htbp]
  \input{mvv_MidY_LineX_387000.pslatex}
  \vspace{-0.6cm}
  $$ $$
  \input{mvv_MidX_LineY_387000.pslatex}
  \caption{In the mid-plane $z/h=0$: $\vrms$ on the horizontal centerline $y/h=0$ (top), $\vrms$ on the vertical centerline $x/h=0$ (bottom); Experiment (crosses), DNS (solid line), LES-DSM (dashed line), LES-DMM (dotted line)}\label{fig:1D-mvv}
\end{figure}

\begin{figure}[htbp]
  \input{muv_MidY_LineX_387000.pslatex}
  \vspace{-0.6cm}
  $$ $$
  \input{muv_MidX_LineY_387000.pslatex}
\caption{In the mid-plane $z/h=0$: $\m{uv}$ on the horizontal centerline $y/h=0$ (top), $\m{uv}$ on the vertical centerline $x/h=0$ (bottom); Experiment (crosses), DNS (solid line), LES-DSM (dashed line), LES-DMM (dotted line)}\label{fig:1D-muv}
\end{figure}

A discussion on the comparisons between the DNS reference results and the experimental ones is available in \cite{leriche00:_direc}. In the sequel, we will focus on comparing the LES-DSM and LES-DMM results with the DNS and experimental ones. A rapid overview of Figs. \ref{fig:1D-mU}--\ref{fig:1D-muv} indicates that both LES models provide results very close to the DNS references, even for the rms-fluctuations in Figs. \ref{fig:1D-muu} and \ref{fig:1D-mvv} and above all for the component $\m{uv}$ of the Reynolds stress in Fig. \ref{fig:1D-muv}. The differences between the profiles of the two LES models and the DNS generally coincide with the existence of local extrema; maxima tend to be slightly over-estimated in the LES whereas minima are somewhat under-estimated. These two effects can be partly justified by the reduced sampling in the LES-DSM and LES-DMM compared to the sampling of the DNS. This phenomenon have already been encountered and studied by Leriche in \cite{leriche99:_direc_cheby}. From these results it is not possible to rank between themselves the performances of the LES-DSM and the LES-DMM.

\subsubsection{Two-dimensional profiles}\label{sec:2D-profiles}
\begin{figure*}[htbp]
\includegraphics[width=5cm]{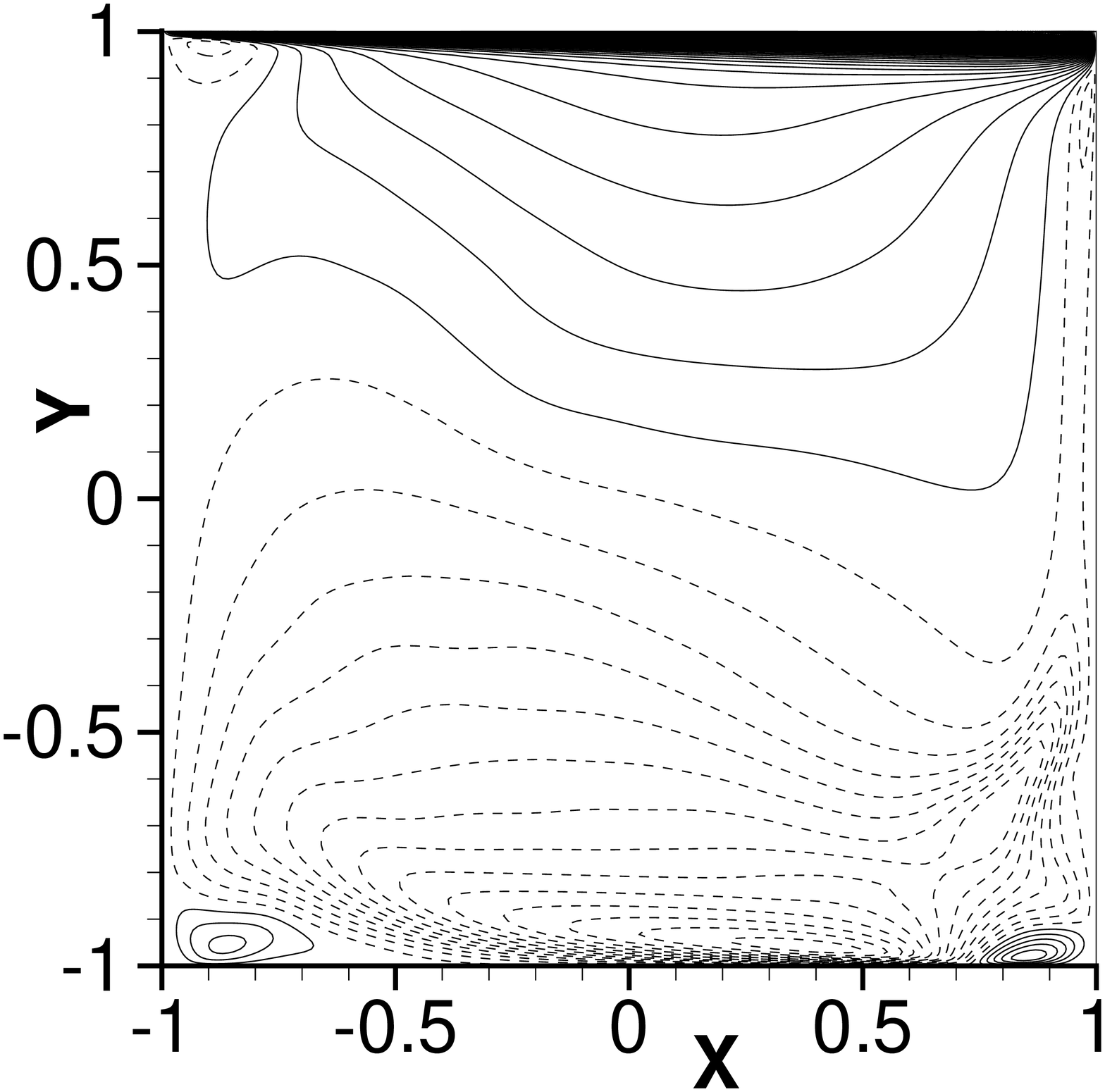}
\includegraphics[width=5cm]{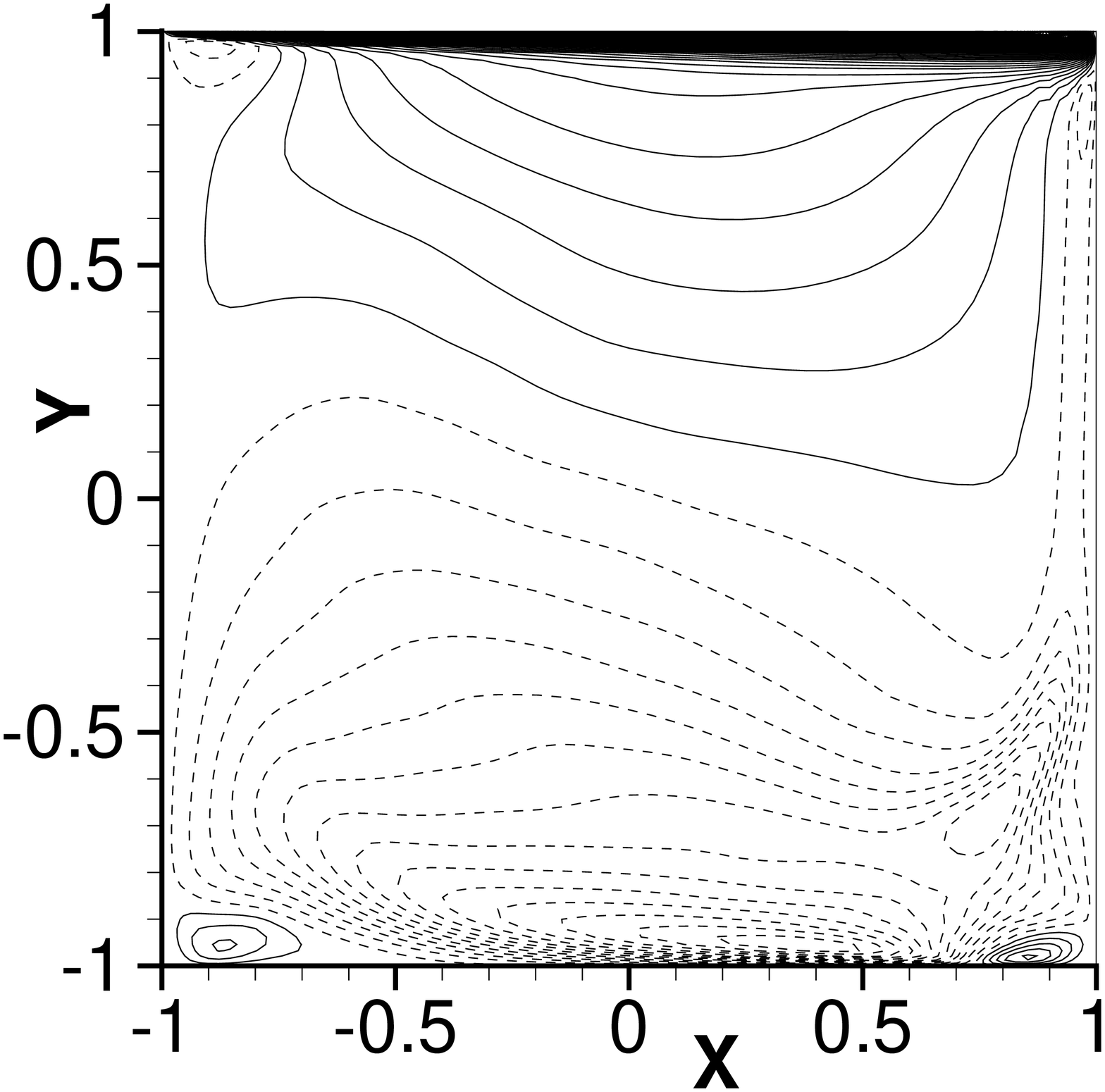}
\includegraphics[width=5cm]{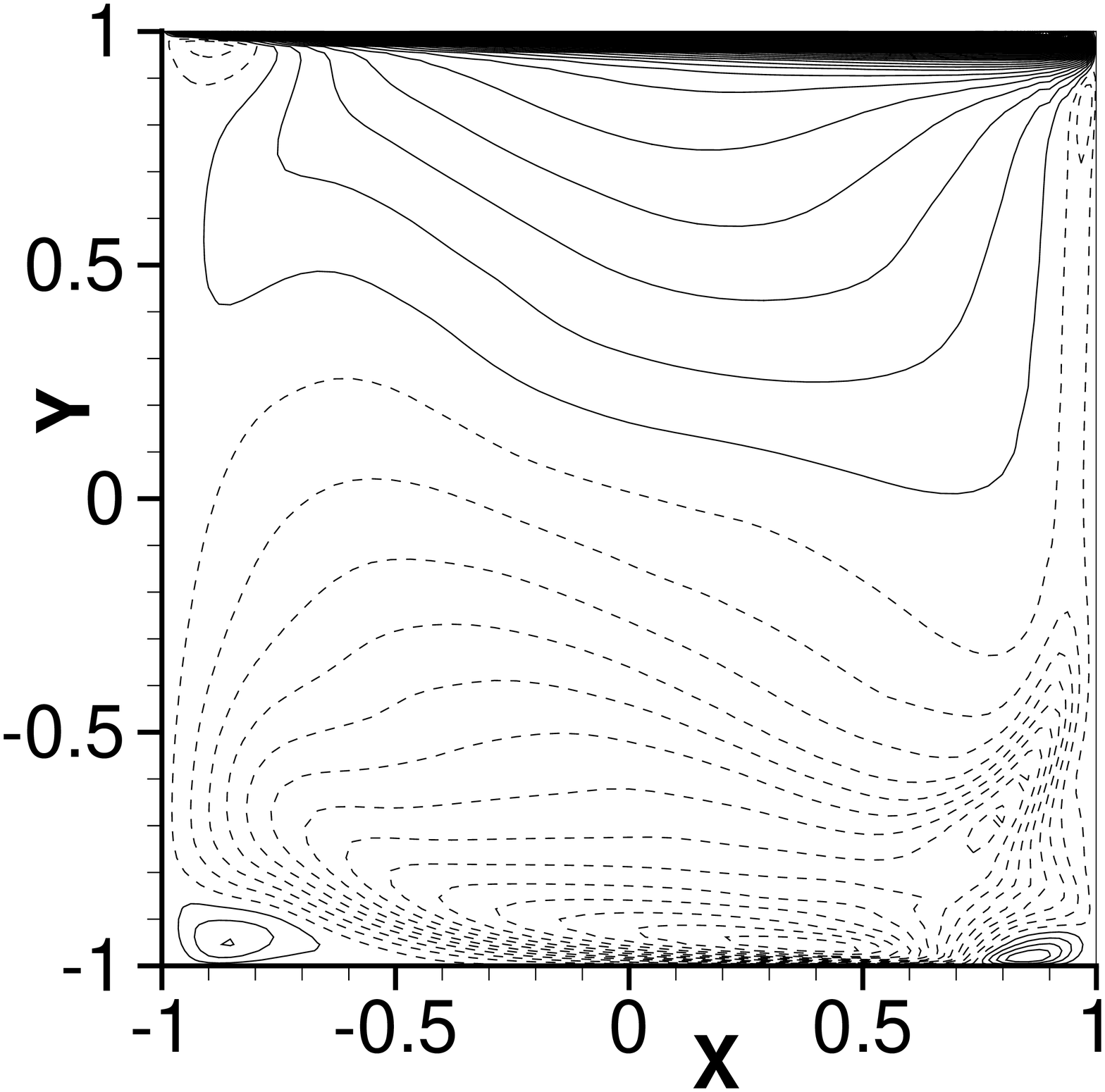}
\includegraphics[width=5cm]{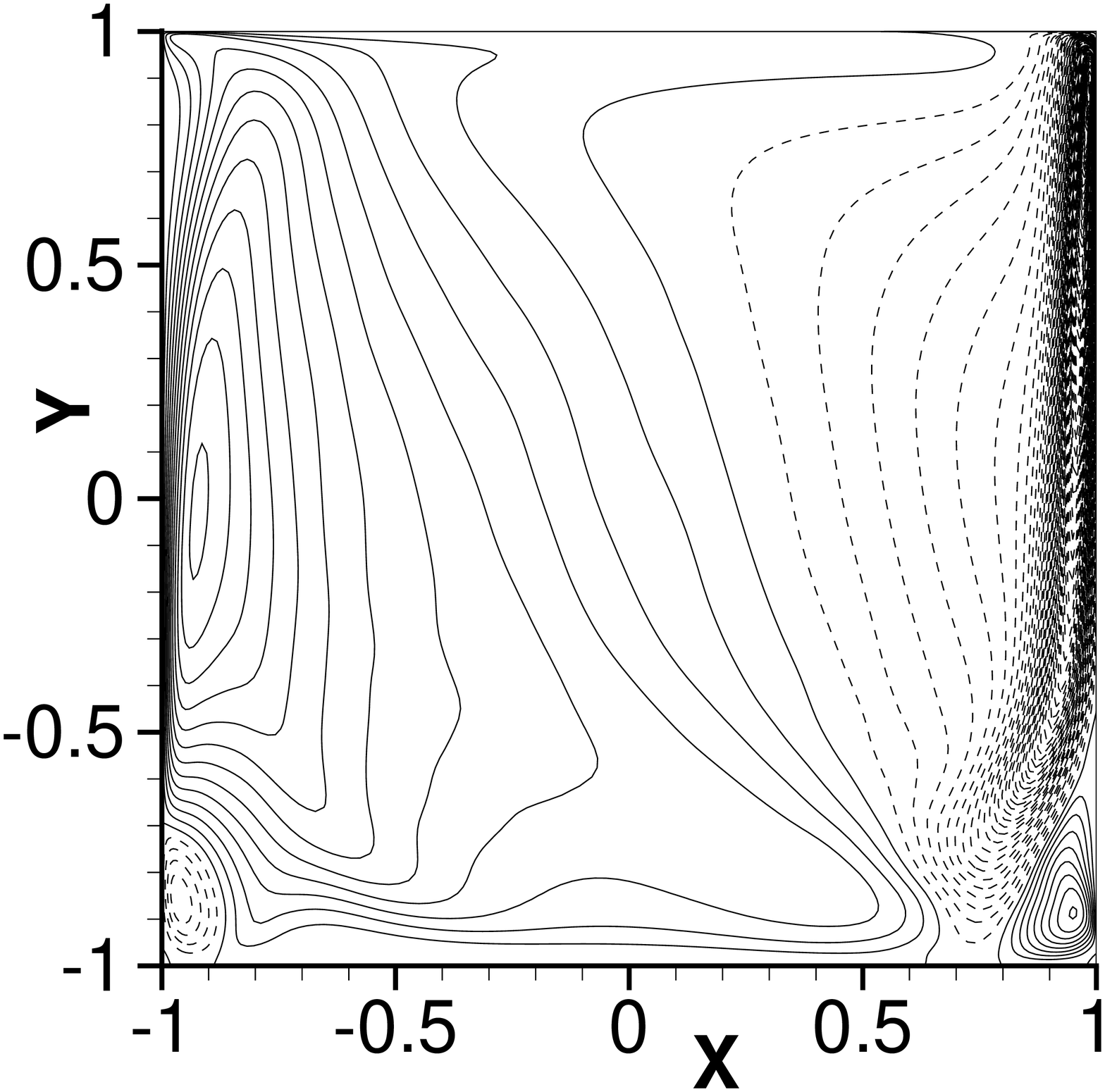}
\includegraphics[width=5cm]{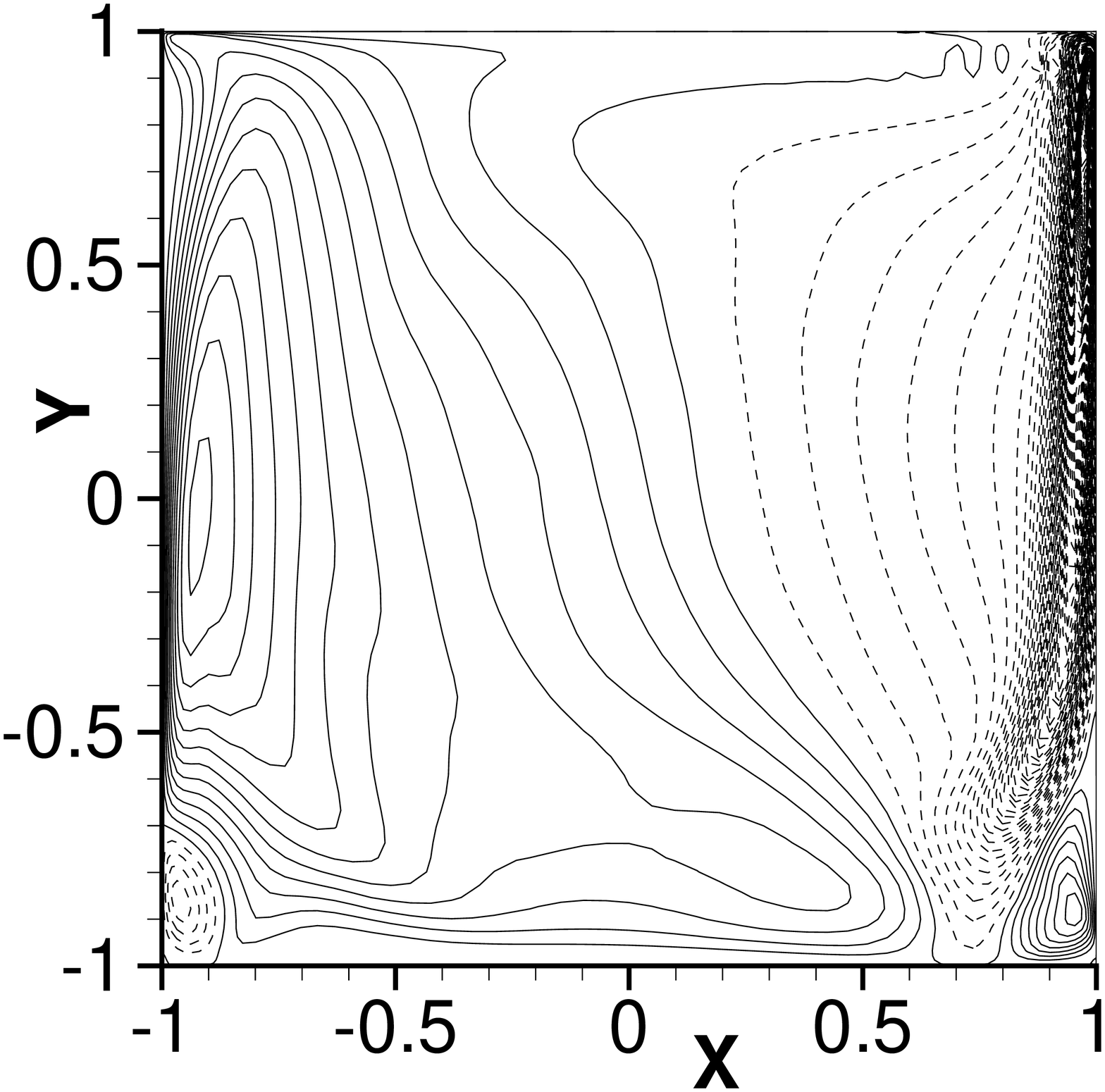}
\includegraphics[width=5cm]{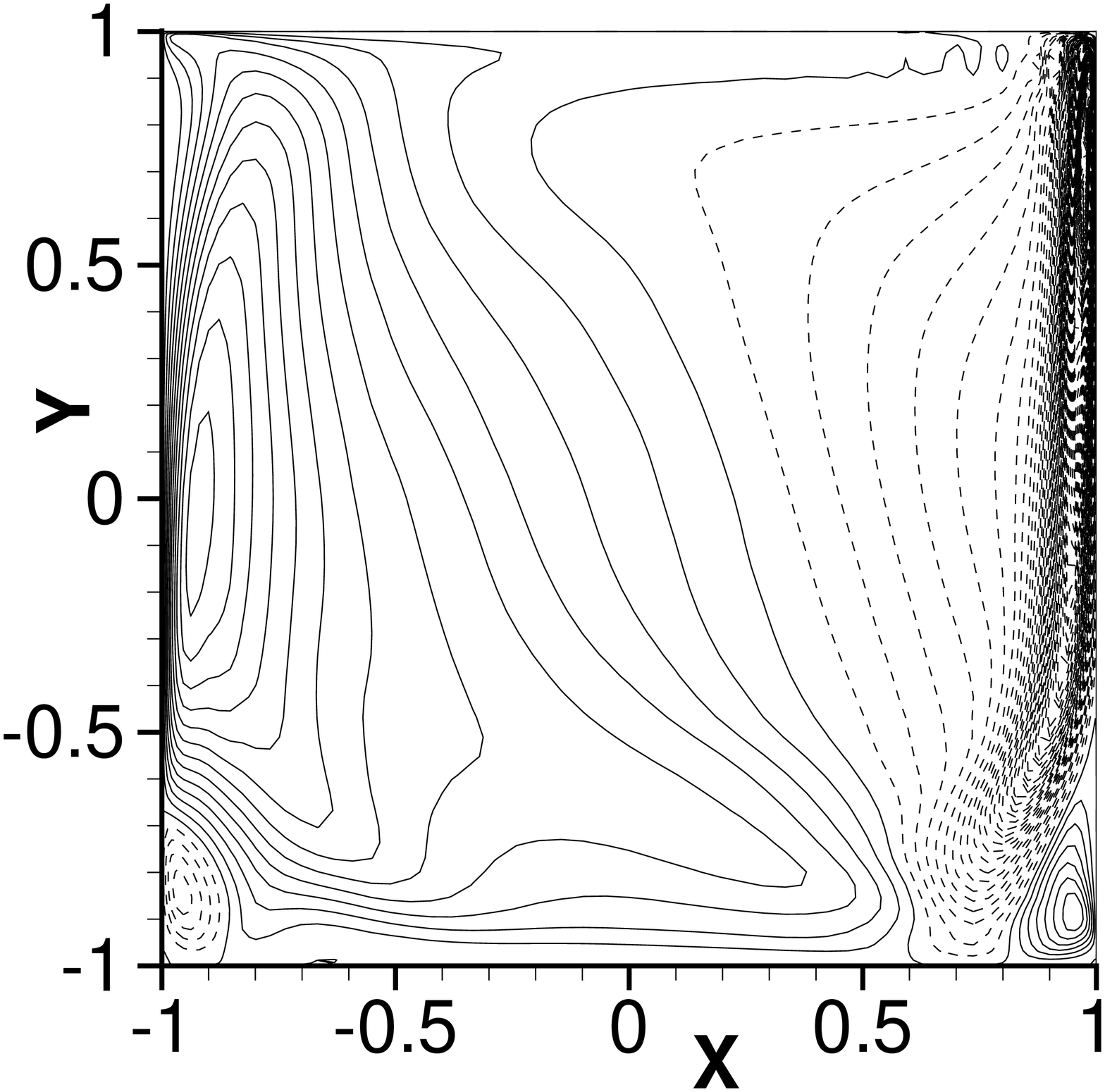}
\caption{Contours of average velocity in the mid-plane $z/h=0$; DNS (left), LES-DSM (center) and LES-DMM (right)---100 contours levels taken between $-0.4$ and $1$ for $\mU$ (top) and between $-0.7$ and $0.2$ for $\mV$ (bottom). Dashed contours lines correspond to negative levels}\label{fig:mU-mV-Zmidplane}
\end{figure*}

\begin{figure*}[htbp]
\includegraphics[width=5cm]{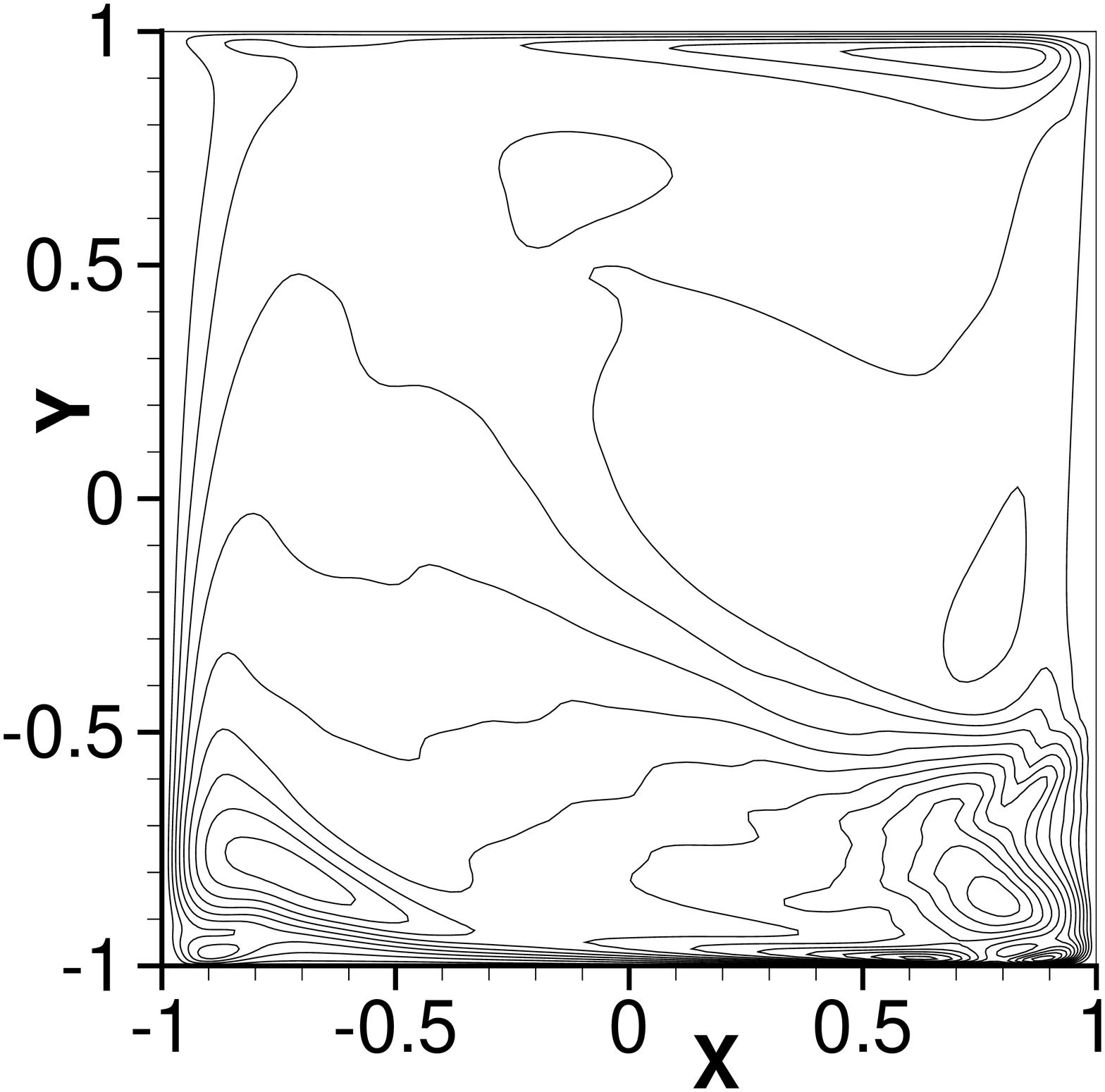}
\includegraphics[width=5cm]{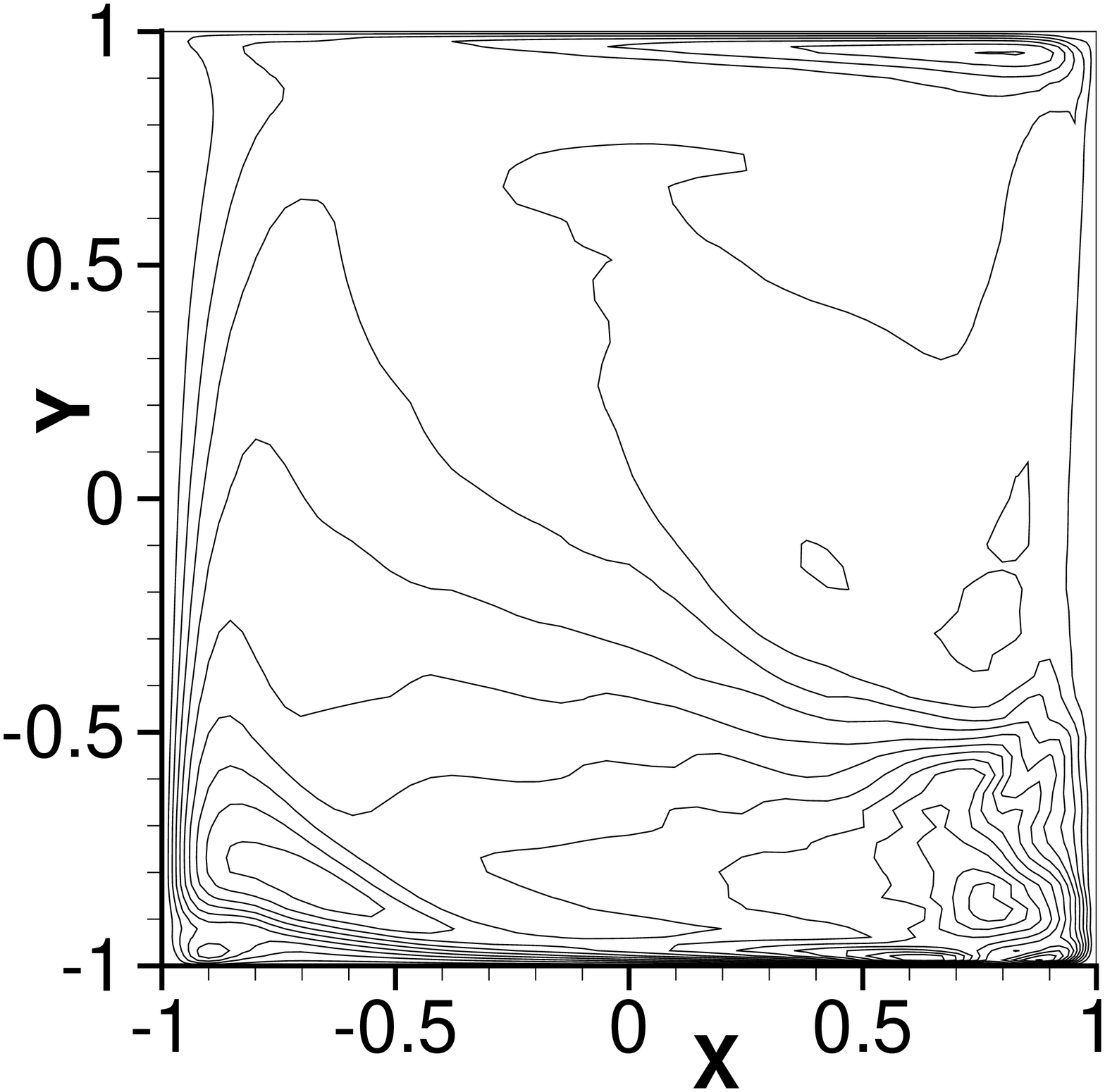}
\includegraphics[width=5cm]{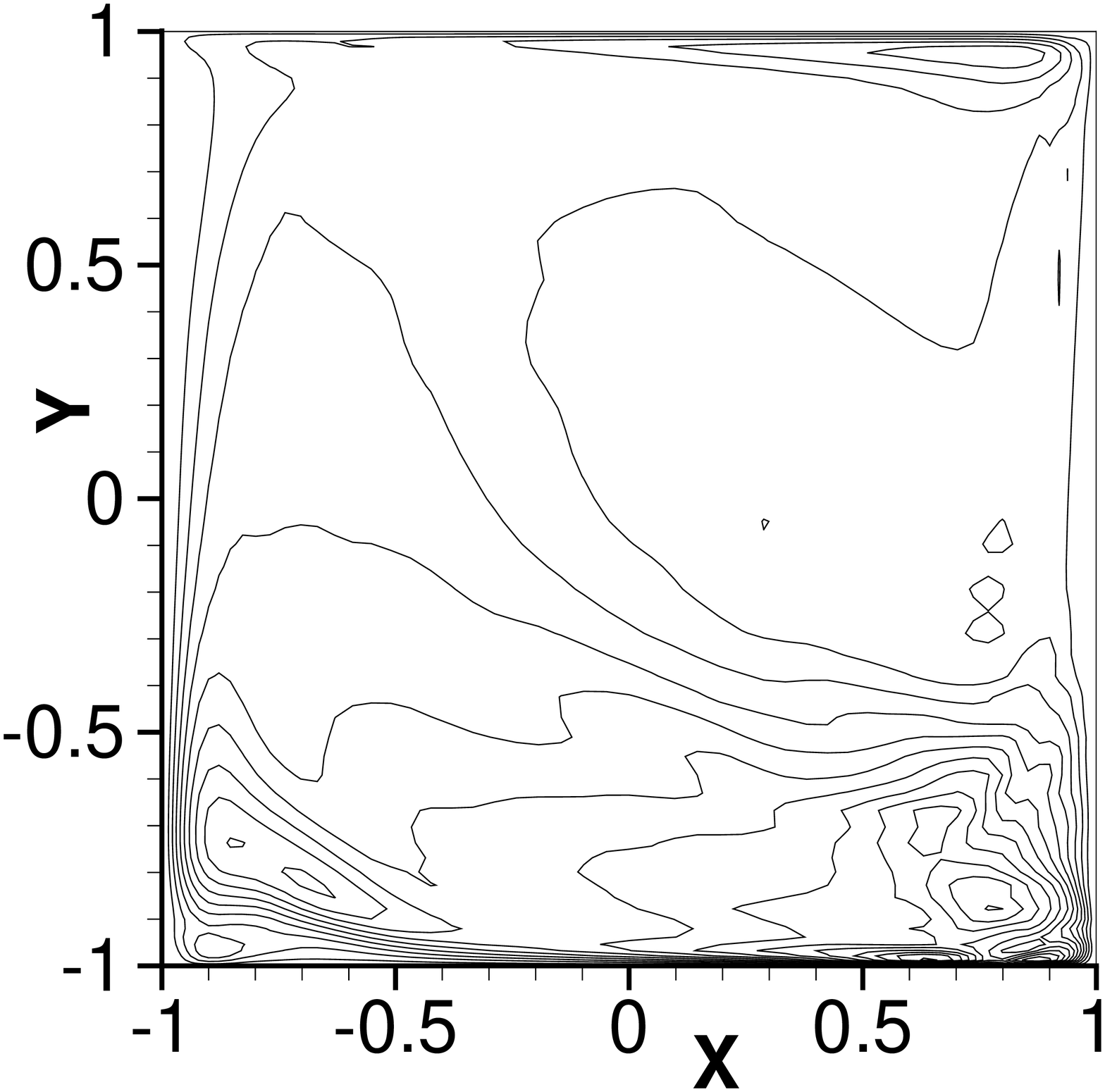}
\includegraphics[width=5cm]{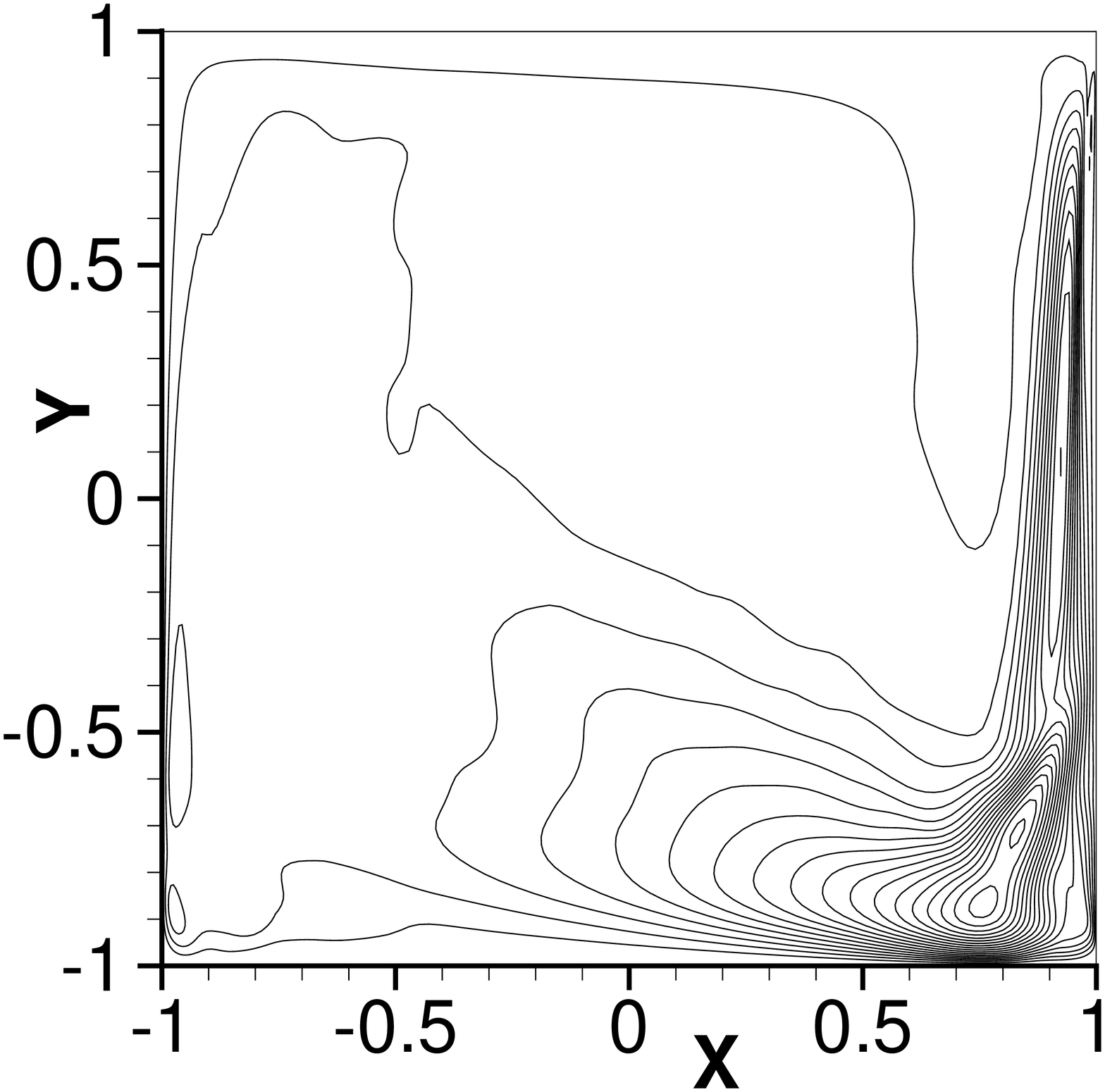}
\includegraphics[width=5cm]{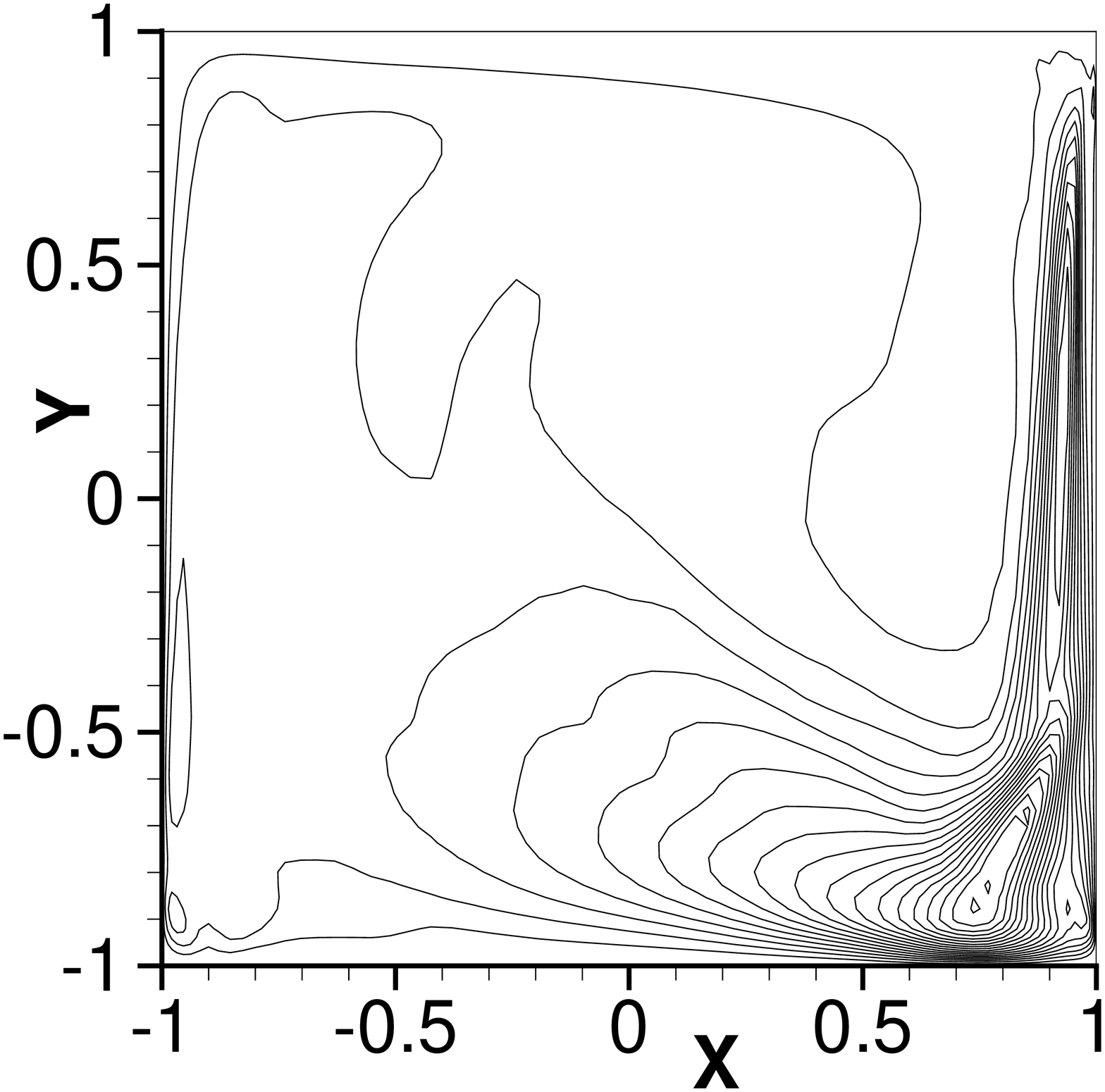}
\includegraphics[width=5cm]{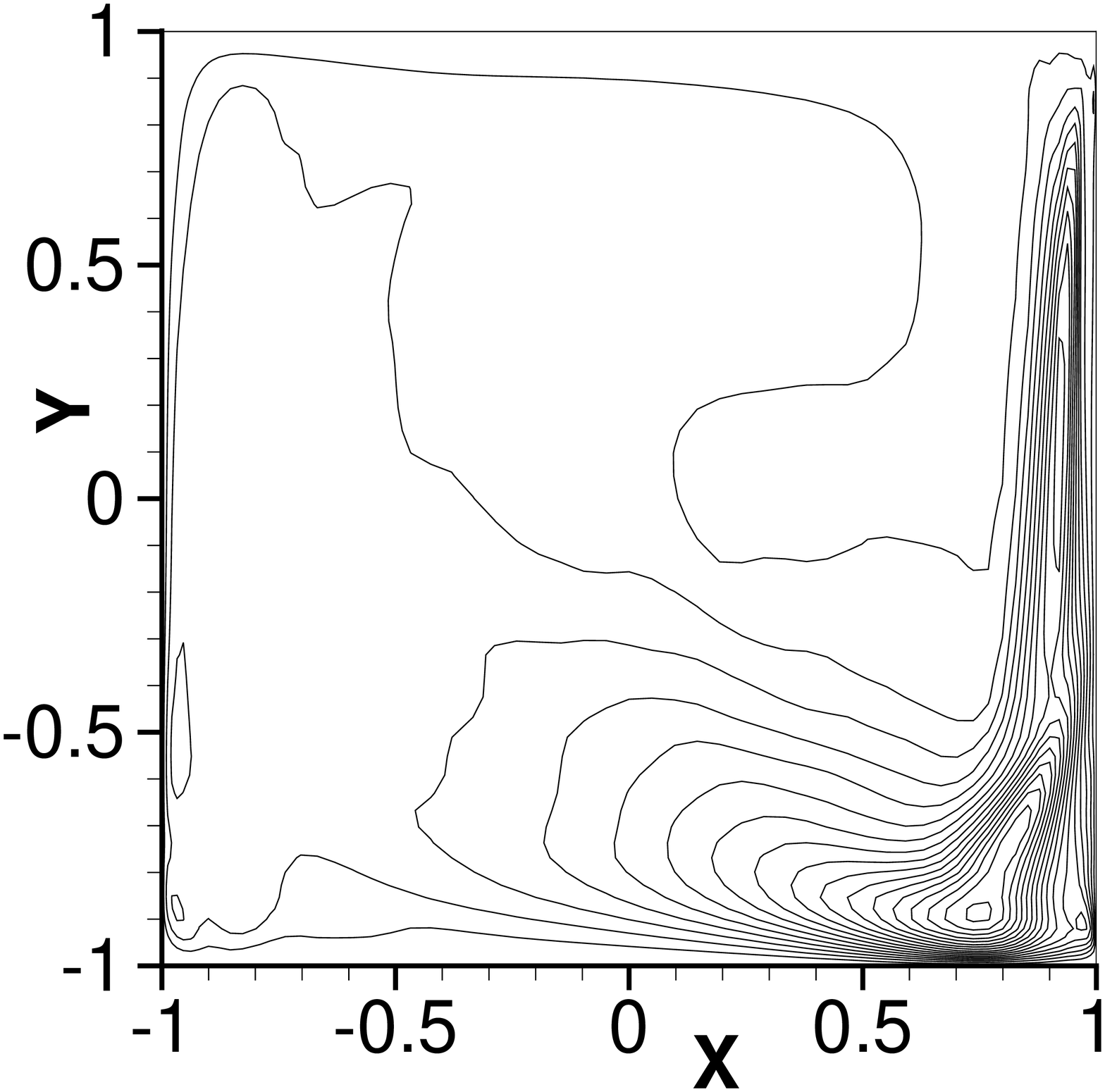}
\caption{Contours of rms-fluctuations of the velocity in the mid-plane $z/h=0$; DNS (left), LES-DSM (center) and LES-DMM (right); 20 contours equally spaced between 0 and $0.1$ for $\urms$ (top) and between $0$ and $0.15$ for $\vrms$ (bottom)}\label{fig:urms-vrms-Zmidplane}
\end{figure*}

The comparisons with the DNS results started in Sec. \ref{sec:1D-profiles} are now extended to the whole mid-plane $z/h=0$ by plotting identical series of contour levels of average velocity components in Fig. \ref{fig:mU-mV-Zmidplane} and of rms-fluctuations of the velocity in Fig. \ref{fig:urms-vrms-Zmidplane}, for the DNS (left column), the LES-DSM (central column) and the LES-DMM (right column).

As previously noted with the one-dimensional profiles, the results provided by the LES-DSM and LES-DMM are both very close to the reference DNS results. Secondary corner eddies located above the bottom wall and below the lid next to the upstream wall are correctly captured in the mean flow. Other finer structures visible in Fig. \ref{fig:urms-vrms-Zmidplane} (bottom), for $\vrms$ near the upstream wall are also correctly captured by both LES modelings. The rms-fluctuations of the $x$-component $u$ of the velocity field is accurately resolved just below the lid which is a high-gradient region for the mean flow. Moreover, in the region near the downstream wall where the wall jet---separated into two elliptical jets---are impinging on the bottom wall, the high gradients of velocity fluctuations are well reproduced. As it will be presented in the following sections, the maximum of turbulence production belongs to this region of the flow domain which will be indeed of particular interest in the remaining. 

The flow below the lid and near the corner with the downstream wall presents wiggles in the LES contours for $\mV$, see Fig. \ref{fig:mU-mV-Zmidplane} (bottom). Although less intense, these wiggles are also noticeable on the contours for $\vrms$ at the same location, see Fig. \ref{fig:urms-vrms-Zmidplane} (bottom). More limited effects are noticeable for the equivalent $x$-component fields. These very limited defects in both simulations find their origin in a slight under-resolution of the spectral-element grid in this small region of the cavity where high gradients are present.

%-------- 4 Physical parameters of the LES modeling -------------
\subsection{Physical parameters of the LES modeling} \label{sec:parameters-LES}
The LES modeling for both the LES-DSM and LES-DMM involves the calculation of two scalar fields, namely the dynamic parameter $\Cd$ and the eddy viscosity $\nu_{\textrm{sgs}}$ which are inter-related. As some values of $\Cd$ produced by the two dynamic procedures \eqref{eq:Cdyn} and \eqref{eq:Cdynmix} may locally reach relatively ``high values'' destabilizing the time-integration procedure for the filtered Navier--Stokes computations. Hence it is common to use \textit{ad hoc} averaging or limiting of the dynamic parameter $\Cd$ to ensure stability. Various procedures are reported in the literature: averaging in homogeneous directions \cite{germano91,kravchenko97}, temporal smoothing \cite{breuer98:_large}, integral constraint \cite{ghosal95}, Lagrangian averaging \cite{meneveau96:_lagran} and clipping \cite{zang93,blackburn03:_spect}. In the present work the latter procedure of clipping is used. First the maximum admitted value of the dynamic parameter was ${\Cd}_{\textrm{max}}=(0.18)^2$, $0.18$ corresponding to the theoretical value of the Smagorinsky constant---see \cite{sagaut03:_large}. The negative values of $\Cd$ are also clipped and set to zero for the LES-DSM and LES-DMM. The amount of grid points clipped is indeed very limited and correspond to $0.2 \%$ and $0.08 \%$ of the total number of grid points for LES-DSM and LES-DMM respectively. It was found that the clipping of $\Cd$ to the interval comprised between $-(0.18)^2$ and $+(0.18)^2$---therefore allowing for local negative values of the eddy viscosity---was not affecting at all the stability of the spectral-element filtered Navier--Stokes computation. The difference between the results with or without the negative values of $\Cd$ was found to be negligible in the particular context of the lid-driven cubical cavity flow which is related to the limited amount of backscattering for this flow at a Reynolds number of $12\,000$.

\begin{figure}[htbp]
  \includegraphics[width=5cm]{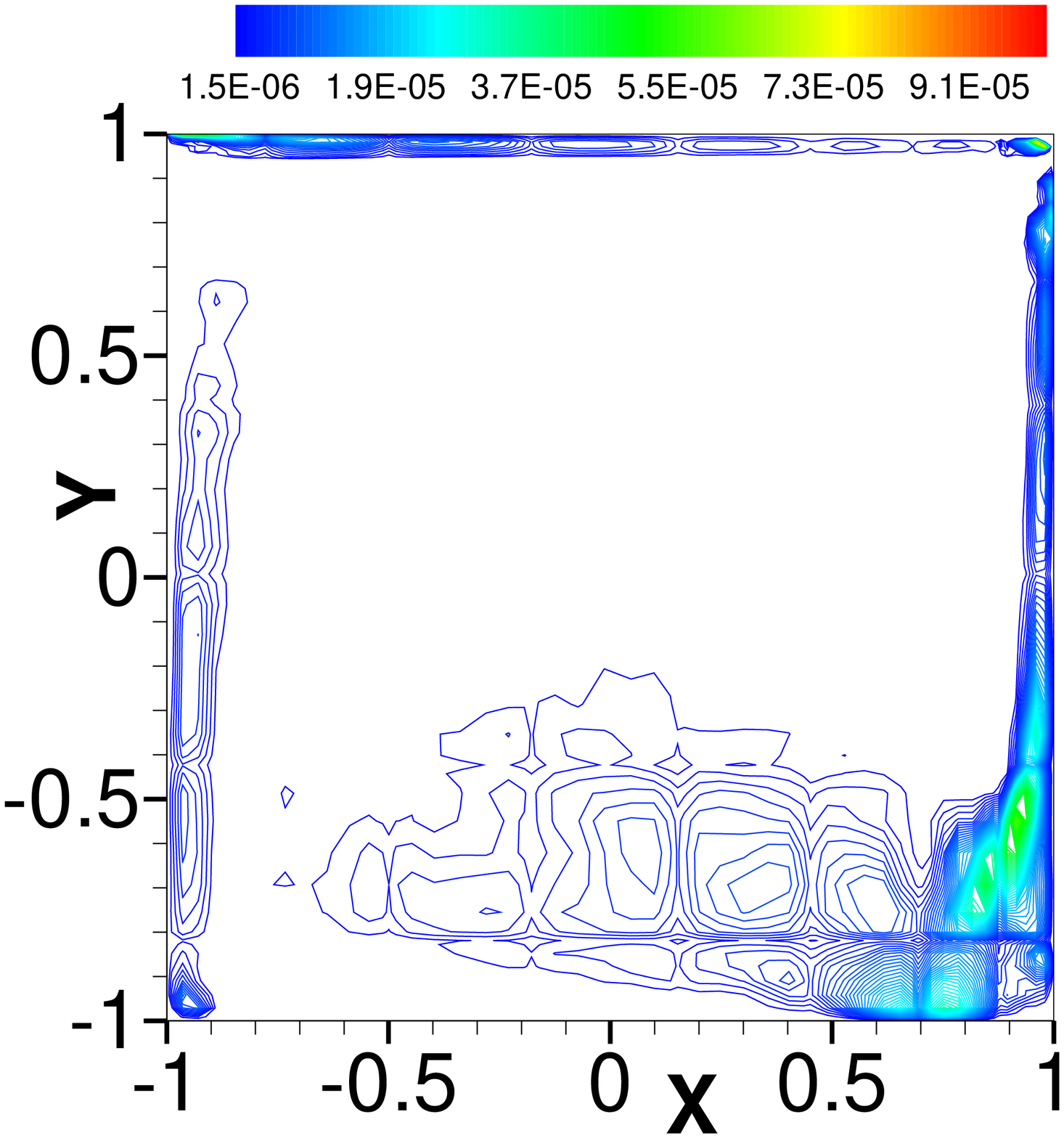}
  \includegraphics[width=5cm]{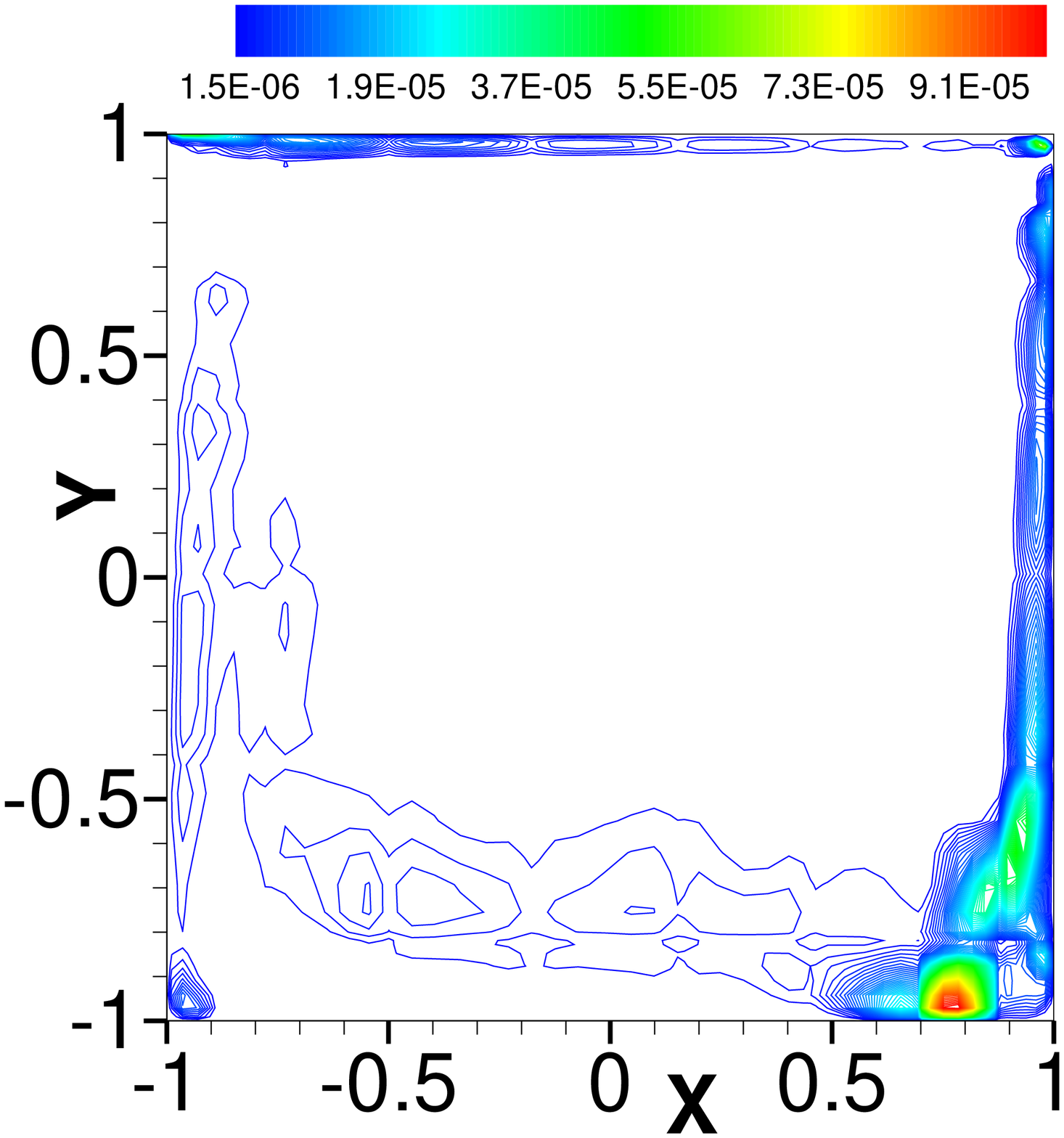}
  \caption{Contours of the average eddy viscosity $\m{\nu_{\textrm{sgs}}}$ for the LES-DSM in the mid-plane $z/h=0$ (left) and in the plane $z/h=0.241$ (right); same series of contour levels is used in the two planes}
  \label{fig:nuT-Zmidplane}
\end{figure}

Fig. \ref{fig:nuT-Zmidplane} displays contour lines of the average eddy viscosity for the LES-DSM in the mid-plane $z/h=0$ and in the plane $z/h=0.241$ where the maximum of average turbulent energy dissipation rate was localized---cf. Sec. \ref{sec:small-scales-turbulence} for greater details. First, the $\cC^0$-continuity breakage in the inter-element continuity is obvious---see Fig. \ref{fig:grid-LES} to compare with the spectral element grid in the mid-plane---and is directly related to the discontinuous nature of the filter length field $\Delta$ defined in Sec. \ref{sec:modal-filter}, Eq. \eqref{eq:filter-length}. The effect of such discontinuity of the subgrid viscosity has been analyzed and discussed by Blackburn and Schmidt in \cite{blackburn03:_spect} using the same numerical framework as ours, namely the SEM. They found that the inter-element discontinuity of the subgrid term does not have a noticeable effect on their physical results which is confirmed by the present work. Finally, it appears clearly that the reasons for resorting to a dynamic procedure are fully justified by Fig. \ref{fig:nuT-Zmidplane}. Indeed, the dynamic procedure automatically turns on the dynamic parameter $\Cd$ which in turn activates subgrid-scale viscous effects in the regions of the flow where turbulent dissipation at the small-scales level occurs---see Sec. \ref{sec:small-scales-turbulence}. 

Similar results are obtained for the eddy viscosity and the dynamic parameter in the case of the dynamic mixed model LES-DMM. The same clipping procedure, with the same clipping values as described earlier for LES-DSM was implemented.

%================ 4 Characterization of turbulence in the flow ====================================
\section{Characterization of turbulence in the flow}\label{sec:turbulence}
This section is devoted to a thorough analysis of some specific features of the flow in the region of the cavity where turbulence occurs. The aims are to ensure that the LES-DSM and LES-DMM are both capable of reproducing the fine physics observed in these regions and also to gain insights in the turbulent mechanisms involved.

%-------- 1 Inhomogeneity of turbulence -------------
\subsection{Inhomogeneity of turbulence}\label{sec:inhomo}
It is easily predictable that such a confined flow will produce an inhomogeneous turbulence but it is worth determining in greater details the turbulent inhomogeneous zones in the cavity. In order to access this information we use the average turbulent energy dissipation rate $\m{\eps}$ defined by
\begin{equation}\label{eq:def-meps}
\m{\eps}=\frac{1}{2} \nu \left\langle \left(\ddp{u_i}{x_j}+\ddp{u_j}{x_i}  \right)\left(\ddp{u_i}{x_j}+\ddp{u_j}{x_i}  \right) \right\rangle= 2\nu \langle S_{ij}S_{ij} \rangle.
\end{equation}
Here and in the sequel, we use index notation and the summation convention, where repeated indices imply summation. The velocity fluctuations being divergence-free, one can rewrite
\begin{equation}
\m{\eps}=\nu \left\langle \ddp{u_i}{x_j}\ddp{u_i}{x_j}    \right\rangle + \nu \frac{\partial^2 \m{u_iu_j}}{\partial x_i \partial x_j},
\end{equation}
which in turn can be recast in terms of $\bomega$ the fluctuating vorticity, $\bOmega$ being the total resolved vorticity field
\begin{equation}
\m{\eps}=\nu \m{\omega_i\omega_i} + 2\nu \frac{\partial^2 \m{u_iu_j}}{\partial x_i \partial x_j}.
\end{equation}
We define the average difference $\m{\delta}$ by the difference between the average turbulent energy dissipation rate, divided by $\nu$, and the average fluctuating enstrophy
\begin{equation}
\m{\delta}=\left\langle\frac{\eps}{\nu}\right\rangle - \m{\omega_i\omega_i}=2\frac{\partial^2 \m{u_iu_j}}{\partial x_i \partial x_j}.
\end{equation}
\begin{figure}[htbp]
\includegraphics[width=8cm]{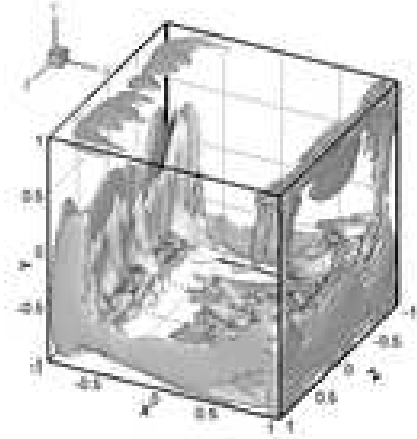}
\caption{Region of the cavity where the turbulent flow is inhomogeneous according to the criterion $|\m{\delta}|/|\m{\delta}_{\text{max}}|>1/100$; LES-DMM}\label{fig:Inhomogeneity_3D}
\end{figure}

For homogeneous flow, the spatial derivatives of the Reynolds stress components $\m{u_i u_j}$ are zero, and subsequently $\m{\delta}=0$. The average difference $\m{\delta}$ was calculated for both databases LES-DSM and LES-DMM. Fig. \ref{fig:Inhomogeneity_3D} displays a 3D view of the volume of the cavity where the flow is inhomogeneous according to the following heuristic criterion: $|\m{\delta}|/|\m{\delta}_{\text{max}}|>1/100$, where $|\m{\delta}_{\text{max}}^{\textrm{DSM}}|=159.8\,U_0^2/h^2$ and $|\m{\delta}_{\text{max}}^{\textrm{DMM}}|=155.6\,U_0^2/h^2$. In other words, it shows the region of the flow where the inhomogeneity of the turbulence---measured by $\m{\delta}$---is above 1\% of its maximum absolute value. 

\begin{figure}[htbp]
\includegraphics[width=4.2cm]{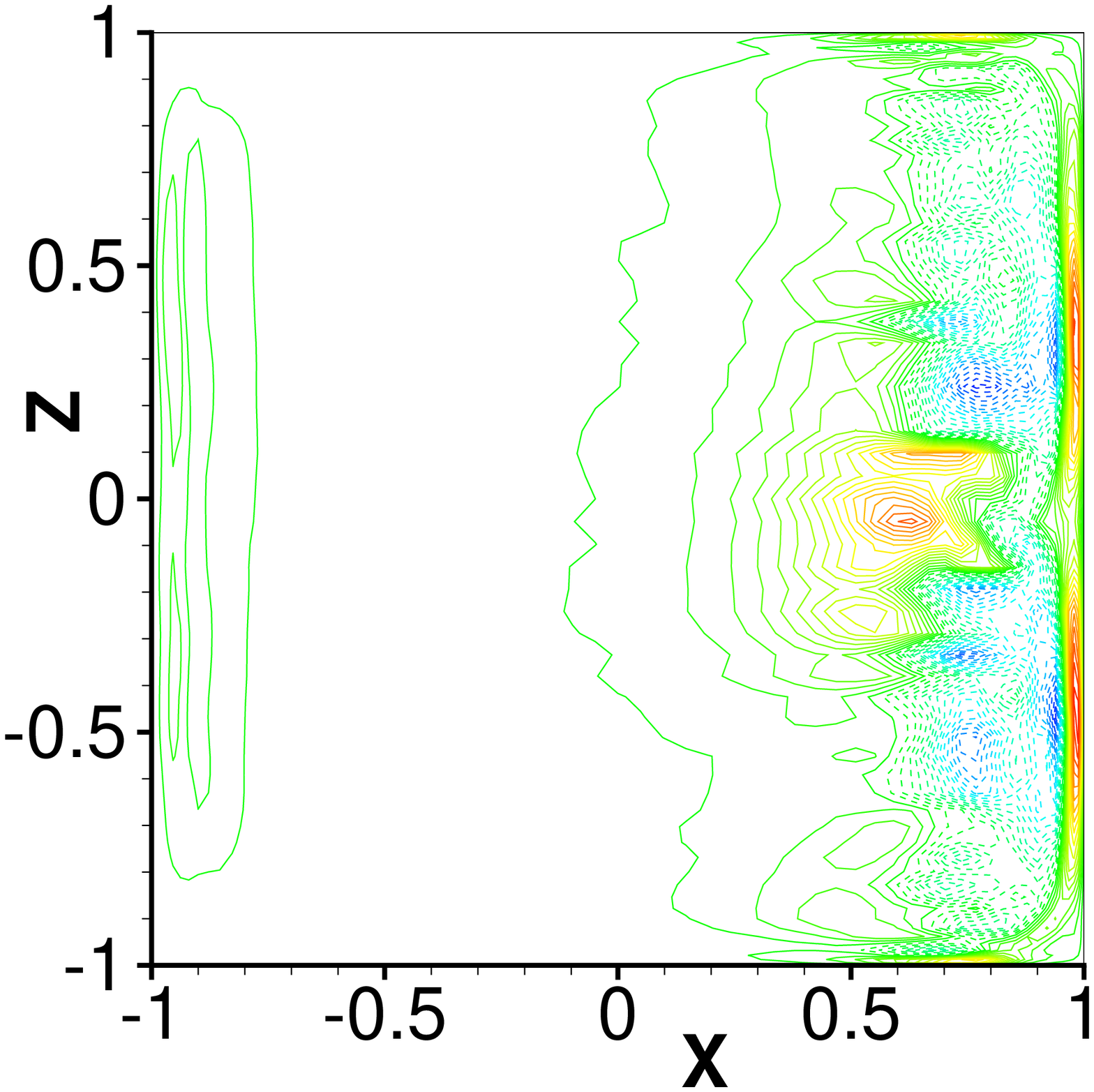}
\includegraphics[width=4.2cm]{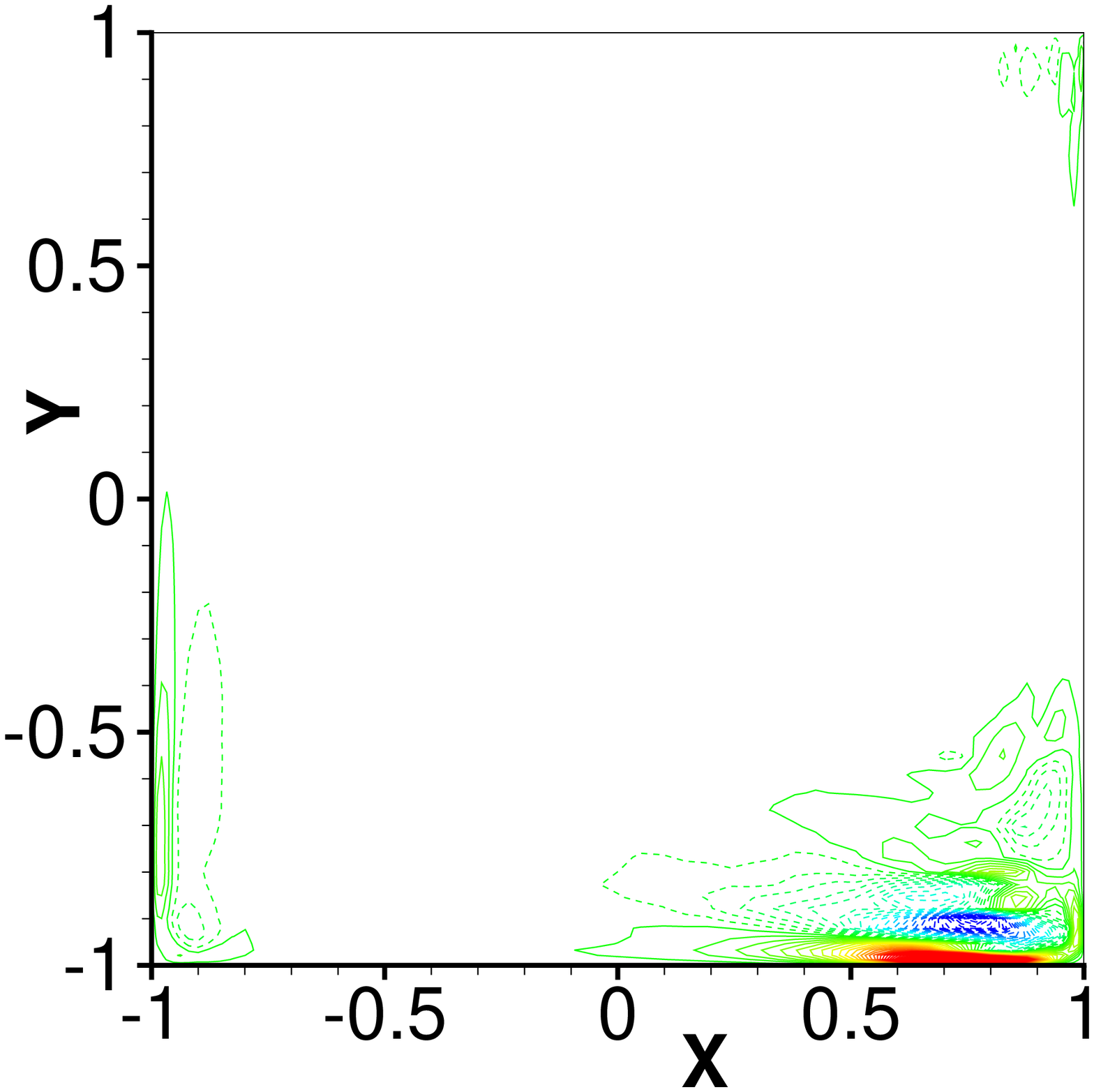}
\caption{Contours of $\m{\delta}$ in the plane $y/h=-0.968$ just above the bottom wall (left) and in the mid-plane $z/h=0$ (right); 100 equally spaced contours corresponding to levels between the threshold $0.01|\m{\delta}_{\textrm{max}}|$; dashed contours correspond to negative levels with a colormap ranging from blue to red; LES-DMM}\label{fig:Inhomogeneity_2D}
\end{figure}

As expected, one can observe in Fig. \ref{fig:Inhomogeneity_2D} that in the region near the downstream wall where the two primary elliptical jets are impinging on the bottom wall, the flow is highly inhomogeneous. More specifically, the inhomogeneity is more important in the zone in between the two elliptical jets where the flow is ejected and recirculating. Likewise similar patterns with lower magnitudes are detected in the regions where the secondary jets and the tertiary jets are impinging. The secondary jets are impinging on the bottom of the upstream wall producing an inhomogeneous turbulence visible in Fig.  \ref{fig:Inhomogeneity_2D} for values of $x/h$ close to $-1$. For the tertiary jets, impinging on the upstream part of the lid, the inhomogeneity is only visible in the 3D view in Fig. \ref{fig:Inhomogeneity_3D}.

%-------- 2 The turbulence production near the bottom wall --------------
\subsection{The turbulence production near the downstream wall}
As mentioned by Leriche and Gavrilakis in \cite{leriche00:_direc}, the largest turbulence production rates in the cavity are to be found in the primary elliptical jets parallel to the downstream wall, near the impact points just above the bottom wall. The budget equations of the resolved second-order moments $\m{u_iu_j}$ governing the turbulence energetics---see \cite{mathieu00:_introd_turbul_flow} and \cite{pope00:_turbul_flows} for greater details---comprise a term named here $P_{ij}$, defined by
\begin{equation}\label{eq:Pij}
P_{ij} = - \m{u_iu_k} \frac{\partial \m{U_j}}{\partial x_k} - \m{u_ju_k} \frac{\partial \m{U_i}}{\partial x_k},
\end{equation}
and corresponding to the interaction of the resolved mean flow and the resolved Reynolds stress tensor. $P_{ij}$ can be interpreted as responsible for the production of Reynolds stresses or in other words for the production of turbulence. 

\subsubsection{Maximum of turbulence production near the downstream wall}\label{sec:max-production}
\begin{figure}[htbp]
\includegraphics[width=6cm]{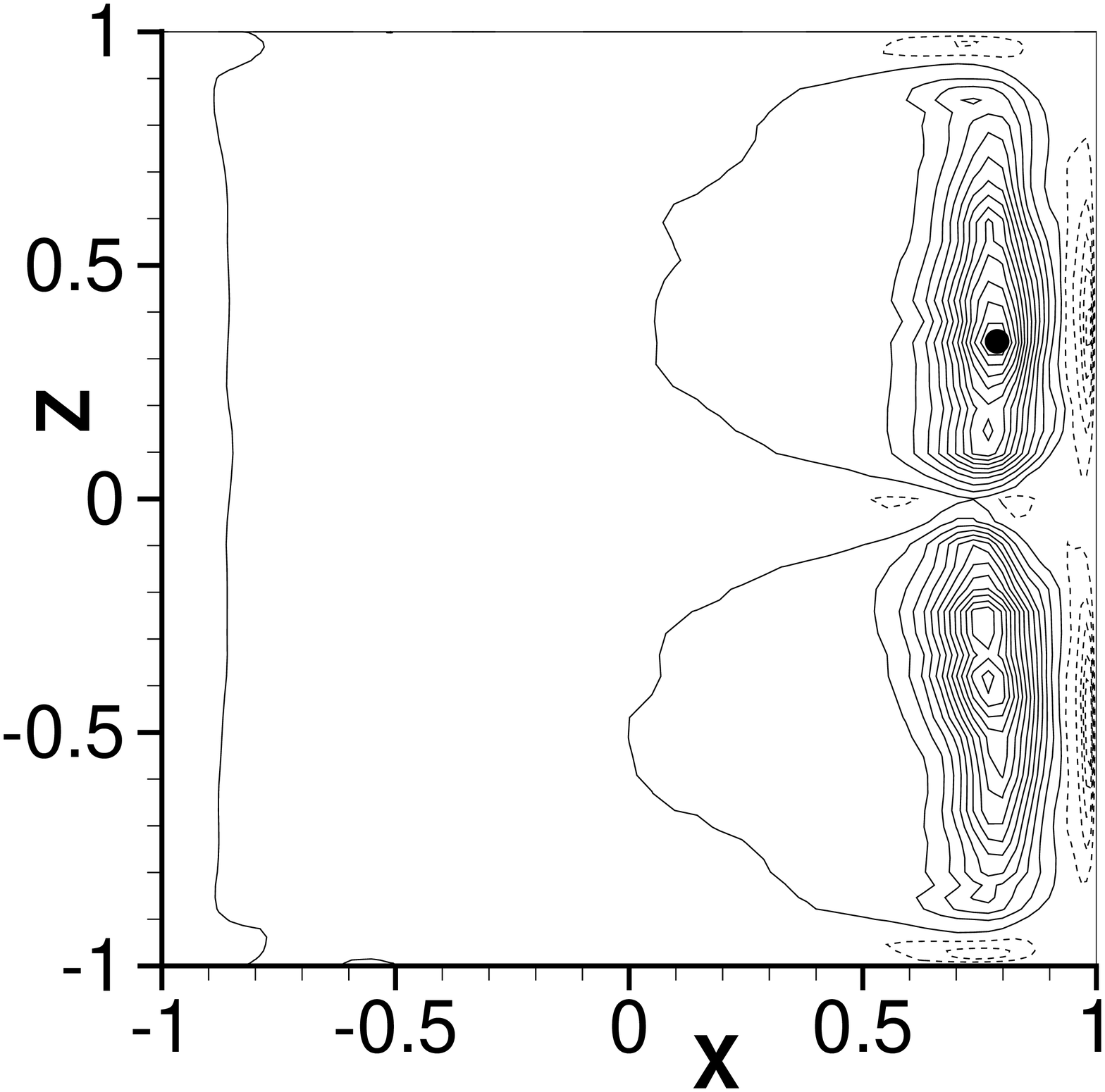}\\
\includegraphics[width=6cm]{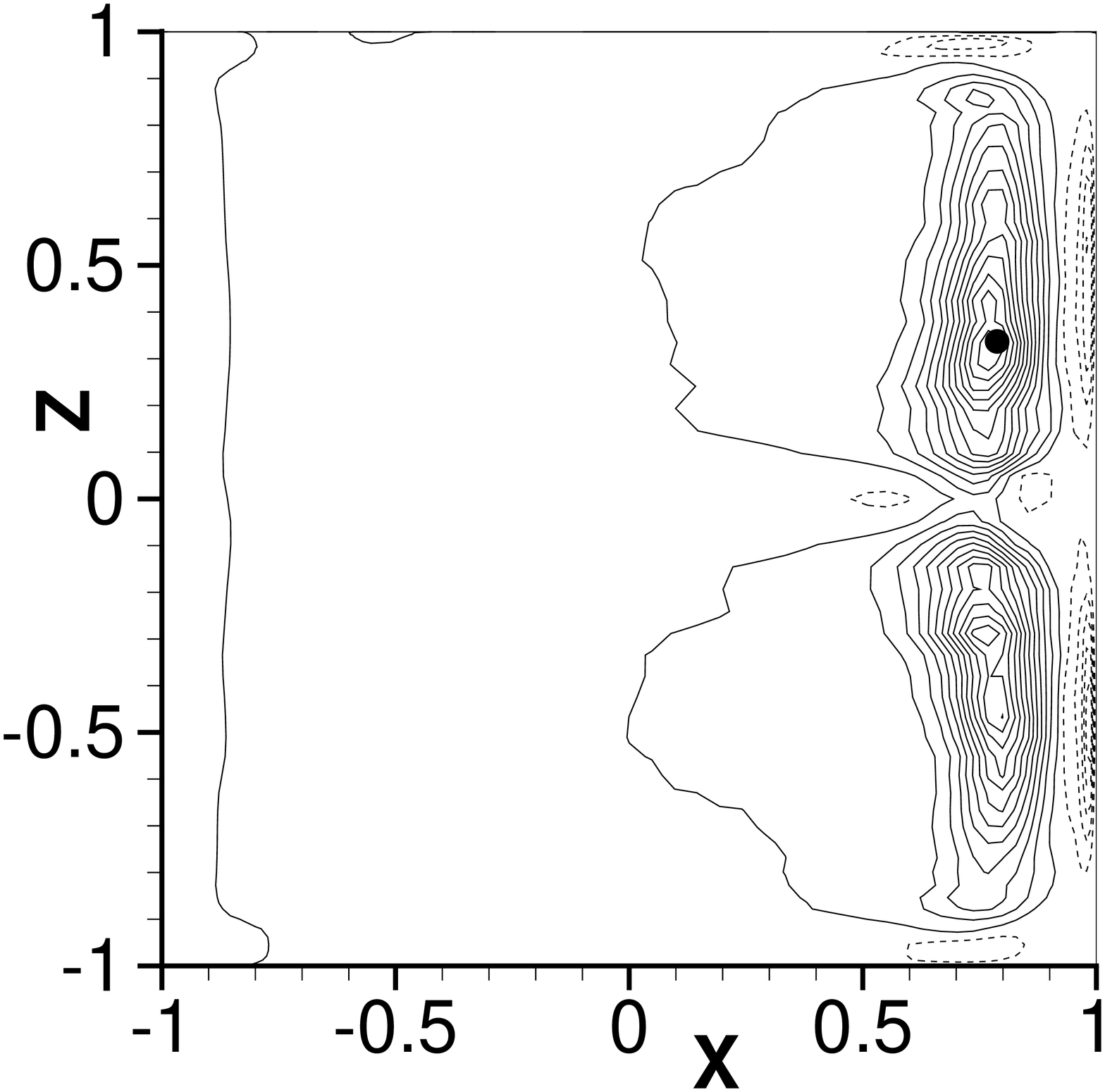}
\caption{Contours of resolved the production of turbulence term $P_{22}$ in the plane $y/h=-0.9388$; LES-DSM (top) and LES-DMM (bottom); 20 contour levels equally spaced between $-0.025\,U_0^3/h$ and $0.070\,U_0^3/h$; dashed lines refer to negative contour levels; bullet points $\bullet$ refer to the same grid node of coordinates ($x/h=0.7874$, $y/h=-0.9388$, $z/h=0.3371$)}\label{fig:P22_above_bottom}
\end{figure}

In the specific case of the separated downstream-wall jet, the term $P_{22}$ is the largest out of the set of turbulence production terms $\{P_{ij}\}$. After probing in the cavity, the maxima of the resolved field $P_{22}$ is to be found in the plane $y/h=-0.9388$ just at a very short distance above the bottom wall. The maximum values obtained are $P_{{22}_{\textrm{max}}}^{\textrm{DSM}}=0.070\,U_0^3/h$ and $P_{{22}_{\textrm{max}}}^{\textrm{DMM}}=0.064\,U_0^3/h$. The contours of the turbulence production term $P_{22}$ in this plane $y/h=-0.9388$ are shown in Fig. \ref{fig:P22_above_bottom} for both LES models. First, it can be noted that these contours are qualitatively very close to the ones obtained by Leriche and Gavrilakis in \cite{leriche00:_direc}. For $x/h>0.5$, the distribution of contours of the production of turbulence allow to clearly visualize the trace of the separated elliptical jets.

\subsubsection{Time histories and power spectra at the maximum of turbulence production}
In Fig. \ref{fig:P22_above_bottom}, one can notice that for each LES model, one grid point---identical for both DSM and DMM---has been highlighted with a bullet point $\bullet$. This point denoted by $\Theta_0$ whose coordinates are $x/h=0.7874$, $y/h=-0.9388$, $z/h=0.3371$, is the closest grid point to the two maxima of $P_{22}$ for LES-DSM and LES-DMM. The point $\Theta_0$ provides the optimal search position for probing time histories of various turbulent fields in the sequel.

\begin{figure}[htbp]
\input{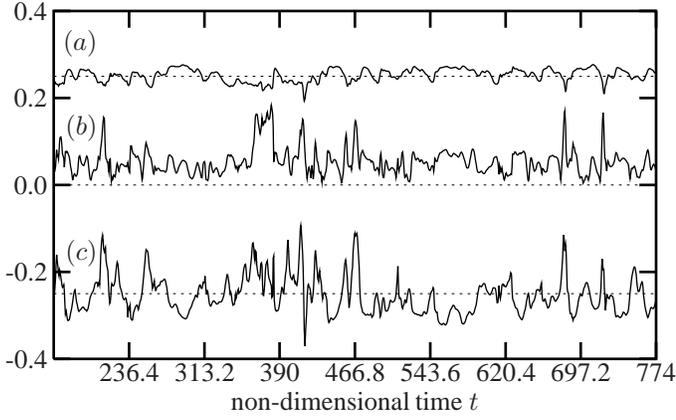}
\caption{Time histories of $p$ (graph $(a)$ shifted of $+0.25$), the turbulent kinetic energy $k$ (graph $(b)$) and $u$ (graph $(c)$ shifted of $-0.25$), at the point of coordinates $(0.7874,-0.9388,0.3371)$; LES-DSM database} 
\label{fig:time_history}
\end{figure}

First, the values of the $x$-component of the fluctuating resolved  velocity field $u$, of the fluctuating resolved pressure $p$ and the resolved turbulent kinetic energy $k=u_iu_i/2$, have been extracted of the LES-DSM database for each and every sample. These time histories are shown in Fig. \ref{fig:time_history}---note that only the last $1\,024$ samples out of the total of $1\,290$ that constitutes the database are presented.
\begin{figure}[htbp]
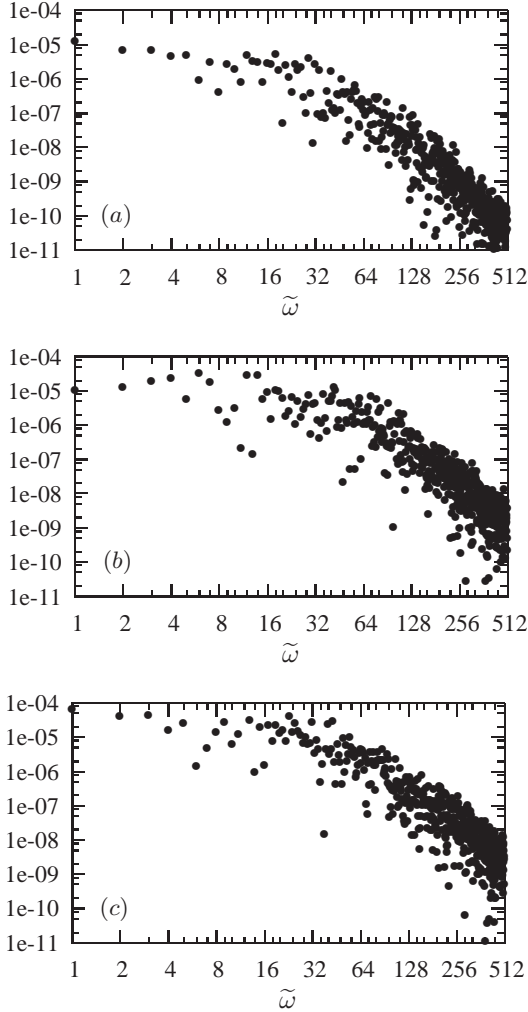

\input{data_p_energy_spectrum.pslatex}\\
\input{data_k_energy_spectrum.pslatex}\\
\input{data_u_energy_spectrum.pslatex}
\caption{Power spectra for the fluctuating pressure $p$ $(a)$, the turbulent kinetic energy $k$ $(b)$, and the fluctuating velocity field $u$ $(c)$, obtained from the time histories in Fig. \ref{fig:time_history}; LES-DSM database}\label{fig:power-spectra}
\end{figure}

\begin{figure}[htbp]
\includegraphics[width=6cm]{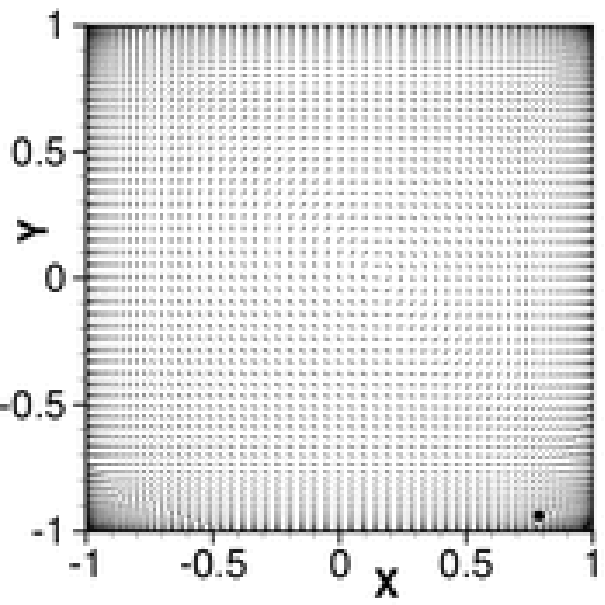}\\
\includegraphics[width=6cm]{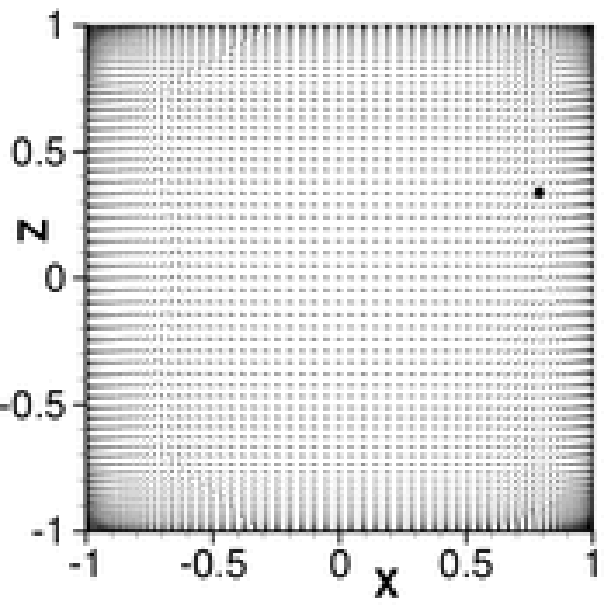}
\caption{Two-dimensional projected average resolved velocity vectors in the mid-plane $z/h=0$ (top) and in the plane $y/h=-0.9388$ (bottom); bullet point refers to $\Theta_0$; LES-DSM; bullet points $\bullet$ refer to the same grid node of coordinates ($x/h=0.7874$, $y/h=-0.9388$, $z/h=0.3371$)}\label{fig:mean-flow}
\end{figure}

Based on these results the corresponding power spectra have been computed by fast Fourier transform---\textit{a posteriori} justifying the choice of $1\,024$ samples in the previous time histories---and are presented in Fig. \ref{fig:power-spectra}. The scattering of points in the high-wavenumber $\widetilde{\omega}$ zone is expected for spectra of non-spatially average fields. For such inhomogeneous flow with highly localized turbulent effects, averaging in space the fields not only pleasantly reduce the scattering of points but concurrently strongly modifies the high-wavenumber scaling which is the main source of information brought by the spectra. Nevertheless, the spectra offer a qualitative information regarding the Eulerian time scales of the spatial structures of turbulence convected past the point $\Theta_0$. The resolved mean flow depicted in Fig. \ref{fig:mean-flow}---highlighting the presence of the core central primary vortex and secondary corner vortices---serves merely to convect turbulence at the bullet-spotted point $\Theta_0$. A careful scrutinizing of the resolved mean flow in the vicinity of $\Theta_0$ shows that this point is exactly positioned at the ``interface'' between the core central vortex and the bottom corner vortex.
The power spectra in Fig. \ref{fig:power-spectra} feature the distribution of frequencies---or equivalently of time scales. These frequencies $\widetilde{\omega}$ refer to the convection past $\Theta_0$ of turbulent structures of size of order $\ell$ at a velocity of order $\mU(\Theta_0)$, leading to the relation
\begin{equation}\label{eq:frequency}
 \widetilde{\omega}=\frac{\mU(\Theta_0)}{\ell}.
\end{equation}
The average resolved velocity field at $\Theta_0$ being given, the spectra hence instruct us on the distribution of spatial scales of resolved turbulent structures convected by the mean flow at this point where the production of turbulence is maximum. Unfortunately, the relatively low sampling of the LES-DSM database and the not-long-enough simulation range interval does not permit to reach the highest frequencies of the order of $\mU(\Theta_0) /\Delta$ where $\Delta$ is the filter length---see Eq. \eqref{eq:filter-length}---defining the LES scale separation.

\subsubsection{Determination of coherent structures responsible for the peaks of turbulence production} \label{sec:coherent-structures}
\begin{figure}[htbp]
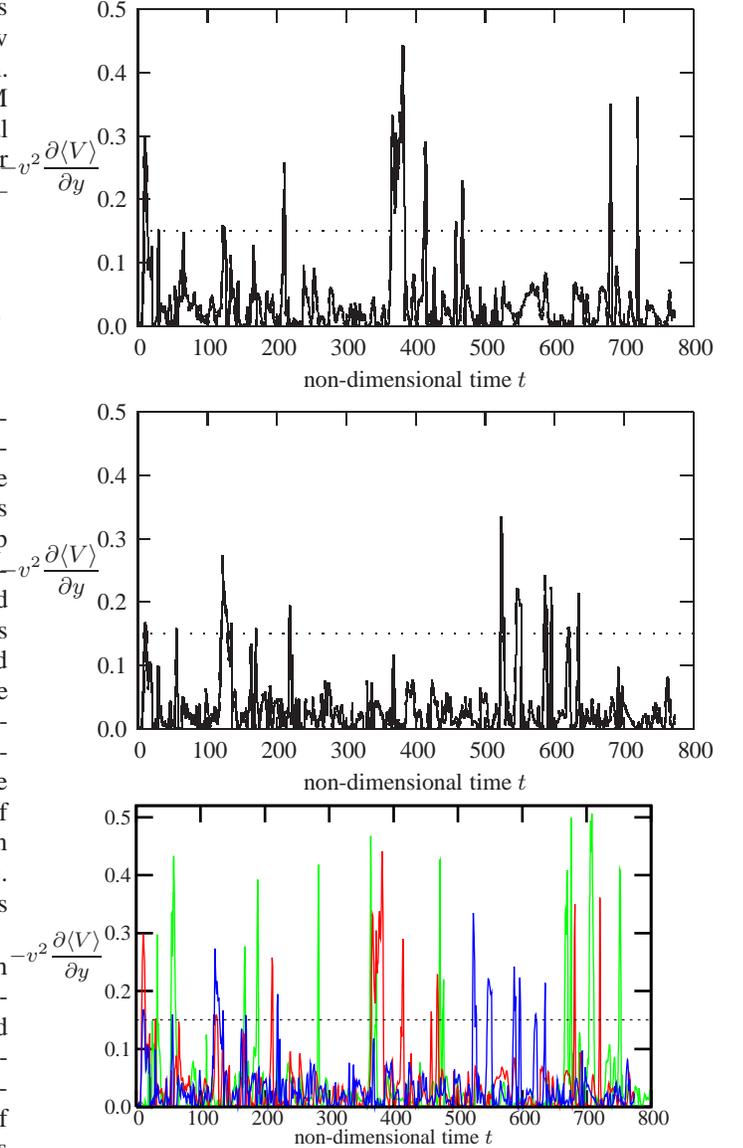

\input{Time_History_P22_DYN.tex}
\input{Time_History_P22_Dmixed.tex}
\input{Time_History_P22.pslatex}
\caption{Time histories of the resolved term $-v^2\partial \mV  /\partial y$ in $U_0^3/h$ units for the DNS (green),  LES-DSM (red) and the LES-DMM (blue); the dotted lines represent the threshold value $0.15\, U_0^3/h$}\label{fig:Time_History_P22}
\end{figure}

In an attempt to provide a comprehensive and thorough assessment of the performances of both LES models, the determination of the coherent structures responsible for the intense turbulence production at the point $\Theta_0$ has been envisaged as an ultimate challenge for both SGS modeling. The first step towards this goal necessitates to study the instantaneous distribution of the resolved term $-v^2\partial \mV /\partial y$ which was found to be the predominant term in $P_{22}$, see \cite{leriche00:_direc}. Fig. \ref{fig:Time_History_P22} displays the time histories of this term for the DNS, LES-DSM and LES-DMM. Both LES present a limited number of high-value peaks which are assumed to be engendered by specific coherent vortices or large eddies. The intensity of the peaks produced by the LES-DMM is lower than those generated by the LES-DSM. This is supposed to be due to an over-evaluation of the eddy viscosity by the dynamic procedure of the DMM. In addition, this is consistent with the observation made in Sec. \ref{sec:max-production} where $P_{{22}_{\textrm{max}}}^{\textrm{DMM}}<P_{{22}_{\textrm{max}}}^{\textrm{DSM}}$ was found and with the values of resolved $\m{\cK}$ and $\m{\kappa}$ in Table \ref{tab:cKkappa}.

\begin{figure}[htbp]
\includegraphics[width=4.2cm]{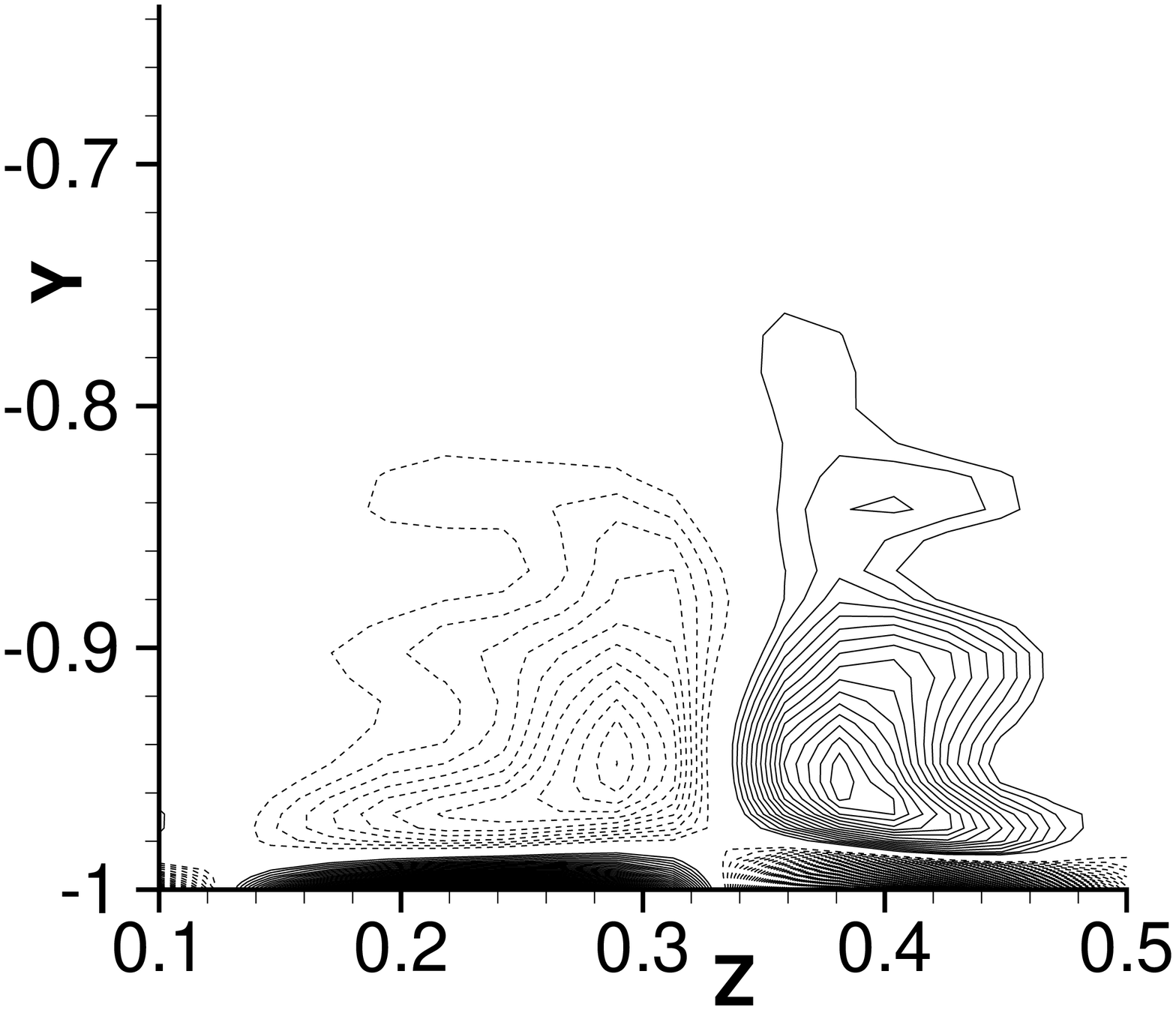}
\includegraphics[width=4.2cm]{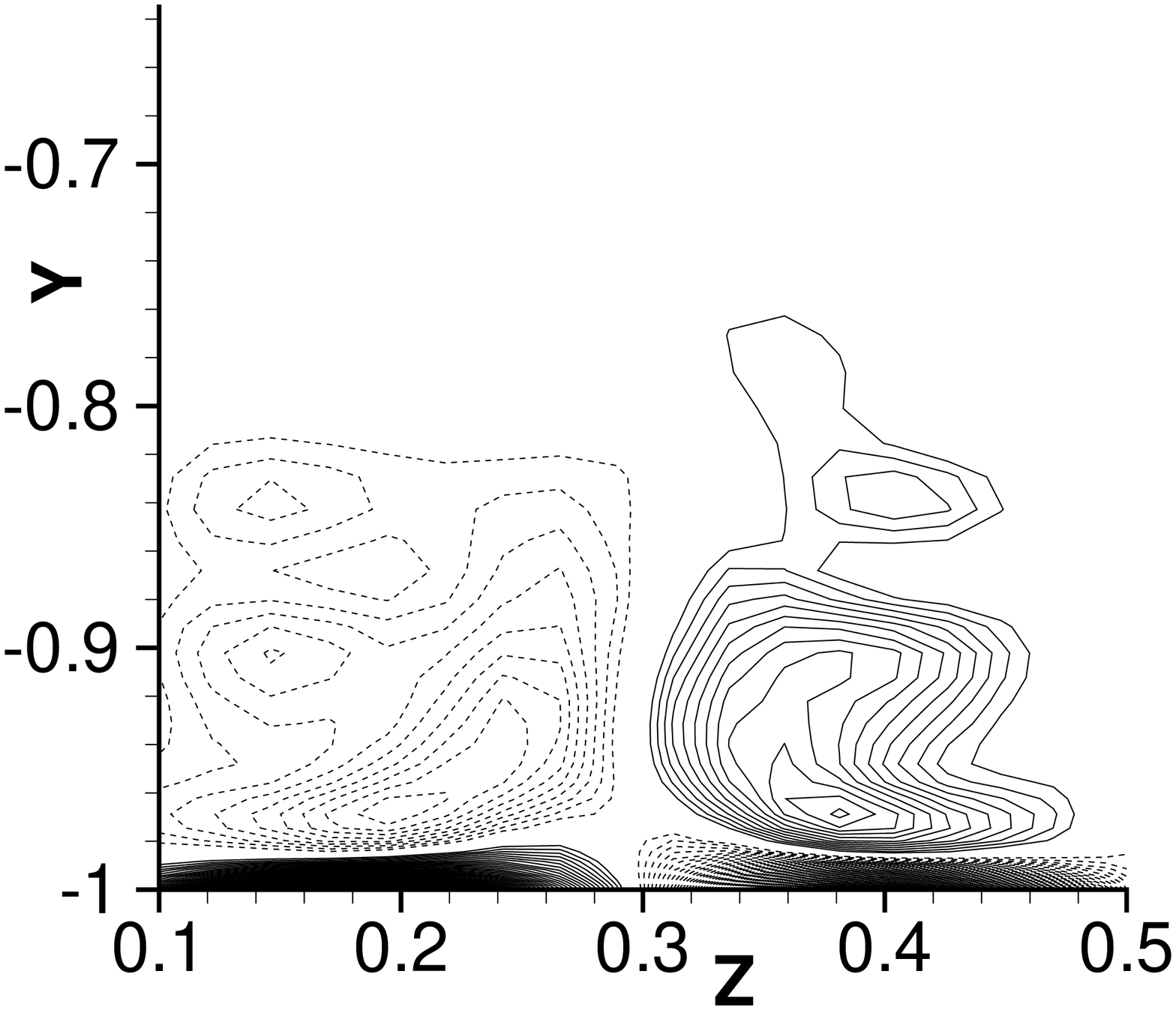}
\caption{Contours of the conditionally averaged resolved vorticity field $\m{\Omega_x}_{\textrm{ca}}$ in the plane $x/h=0.7874$ for the LES-DSM (left) and the LES-DMM (right); dashed lines refer to negative contour levels}\label{fig:Conditional_Averaging}
\end{figure}

In order to finally characterize possible large eddies which would be responsible for these peaks, database samples producing a resolved term $-v^2\partial \mV /\partial y$ above the threshold value $0.15\,U_0^3/h$ were put aside to form a subset of the complete databases. The size of the subset of samples for LES-DSM (resp. LES-DMM) is approximately $6\%$ (resp. $5\%$) of the size of the complete database. Based on these two subsets a conditional averaging of the streamwise resolved vorticity field $\Omega_x$ is performed. Fig. \ref{fig:Conditional_Averaging} displays the contours of this quantity in the vicinity of $\Theta_0$---the domain represented corresponds to only $4\%$ of the surface of a normal section of the cavity---for both models. Two counter-rotating vortices are clearly exhibited by both models, together with the intense influenced shear layers laying on the bottom wall at $y/h=-1$. This vortex pair is identified as the coherent eddy responsible for the turbulence peaks and production in this region. The characteristic length scale of this large eddy is of the order of $0.1h$. Having identified this vortex pair we can further analyze the time histories at $\Theta_0$ of the resolved pressure and the resolved turbulent kinetic energy depicted on Fig. \ref{fig:time_history}, graph $(a)$ and $(b)$ respectively. One can notice that the intense peaks of the resolved term $-v^2\partial \mV /\partial y$ on Fig. \ref{fig:Time_History_P22} correspond to intense peaks of resolved turbulent kinetic energy on Fig. \ref{fig:time_history} $(b)$ and to low-pressure peaks on Fig. \ref{fig:time_history} $(a)$. The vortex pairs generated by this turbulent flow are responsible for the low pressures and the high turbulent kinetic energy thereby justifying the observed correlations between these three time histories. Moreover the intensity of the vortex pair calculated by the LES-DSM is again higher than the one from the LES-DMM: $\m{\Omega_{x}^{\textrm{DSM}}}_{\textrm{ca}_{\textrm{max}}}=18.6 \, U_0/h$ and $\m{\Omega_{x}^{\textrm{DMM}}}_{\textrm{ca}_{\textrm{max}}}=14.0 \, U_0/h$, where the subscript ``ca'' stands for conditionally averaged. The intensity, the more regular structure and the localization of the vortex pair are three features suggesting that the dynamic Smagorinsky model provides a better SGS modeling than the dynamic mixed model. Nevertheless, the DMM performances in terms of SGS modeling are more than satisfactory. When considering the complete averaging of the $x$-component of the resolved vorticity field in the region where the vortex pair has been localized by conditional averaging, see Fig. \ref{fig:Conditional_Averaging}, $\m{\Omega_x}$ was found almost constant and of magnitude approximately $0.9\,U_0/h$.

%-------- 3 Localization of small-scale turbulent structures-------------
\subsection{Small-scales turbulent structures}\label{sec:small-scales-turbulence}
A characteristic of high-Reynolds-number turbulence is that the vorticity possesses intense small-scale, random variations in both space and time. The spatial scale for vorticity fluctuations is the smallest in the continuum of turbulent scales, i.e the Kolmogorov scale. Analogously to the vorticity fluctuations, for large-Reynolds-number turbulence velocity gradients $\partial{u_i}/\partial x_j$ are also dominated by the small scales of turbulence and the overall energy dissipation rate of kinetic energy is dominated by the average turbulent energy dissipation rate $\m{\eps}$ defined in Eq. \eqref{eq:def-meps}. In the LES framework, the interest for small scales is twofold. First, small scales fall into the range of subgrid scales and therefore are not simulated but wholly modeled to properly reproduce their interactions with larger resolved scales of the flow. Second, the small scales have the crucial role to terminate the turbulent energy cascade by dissipating the energy originating from large eddies. An incorrect SGS modeling will produce either an over-dissipation or conversely an under-dissipation of kinetic energy. The time histories of the total kinetic energy of the cavity flow $\cK (t)$ and of the total turbulent fluctuating energy $\kappa (t)$ presented in Fig. \ref{fig:time-history-energy} are in this framework a precious proof of the correct global prediction of the energy dissipation by the modeled small scales in volume.

\subsubsection{Localization of small-scales structures}
\begin{figure}[htbp]
\includegraphics[width=8cm]{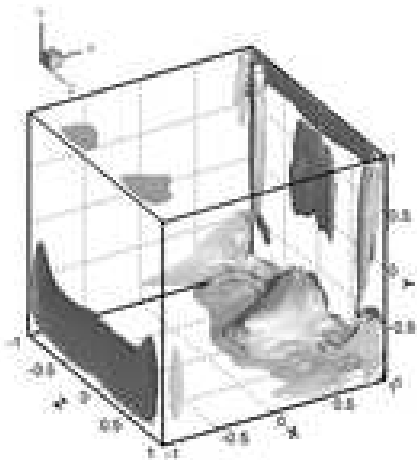}
\caption{Visualization of the region of the cavity where the average turbulent energy dissipation rate $\m{\eps}$ is above $1\%$ of its maximum value $3\,570\,\nu U_0^2/h^2$; LES-DMM}\label{fig:Small_Scales_3D}
\end{figure}
\begin{figure}[htbp]
\includegraphics[width=4.2cm]{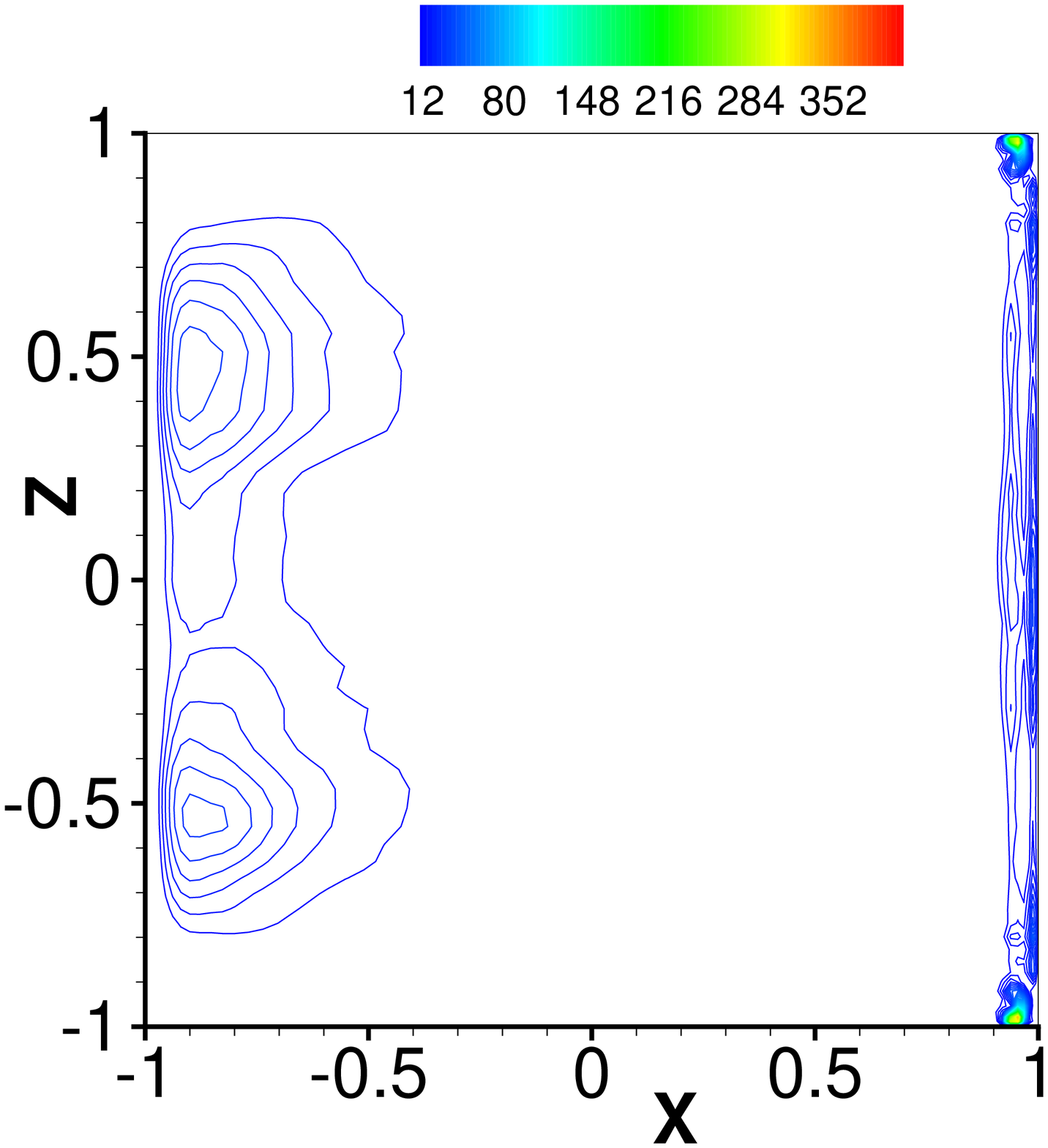}
\includegraphics[width=4.2cm]{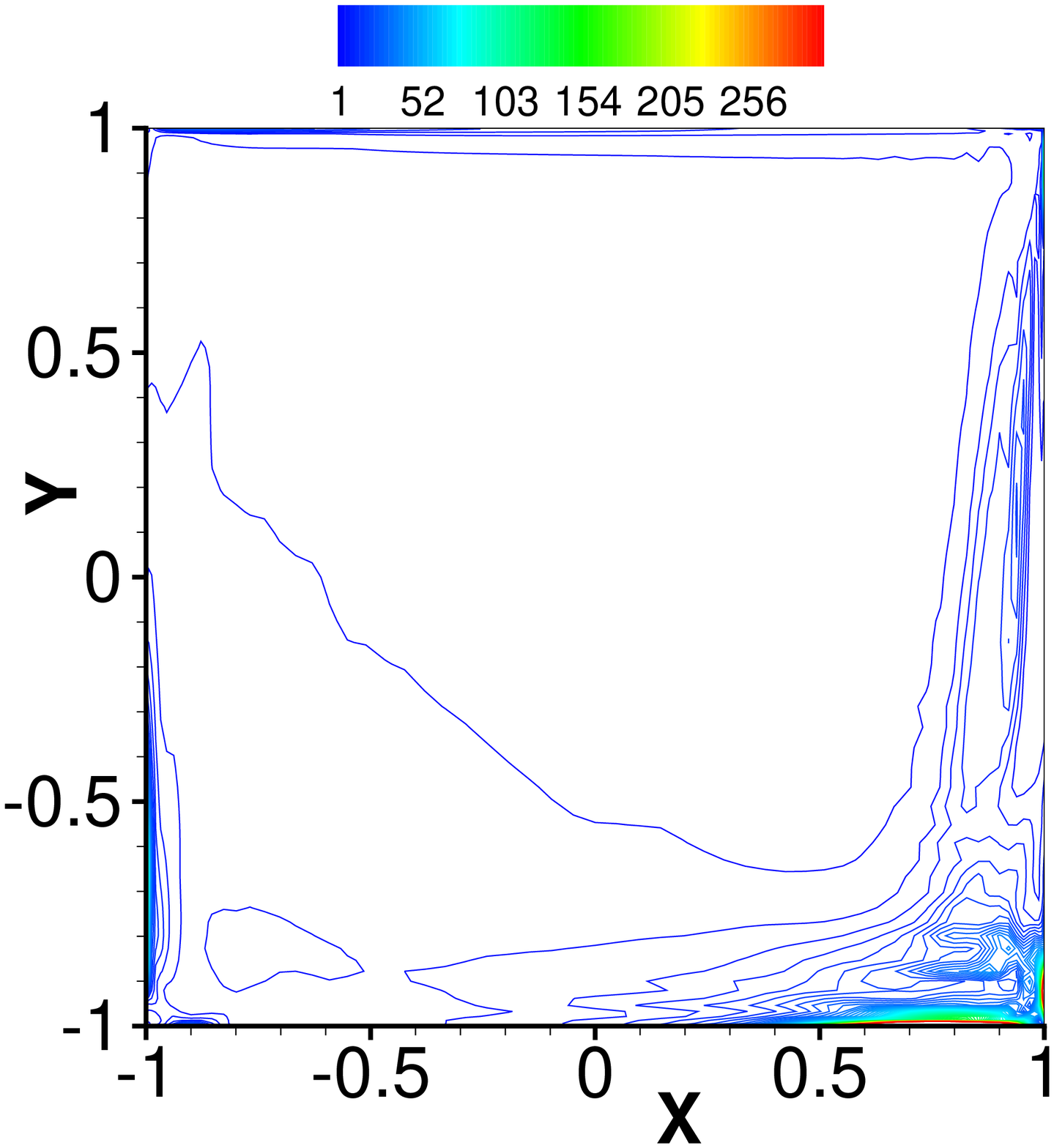}\\
\includegraphics[width=4.2cm]{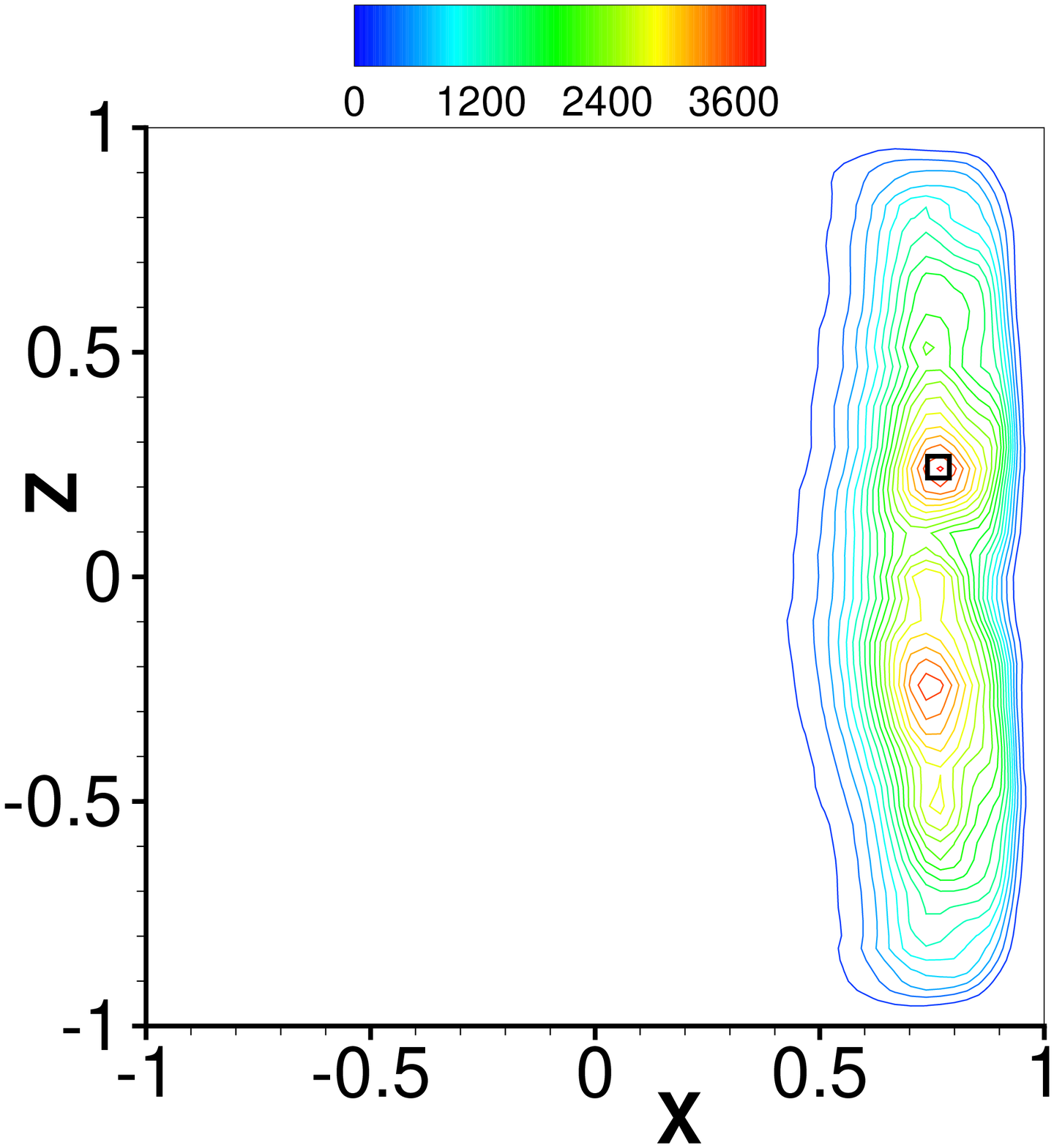}
\includegraphics[width=4.2cm]{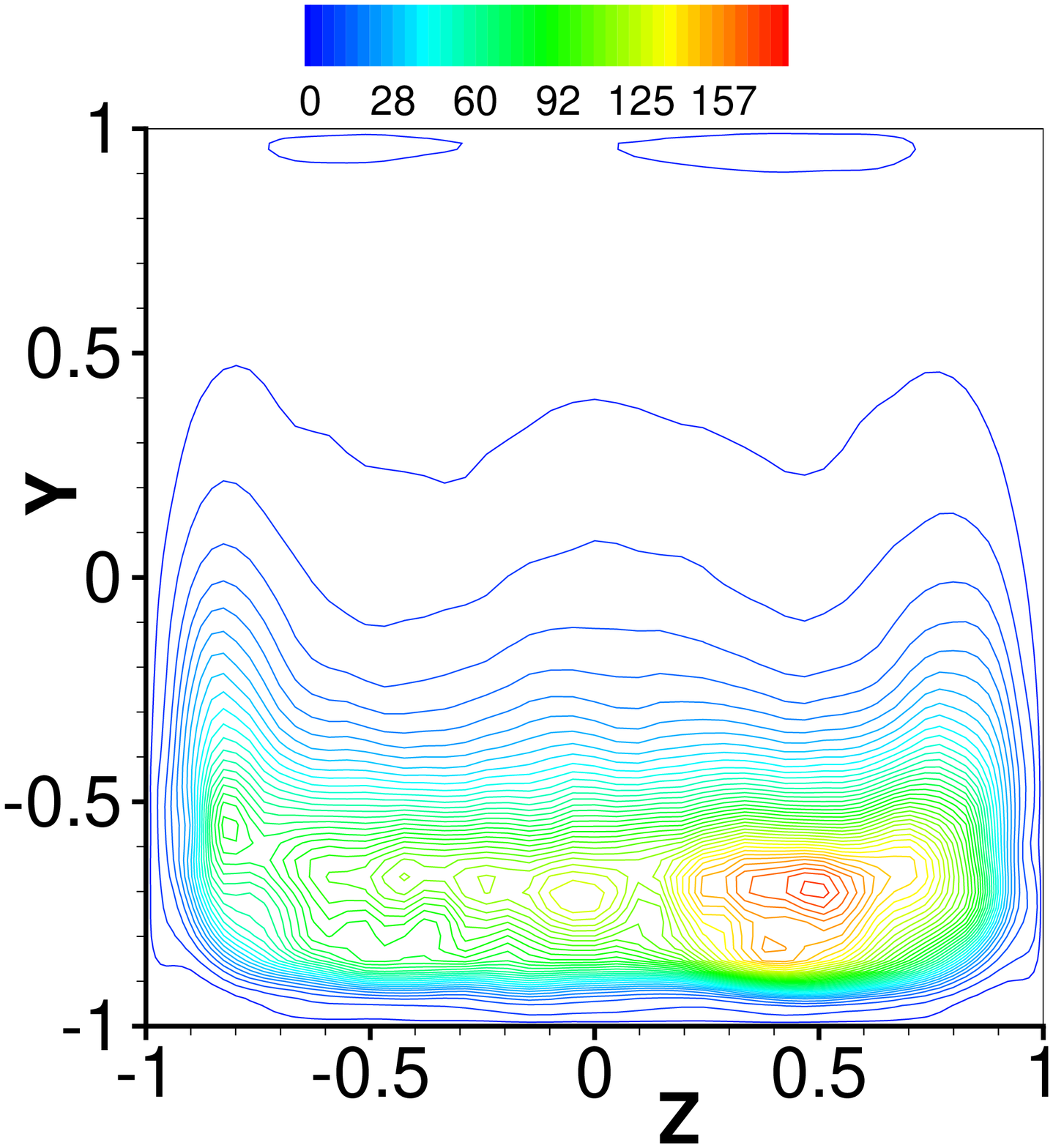}
\caption{Two-dimensional contour lines of $\m{\eps}$ in the following planes: lid-plane $y/h=1$ (top-left), plane $z/h=0.241$ (top-right), bottom-plane $y/h=-1$ (bottom-left), upstream-plane $x/h=-1$ (bottom-right); LES-DSM; black rectangle $\square$ refers to the grid node of coordinates ($x/h=0.7685$, $y/h=-1$, $z/h=0.2410$)}\label{fig:Small_Scales_2D}
\end{figure}

In this context, it appears relevant to first locate small-scales turbulent structures in the cavity and afterwards to check the correlation between the small-scales positioning and the activation of the SGS modeling represented in Fig. \ref{fig:nuT-Zmidplane}. Small scales can be indirectly localized by investigating the zones of intense average turbulent energy dissipation rate. Indeed $\m{\eps}$ involves products of fluctuating velocity gradients, see Eq. \eqref{eq:def-meps}. First qualitatively, the region of the cavity flow corresponding to values of $\m{\eps}$ above $1\%$ of its maximum value is shown in Fig. \ref{fig:Small_Scales_3D} for the LES-DMM. As foreseen, the wall-jet-impinging regions are subject to intense turbulent energy dissipation at the small-scales level. The two-dimensional cuts in Fig. \ref{fig:Small_Scales_2D} offer a more detailed information regarding the intensity of $\m{\eps}$ in four different planes of specific interest. It is worth keeping in mind that the more intense $\m{\eps}$ the more small scales are involved in the dissipation process. Fig. \ref{fig:Small_Scales_2D} displays with decreasing intensity, the dissipation due to the impingements of the separated wall jets on the bottom wall (bottom-left), on the upstream wall (bottom-right) and on the lid-plane (top-left). It appears that the LES-DSM is not able to properly reproduce the same intensity for the two symmetric jets impinging on the upstream wall (bottom-right). The same asymmetry in the intensity of $\m{\eps}$ is observed for the LES-DMM which could presumably be due to the observed asymmetry---with respect to the mid-plane---of the eddy viscosities generated by the dynamic procedures of both SGS modeling. 

In Fig. \ref{fig:Small_Scales_2D} (bottom-left), one can notice that one grid point has been highlighted with a black rectangle $\square$. This point denoted by $\Xi_0$ whose coordinates are $x/h=0.7685$, $y/h=-1$, $z/h=0.2410$, is the closest grid point to the maximum of $\m{\eps}$ for LES-DSM. The point $\Xi_0$ provides the optimal search position for probing small-scales related fields. The plane-cut $z/h=0.241$---passing by $\Xi_0$---of $\m{\eps}$ in Fig. \ref{fig:Small_Scales_2D} (top-right) exhibits a qualitative correlation with the same plane-cut for the average eddy viscosity $\m{\nu_{\textrm{sgs}}}$ in Fig. \ref{fig:nuT-Zmidplane} (right). 

\subsubsection{Correlation between small-scales localization and eddy viscosity}
Such correlation between the small-scales localization and the activation of the LES dynamic Smagorinsky modeling is important in practice to ensure the effectiveness of the SGS modeling. Therefore a more quantitative approach is required, which relies on the calculation of a correlation field based on the instantaneous values of $\eps$ and $\nusgs$. The following correlation coefficient $\cC$, defined by
\begin{equation}\label{eq:correlation}
  \cC = \cC(\eps,\nusgs ) = \frac{\m{\eps\,\nusgs } - \m{\eps}\mnusgs }{(\m{\eps ''^2} \m{\nusgs ''^2})^{1/2}},
\end{equation}
where $''$ stands for the fluctuating part of the considered field, was calculated for the complete set of samples LES-DSM. 
\begin{figure}[htbp]
\includegraphics[width=6cm]{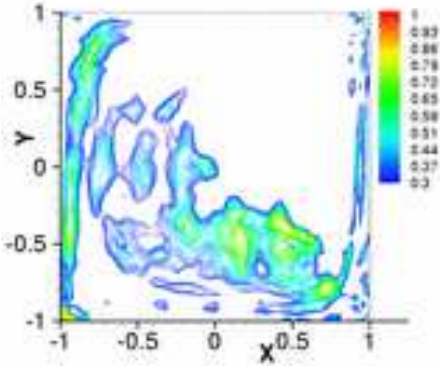}
\caption{Contours of the correlation coefficient $\cC$ in the plane $z/h=0.241$ containing the point $\Xi_0$; LES-DSM}\label{fig:contours-correlation}
\end{figure}

Contours of $\cC$ in the plane $z/h=0.241$, passing by $\Xi_0$ are presented in Fig. \ref{fig:contours-correlation}. The high-correlation zones reproduce in essence the turbulent-dominated regions of the cavity and even suggest the mean-flow convective effect of the central core vortex and other secondary corner vortices on the turbulent pockets. Nevertheless, higher correlations would have been expected in the vicinity of $\Xi_0$. Such low correlations are evidences of the limitations of the LES in this region of the cavity flow. Conversely, the high correlations near the upstream wall are in good agreement with the small-scales localization. More precisely the poor correlation in the vicinity of $\Xi_0$ is imputed to the fact that in this region, the term $S_{ij}S_{ij}$ in the turbulent energy dissipation rate---see Eq. \eqref{eq:def-meps}---varies very rapidly in space likewise the eddy viscosity. At this point, the information provided by the analysis of the subgrid-scale activity in the next section is a good complement to the previous correlation study.

\subsubsection{Subgrid-scale activity}
The filtered kinetic energy can be decomposed into the kinetic energy of the resolved velocity field and the residual kinetic energy which is equal to $\tau_{ii}/2$. The conservation equation for the kinetic energy of the resolved velocity field---see pp. 585--586 in \cite{pope00:_turbul_flows} for greater details---comprises transport terms as well as source/sink terms which are of prime interest. First is the sink term $\epsilon_\nu=2\nu \overline{S}_{ij}\overline{S}_{ij}=2\overline{S}_{ij}\overline{S}_{ij}/\textrm{Re}$ corresponding to the viscous dissipation associated with the resolved velocity field. The second sink term $\epsilon_{\textrm{sgs}}=-\tau_{ij}^\dd \overline{S}_{ij}$ corresponds to the SGS contribution and represents the rate of transfer of energy from the resolved scales of the flow to the subgrid scales. This term $\epsilon_{\textrm{sgs}}$ is often inappropriately referred to as the SGS dissipation in the literature. Indeed $\epsilon_{\textrm{sgs}}$ does not correspond to any physical dissipation but finds its origin in inertial processes. In addition, it is important to note that locally $\epsilon_{\textrm{sgs}}$ can take negative values.

The SGS activity, denoted by $\cAs$ in the sequel, allows to study the local energy fluxes due to the SGS effects. Following Geurts and Fr\"ohlich in \cite{geurts02} and Meyers \etal\ in \cite{meyers05:_optim}, $\cAs$ is defined as
\begin{equation}
\cAs = \frac{\epsilon_{\textrm{sgs}}}{\epsilon_{\textrm{sgs}}+\epsilon_\nu}=\frac{-\tau_{ij}^\dd \overline{S}_{ij}}{-\tau_{ij}^\dd \overline{S}_{ij}+2\nu \overline{S}_{ij}\overline{S}_{ij}}.
\end{equation}
The SGS activity $\cAs$ measures the importance of the subgrid scales in the overall dissipation process of the kinetic energy of the resolved velocity field. As mentioned by Meyers \etal\ in \cite{meyers05:_optim}, the SGS activity varies between zero and one where a value of zero corresponds to DNS and $\cAs=1$ is associated with LES at infinite Reynolds number. Moreover the value of $\cAs$ is directly related to the filter width $\Deltaf$ and measures the ``distance'' between a DNS resolving all flow features at sufficiently high spatial resolution and an actual LES corresponding to a specific filter width and mesh spacing. In the particular case of the LES-DSM, the SGS sink term is $\epsilon_{\textrm{sgs}}=2\nusgs \overline{S}_{ij}\overline{S}_{ij}=2\Cd \Deltaf^2 |\bSf|\,\overline{S}_{ij}\overline{S}_{ij}$, leading to
\begin{equation}
\cAs^{\textrm{DSM}} = \frac{\nusgs}{\nusgs+\nu}=\frac{\Cd \Deltaf^2 |\bSf|/\nu}{1+\Cd \Deltaf^2 |\bSf|/\nu}.
\end{equation}

\begin{figure}[htbp]
\includegraphics[width=5.5cm]{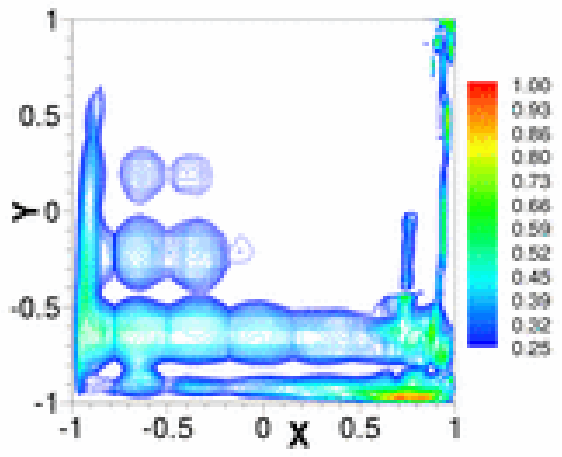}
\includegraphics[width=5.5cm]{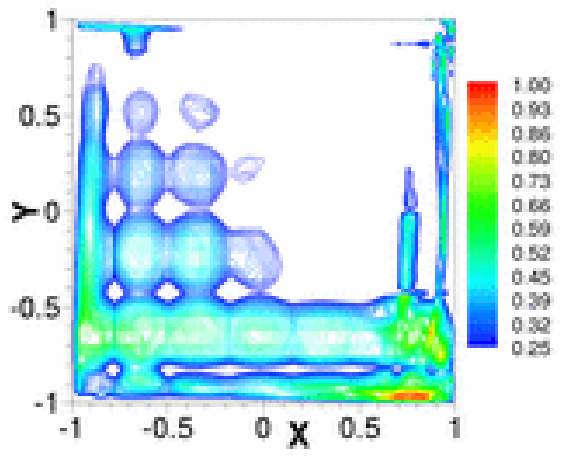}
\caption{Contours of the average SGS activity $\m{\cAs}$ in the plane $z/h=0.241$ containing the point $\Xi_0$: LES-DSM (top) and LES-DMM(bottom); same series of contour levels is used for both models}\label{fig:SGS-activity}
\end{figure}

Fig. \ref{fig:SGS-activity} displays the average value $\m{\cAs}$ of the SGS activity in the plane $z/h=0.241$ containing $\Xi_0$ and where the turbulent energy dissipation rate is maximum. First, it appears that the SGS activity is slightly higher for the LES-DMM than for the LES-DSM. Moreover, it appears very clearly that the SGS modeling is activated in the region of the cavity where the different wall jets are present, with maxima in the impingement zones. The LES-DMM is more effective in activating the subgrid scales in these particular zones. In the zone where the tertiary wall jet is impinging on the lid, SGS dissipation for the LES-DSM is less than $25\%$ of the total dissipation, whereas it is above $45\%$ for the LES-DMM.

%-------- 3 Vorticity and Helicity --------------
\subsection{Helical properties of the cavity flow}

The helicity $\cH$ of the fluid flow confined in the cavity $\cV$ at instant $t$ is defined by
\begin{equation}\label{eq:helicity-defintion}
\cH (t) = \int_{\cV} \bu \cdot \bomega \,\dd \cV,
\end{equation}
and is a measure of linkages and knots between the vorticity lines of the flow. The quantity $h(\xx,t)=\bu \cdot \bomega$ is the helicity density and is a pseudo-scalar quantity just like $\cH$. The helicity is an important flow quantity because just like the total energy of the flow $\cK$, it is an invariant of three-dimensional homogeneous turbulence \cite{moffat92:_helic}. The study of the resolved helicity $\cH$ and the average resolved helicity density $\m{h}$ in the particular context of the lid-driven cavity flow in a locally-turbulent regime allows to gain insights into very important features of the turbulence dynamics \cite{moffat92:_helic}. For instance TGL vortices and secondary corner eddies are structures encountered in the lid-driven cavity flow which are well known as typical helical structures.

\begin{figure}[htbp]
\includegraphics[width=4.2cm]{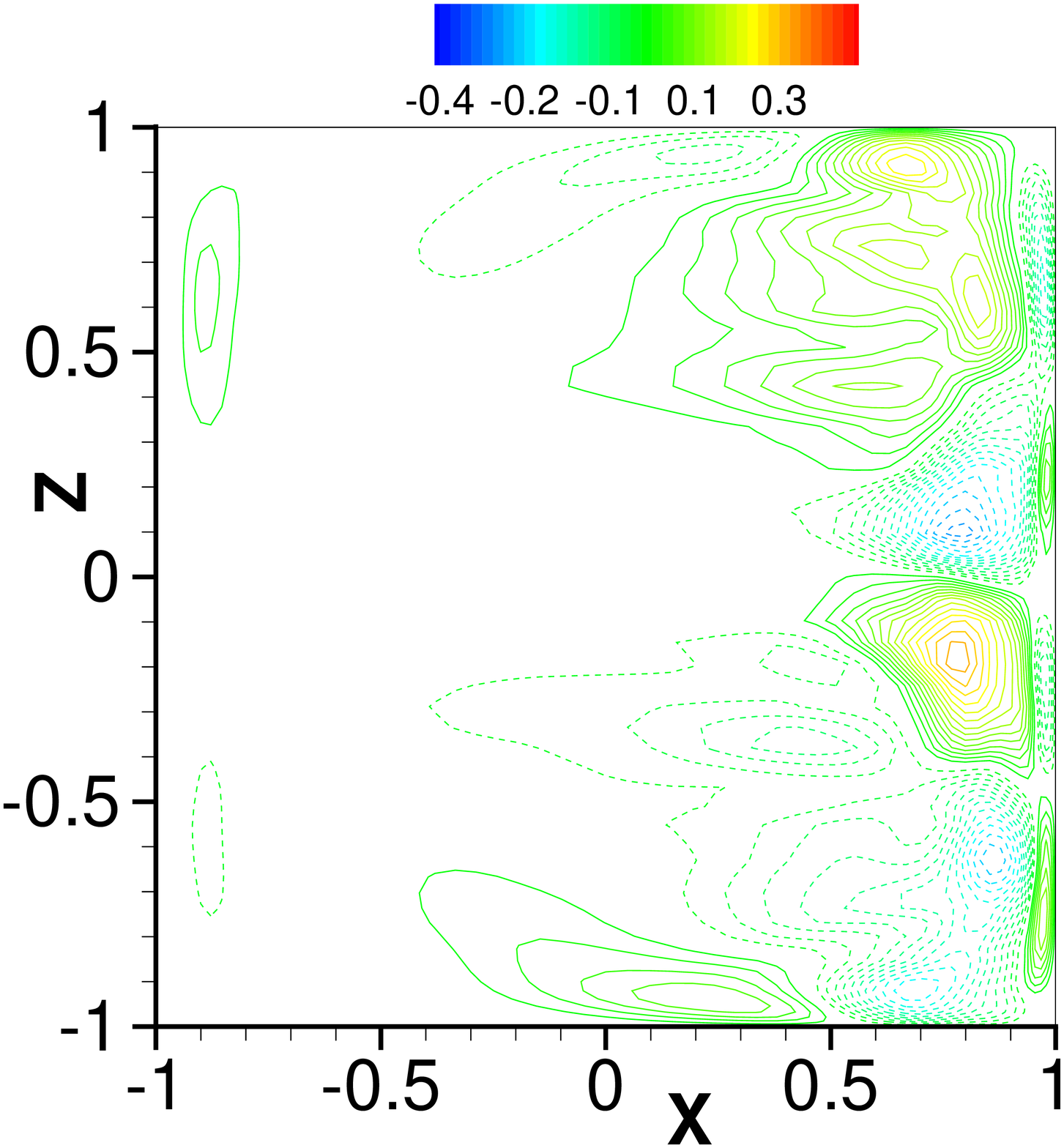}
\includegraphics[width=4.2cm]{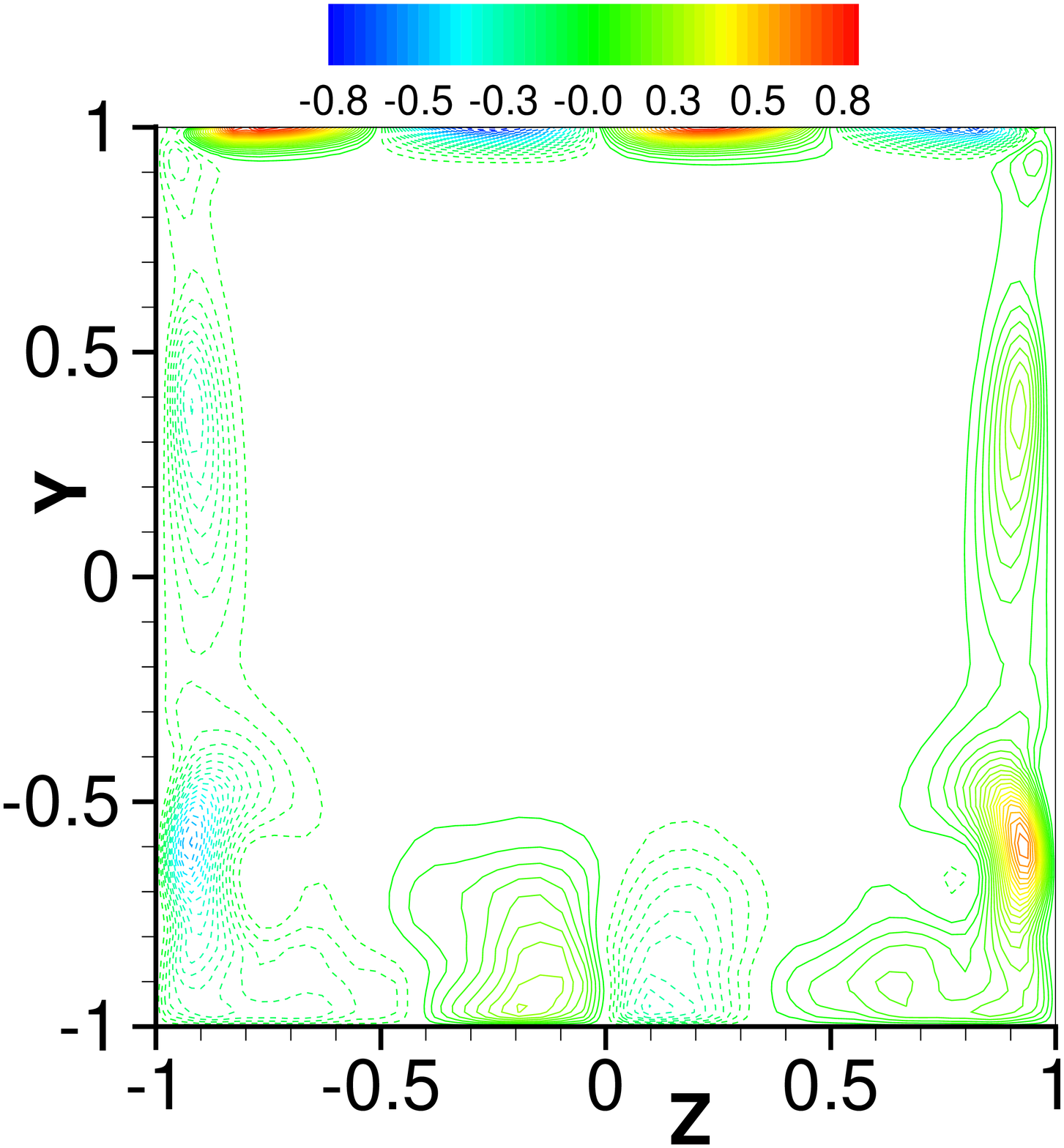}
\caption{Contours of the average resolved helicity density $\m{h}$ in the bottom plane $y/h=-1$ (left) and in the plane $x/h=0.7874$ containing $\Theta_0$; LES-DSM}\label{fig:helicity-density}
\end{figure}
Mappings of the average resolved helicity density in Fig. \ref{fig:helicity-density} allows to locate resolved helical coherent structures (HCS). These HCS are particularly intense in the secondary-corner-eddy region and are consistent with the experimentally observed typical HCS, namely streamwise counter-rotating vortices \cite{tsinober83}. This pairing of coherent helical structures correspond to a pairing of coherent vortical structures having opposite vorticity and consequently opposite helicity. Such observation justifies the relatively small---but non-zero---resolved average helicity reached by both LES models: $\m{\cH^{\textrm{DSM}}}=0.00764\,U_0^2h^2$ and $\m{\cH^{\textrm{DMM}}}=-0.00572\,U_0^2h^2$. Smaller HCS have been identified earlier in Sec. \ref{sec:coherent-structures}, where streamwise counter-rotating vortices---cf. Fig. \ref{fig:Conditional_Averaging}---near the bottom wall, have been identified by the conditional averaging as the principal coherent structures responsible for the high-intensity peaks in the production of turbulence in this region of the flow. Finally, it is noteworthy to emphasize the strong link between average resolved helicity density contours and average resolved turbulent kinetic energy dissipation rate ones in Fig. \ref{fig:Small_Scales_2D}.

\begin{figure}[htbp]
\input{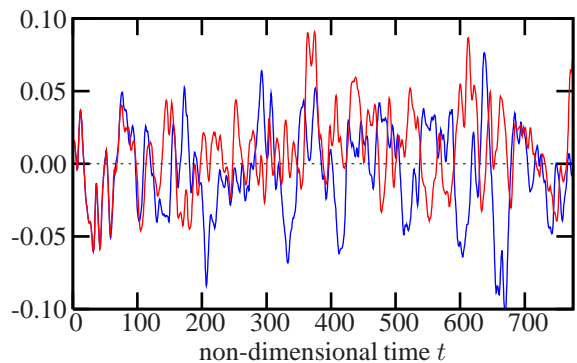}
\caption{Time histories of $\cH(t)$ for the LES-DSM (red lines) and the LES-DMM (blue lines)}\label{fig:helicity-time-histories}
\end{figure}

As mentioned previously the average resolved helicity of the flow is non-zero. The time histories of the resolved helicity for the whole simulations are shown in Fig. \ref{fig:helicity-time-histories} for both LES-DSM and LES-DMM. In Sec. \ref{sec:statistical-ensemble-averaging} was mentioned that both SGS models start being effective and producing different global results after a transient period of about $80\, h/U_0$ time units, and likewise the helicity as can be seen in Fig. \ref{fig:helicity-time-histories}. Moreover, the amplitude of the resolved helicity fluctuations is not decaying during the simulation and the LES-DMM qualitatively produces more high-amplitude negative helical values therefore justifying its negative average value.

\begin{figure}[htbp]
\input{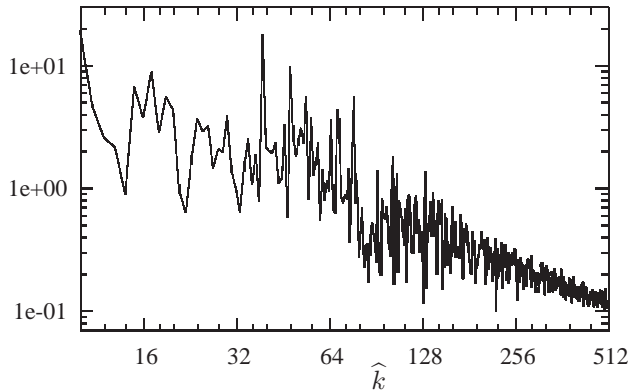}
\caption{One-dimensional relative helicity spectrum $\alpha(\widehat{k})$; LES-DMM}\label{fig:helicity_spectra}
\end{figure}

Helicity, like energy, is cascaded from large scales down to the Kolmogorov dissipation scale, where it is destroyed. Unfortunately, the relatively low Reynolds number of both LES does not permit the determination of quantitative scalings of energy and helicity spectra which could be compared to the Kolmogorov scalings in $\widehat{k}^{-5/3}$ in helical three-dimensional homogeneous isotropic turbulence, as mentioned by Borue and Orszag in \cite{borue97:_spect}. In the same paper, Borue and Orszag conclude that helicity is inherently a large-scale quantity which behaves similarly to a passive scalar. Consequently the one-dimensional relative helicity spectrum defined by
\begin{equation}
\alpha(\widehat{k})=\frac{\widehat{\cH}(\widehat{k})}{2\widehat{k}\,\widehat{\cK}(\widehat{k})},
\end{equation}
where $\widehat{\cH}$ (resp. $\widehat{\cK}$) is the one-dimensional resolved helicity (resp. energy) spectrum, decreases at small scales. Even if in our context, the turbulence is not homogeneous nor isotropic, the previous assertion is undeniably verified by both LES as can be seen only for the LES-DMM in Fig. \ref{fig:helicity_spectra}, for high values of $\widehat{k}$ corresponding to small scales. Similar relative helicity spectrum is obtained for the LES-DSM. This suggests that the decreasing trend at small scales of the relative helicity spectrum is more general and not only limited to the homogeneous and isotropic turbulence theoretical framework, just like the Kolmogorov scale in the inertial range.

%================ Conclusions =============================================================
\section{Conclusions}
\label{sec:conclusions}
The long-integration results of two LES of the lid-driven cubical cavity flow at a Reynolds number of $12\,000$ have been presented for two dynamic subgrid-scale models, namely a dynamic Smagorinsky model and a dynamic mixed model. These simulations were based on an accurate spectral-element spatial discretization, having two times less points per space direction than the direct numerical simulation reference result from Leriche and Gavrilakis \cite{leriche00:_direc}. All filtering levels introduced in both SGS modelings rely on explicit modal filters in the spectral space, retaining $\cC^0$-continuity of the numerical solution of the filtered Navier--Stokes equations. An additional nodal filter was used to stabilize both LES. Time-averaging was shown to be equivalent to ensemble-averaging, with respect to the global precision level of the numerical integration.

Partial simulation results using the UDNS and the Smagorinsky model as subgrid-scale models, have served to prove the necessity of a dynamic SGS procedure. Full LES results for both dynamic models have shown very good agreement with the DNS reference results. The agreement with the experimental reference results from Prasad and Koseff \cite{prasad89:_reynol} is qualitatively good.

At a Reynolds number of $12\,000$, the lid-driven cavity flow is placed in a locally turbulent regime and such turbulent flow is proved to be highly inhomogeneous in the secondary-corner regions of the cavity where turbulence production and dissipation are important. The maximum production of turbulence was found to be located in the downstream-corner-eddy region just above the bottom wall. An analysis of the spectra of turbulent quantities at this point allowed us to determine the distribution of the scales of the turbulent structures convected past this maximum. Moreover, both LES were able to capture the coherent counter-rotating pair of vortices which are mainly responsible for the peaks of turbulence production still at this point. LES-DSM have shown globally more intense and better results than the LES-DMM in this matter. 

Small-scales turbulent structures were located indirectly by studying the regions of intense turbulent energy dissipation rate $\eps$. The eddy-viscosity field was shown to be strongly correlated to $\eps$ in the turbulent areas of the flow, but the clipping procedure---necessary for stabilizing the numerics---imposed to the dynamic parameters strongly diminishes this correlation in the intense turbulent zones. Subgrid-scales activity has been analyzed and the higher SGS activity is associated with the LES-DMM.

Helical properties of the flow were investigated. Typical helical coherent structures were identified in the secondary-corner regions. These structures appear to be correlated to the turbulent energy dissipation rate $\eps$. The relative helicity spectra is shown to be decreasing at small scales, which is in agreement with the theoretical results from Borue and Orszag \cite{borue97:_spect} for the three-dimensional isotropic inhomogeneous turbulence.
%================ Acknowledgments==========================================================
\begin{acknowledgments}
The authors would like to thank Prof. Pierre Sagaut for fruitful discussions.

This research is being partially funded by a Swiss National Science Fundation Grant (No. 200020--101707), whose support is gratefully acknowledged. 

The DNS data were obtained on supercomputing facilities at the Swiss National Supercomputing Center CSCS and the LES data on Pleiades cluster at EPFL--ISE.
\end{acknowledgments}


\begin{thebibliography}{10}

\bibitem{shankar00:_fluid_mechan_driven_cavit}
P~N Shankar and M~D Deshpande.
\newblock Fluid {M}echanics in the {D}riven {C}avity.
\newblock {\em Annu. Rev. Fluid Mech.}, 32:93--136, 2000.

\bibitem{albensoeder05:_accur}
S~Albensoeder and H~C Kuhlmann.
\newblock Accurate three-dimensional lid-driven cavity flow.
\newblock {\em J. Comp. Phys.}, 206:536--558, 2005.

\bibitem{iwatsu89:_numer}
R~Iwatsu, K~Ishii, T~Kawamura, K~Kuwahara, and J~M Hyun.
\newblock Numerical simulation of three-dimensional flow structure in a driven
  cavity.
\newblock {\em Fluid. Dyn. Res.}, 5(3):173--189, 1989.

\bibitem{leriche99:_direc_cheby}
E~Leriche.
\newblock {\em Direct numerical simulation of lid-driven cavity flow by a
  {C}hebyshev spectral method}.
\newblock PhD thesis, no. 1932, \'Ecole {P}olytechnique {F}\'ed\'erale de
  {L}ausanne, 1999.

\bibitem{leriche00:_direc}
E~Leriche and S~Gavrilakis.
\newblock Direct numerical simulation of the flow in the lid-driven cubical
  cavity.
\newblock {\em Phys. Fluids.}, 12:1363--1376, 2000.

\bibitem{koseff84}
J~R Koseff and R~L Street.
\newblock The lid-driven cavity flow: A synthesis of qualitative and
  quantitative observations.
\newblock {\em J. Fluids Eng.-Transactions {ASME}}, 106:390--398, 1984.

\bibitem{prasad89:_reynol}
A~K Prasad and J~R Koseff.
\newblock {R}eynolds number and end-wall effects on a lid-driven cavity flow.
\newblock {\em Phys. Fluids.}, 1(2):208--218, 1989.

\bibitem{chen05:_tip4p}
C~K Chen and DTW Lin.
\newblock {T}{I}{P}4{P} potential for lid-driven cavity flow.
\newblock {\em Acta Mechanica}, 178:223--237, 2005.

\bibitem{he04:_bgk}
N-Z He, N-C Wang, B-C Shi, and Z-L Guo.
\newblock A unified incompressible lattice {B}{G}{K} model and its application
  to three-dimensional lid-driven cavity flow.
\newblock {\em Chin. Phys.}, 13:40--46, 2004.

\bibitem{zang92:_applic}
Y~Zang, R~L Street, and J~R Koseff.
\newblock Application of a dynamic subgrid-scale model to turbulent
  recirculating flows.
\newblock In {\em Annual Research Briefs}, volume~85. Center for Turbulence
  Research, Stanford University/{N}{A}{S}{A}--{A}mes, 1992.

\bibitem{zang93}
Y~Zang, R~L Street, and J~R Koseff.
\newblock A dynamic mixed subgrid-scale model and its application to turbulent
  recirculating flows.
\newblock {\em Phys. Fluids A}, 5:3186--3193, 1993.

\bibitem{sagaut03:_large}
P~Sagaut.
\newblock {\em Large eddy simulation for incompressible flows: an
  introduction}.
\newblock Springer, Berlin, 3rd edition, 2005.

\bibitem{smagorinsky63:_gener}
J~S Smagorinsky.
\newblock General circulation experiments with the primitive equations. {I}:
  The basic experiment.
\newblock {\em Month. Weath. Rev.}, 91:99--165, 1963.

\bibitem{germano91}
M~Germano, U~Piomelli, P~Moin, and W~H Cabot.
\newblock A dynamic subgrid-scale eddy viscosity model.
\newblock {\em Phys. Fluids A}, 3:1760--1765, 1991.

\bibitem{bouffanais05:_large}
R~Bouffanais, M~O Deville, P~F Fischer, E~Leriche, and D~Weill.
\newblock Large-eddy simulation of the lid-driven cubic cavity flow by the
  spectral element method.
\newblock {\em J. Sci. Comput.}, 27:151--162, 2006.

\bibitem{bardina83:_improv}
J~Bardina, J~H Ferziger, and W~C Reynolds.
\newblock Improved turbulence models based on large eddy simulation of
  homogeneous, incompressible, turbulent flows.
\newblock Technical Report TF-19, Thermal Sciences Division, Department of
  Mechanical Engineering, Stanford University, Stanford, 1983.

\bibitem{liu94}
S~Liu, C~Meneveau, and J~Katz.
\newblock On the properties of similarity subgrid-scale models as deduced from
  measurements in a turbulent jet.
\newblock {\em J. Fluid Mech.}, 275:83--119, 1994.

\bibitem{vreman94}
B~Vreman, B~Geurts, and H~Kuerten.
\newblock On the formulation of the dynamic mixed subgrid-scale model.
\newblock {\em Phys. Fluids.}, 6:4057--4059, 1994.

\bibitem{salvetti97}
M~V Salvetti and S~Banerjee.
\newblock A priori tests of a new dynamic subgrid-scale model for
  finite-difference large eddy simulations.
\newblock {\em Phys. Fluids.}, 9:2831--2847, 1997.

\bibitem{horiuti97}
K~Horiuti.
\newblock A new dynamic two-parameter mixed model for large-eddy simulation.
\newblock {\em Phys. Fluids.}, 9:3443--3464, 1997.

\bibitem{morinishi01}
Y~Morinishi and O~V Vasilyev.
\newblock A recommended modification to the dynamic two-parameter mixed subgrid
  scale model for large eddy simulation of wall bounded turbulent flow.
\newblock {\em Phys. Fluids.}, 13:3400--3410, 2001.

\bibitem{vreman97:_large}
B~Vreman, B~Geurts, and H~Kuerten.
\newblock Large-eddy simulation of the turbulent mixing layer.
\newblock {\em J. Fluid Mech.}, 339:357--390, 1997.

\bibitem{winckelmans01:_explic_smagor}
G~S Winckelmans, A~A Wray, O~V Vasilyev, and H~Jeanmart.
\newblock Explicit-filtering large-eddy simulation using the tensor-diffusivity
  model supplemented by a dynamic {S}magorinsky term.
\newblock {\em Phys. Fluids.}, 13:1385--1403, 2001.

\bibitem{ghosal96}
S~Ghosal.
\newblock An analysis of numerical error in large-eddy simulation of
  turbulence.
\newblock {\em J. Comp. Phys.}, 125:187--206, 1996.

\bibitem{germano86:_navier}
M~Germano.
\newblock A proposal for a redefinition of the turbulent stresses in the
  filtered {N}avier--{S}tokes equations.
\newblock {\em Phys. Fluids.}, 29:2323--2324, 1986.

\bibitem{deville02:_high}
M~O Deville, P~F Fischer, and E~H Mund.
\newblock {\em High-order methods for incompressible fluid flow}.
\newblock Cambridge University Press, Cambridge, 2002.

\bibitem{maday89:_spect_navier_stokes}
Y~Maday and A~T Patera.
\newblock {\em Spectral element methods for the incompressible
  {N}avier--{S}tokes equations}, pages 71--142.
\newblock State-of-the-Art Survey on Computational Mechanics, A. K. Noor \& J.
  T. Oden Eds. ASME, New-York, 1989.

\bibitem{maday92:_nimes_n_stokes}
Y~Maday, A~T Patera, and E~M R{\o}nquist.
\newblock The $\mathbb{P}_{N}\times\mathbb{P}_{N-2}$ method for the
  approximation of the {S}tokes problem.
\newblock Technical Report 92009, Department of Mechanical Engineering, MIT,
  Cambridge, MA, 1992.

\bibitem{blackburn03:_spect}
H~M Blackburn and S~Schmidt.
\newblock Spectral element filtering techniques for large eddy simulation with
  dynamic estimation.
\newblock {\em J. Comp. Phys.}, 186:610--629, 2003.

\bibitem{karamanos99:_large}
G~S Karamanos, S~J Sherwin, and J~F Morrison.
\newblock {\em Large eddy simulation using unstructured spectral/hp elements}.
\newblock in Recent Advances in DNS and LES, D. Knight \& L. Sakell. Kluwer
  Academic, Dordrecht/Norwell, MA, 1999.

\bibitem{perot93}
J~B Perot.
\newblock An analysis of the fractional step method.
\newblock {\em J. Comp. Phys.}, 108:51--58, 1993.

\bibitem{gomilko03:_stokes}
A~M Gomilko, V~S Malyuga, and V~V Meleshko.
\newblock On steady {S}tokes flow in a trihedral rectangular corner.
\newblock {\em J. Fluid Mech.}, 476:159--177, 2003.

\bibitem{fischer01:_filter_based_stabil_spect_elemen_method}
P~F Fischer and J~S Mullen.
\newblock Filter-based stabilization of spectral element methods.
\newblock {\em Comptes {R}endus de l'{A}cad{\'e}mie des {S}ciences {P}aris},
  332(I):265--270, 2001.
\newblock {A}nalyse {N}um{\'e}rique.

\bibitem{boyd98:_two_cheby_legen}
J~P Boyd.
\newblock Two comments on filtering (artificial viscosity) for {C}hebyshev and
  {L}egendre spectral and spectral element methods: Preserving boundary
  conditions and interpretation of the filter as a diffusion.
\newblock {\em J. Comp. Phys.}, 143:283--288, 1998.

\bibitem{karamanos00:_spect_vanis_viscos_method_large_eddy_simul}
G~S Karamanos and G~E Karniadakis.
\newblock A spectral vanishing viscosity method for large-eddy simulations.
\newblock {\em J. Comp. Phys.}, 163:22--50, 2000.

\bibitem{leriche05:_direc_numer_simul_lid_driven}
E~Leriche.
\newblock {Direct numerical simulation of lid driven cavity at high Reynolds
  numbers}.
\newblock {\em J. Sci. Comput.}, 27:335--345, 2006.

\bibitem{prasad:}
A~K Prasad.
\newblock Personal communication.
\newblock Additional experimental data.

\bibitem{kravchenko97}
A~G Kravchenko and P~Moin.
\newblock On the effect of numerical errors in large eddy simulations of
  turbulent flows.
\newblock {\em J. Comp. Phys.}, 131:310--322, 1997.

\bibitem{breuer98:_large}
M~Breuer.
\newblock Large eddy simulation of the subcritical flow past a cylinder:
  numerical and modeling aspects.
\newblock {\em Int. J. Numer. Methods Fluids}, 28:1281--1302, 1998.

\bibitem{ghosal95}
S~Ghosal, T~S Lund, P~Moin, and K~Akselvoll.
\newblock A dynamic localization model for large-eddy simulation of turbulent
  flows.
\newblock {\em J. Fluid Mech.}, 286:229--255, 1995.

\bibitem{meneveau96:_lagran}
C~Meneveau, T~S Lund, and W~H Cabot.
\newblock A {L}agrangian dynamic subgrid-scale model of turbulence.
\newblock {\em J. Fluid Mech.}, 319:353--385, 1996.

\bibitem{mathieu00:_introd_turbul_flow}
J~Mathieu and J~Scott.
\newblock {\em {A}n {I}ntroduction to {T}urbulent {F}low}.
\newblock Cambridge University Press, Cambridge, 2000.

\bibitem{pope00:_turbul_flows}
S~B Pope.
\newblock {\em Turbulent {F}lows}.
\newblock Cambridge University Press, Cambridge, 2000.

\bibitem{geurts02}
B~J Geurts and J~Fr{\"o}lich.
\newblock A framework for predicting accuracy limitations in large-eddy
  simulation.
\newblock {\em Phys. Fluids.}, 14:41--44, 2002.

\bibitem{meyers05:_optim}
J~Meyers, B~J Geurts, and M~Baelmans.
\newblock Optimality of the dynamic procedure for large-eddy simulations.
\newblock {\em Phys. Fluids.}, 17:Art. 045108, 2005.

\bibitem{moffat92:_helic}
H~K Moffat and A~Tsinober.
\newblock Helicity in laminar and turbulent flow.
\newblock {\em Annu. Rev. Fluid Mech.}, 24:281--312, 1992.

\bibitem{tsinober83}
A~Tsinober and E~Levich.
\newblock On the helical nature of three dimensional coherent structures in
  turbulent flows.
\newblock {\em Phys. Lett.}, 99:321--323, 1983.

\bibitem{borue97:_spect}
V~Borue and S~A Orszag.
\newblock Spectra in helical three-dimensional homogeneous isotropic
  turbulence.
\newblock {\em Phys. Rev. {E}}, 55(6):7005--7009, 1997.

\end{thebibliography}
\end{document}